\documentclass[useAMS,usenatbib]{mn2e}
\usepackage{amsmath}
\usepackage{amssymb}
\usepackage{multirow}
\usepackage{graphicx}
\usepackage{epstopdf}
\usepackage{caption}
\usepackage{float}
\usepackage{subcaption}

\def\simlt{\lower.5ex\hbox{$\; \buildrel < \over \sim \;$}}
\def\simgt{\lower.5ex\hbox{$\; \buildrel > \over \sim \;$}}

\newcommand{\bd}{\begin{displaymath}}
\newcommand{\ed}{\end{displaymath}}
\newcommand{\be}{\begin{equation}}
\newcommand{\ee}{\end{equation}}
\newcommand{\beqa}{\begin{eqnarray}}
\newcommand{\eeqa}{\end{eqnarray}}

\newcommand{\Lya}{Ly$\alpha$}

\title[Charting Parameter Space of 21-cm Power Spectrum] {Charting the Parameter Space of the 21-cm Power Spectrum}
\author[Cohen et al.] {Aviad Cohen$^{1}$\thanks{E-mail:
    aviadc11@gmail.com}, Anastasia Fialkov$^{2}$,
  Rennan Barkana$^{1,3,4,5}$\\
  $^{1}$ Raymond and Beverly Sackler School of Physics and Astronomy,
  Tel Aviv University, Tel Aviv 69978, Israel\\
  $^{2}$ Department of Astronomy, Harvard University, 60 Garden Street, MS−51, Cambridge, MA, 02138 U.S.A.\\
  $^{3}$ Sorbonne Universit\'{e}s, Institut Lagrange de
  Paris (ILP), Institut d'Astrophysique de Paris, UPMC Univ Paris 06/CNRS\\
  $^{4}$ Department of Astrophysics, University of Oxford, Denys Wilkinson Building, Keble Road, Oxford OX1 3RH, UK\\ $^5$ Perimeter Institute for
        Theoretical Physics, 31 Caroline St N., Waterloo, ON N2L 2Y5,
        Canada
  }

\begin{document}
\pagerange{\pageref{firstpage}--\pageref{lastpage}} \pubyear{2017}
\maketitle

\label{firstpage}

\begin{abstract} 
The high-redshift 21-cm signal of neutral hydrogen is expected to be observed within the next decade and will reveal epochs of cosmic evolution that have been previously inaccessible. Due to the lack of observations, many of the astrophysical processes that took place at early times are poorly constrained. In recent work we explored the astrophysical parameter space and the resulting large variety of possible global (sky-averaged) 21-cm signals. Here we extend our analysis to the fluctuations in the 21-cm signal, accounting for those introduced by density and velocity, Ly$\alpha$ radiation, X-ray heating, and ionization. While the radiation sources are usually highlighted, we find that in many cases the density fluctuations play a significant role at intermediate redshifts. Using both the power spectrum and its slope, we show that properties of high-redshift sources can be extracted from the observable features of the fluctuation pattern. For instance, the peak amplitude of ionization fluctuations can be used to estimate whether heating occurred early or late and, in the early case, to also deduce the cosmic mean ionized fraction at that time. The slope of the power spectrum has a more universal redshift evolution than the power spectrum itself and can thus be used more easily as a tracer of high-redshift astrophysics. Its peaks can be used, for example, to estimate the redshift of the Ly$\alpha$ coupling transition and the redshift of the heating transition (and the mean gas temperature at that time). We also show that a tight correlation is predicted between features of the power spectrum and of the global signal, potentially yielding important consistency checks.
\end{abstract}


\section{Introduction}
\label{Sec:Intro}

The most promising probe of the early universe and the epoch of
primordial star formation (cosmic dawn) is the redshifted spectral
line of atomic hydrogen, which has a rest-frame wavelength of
21~cm. This signal is expected to be produced prior to complete
reionization by abundant neutral hydrogen in the intergalactic medium
(IGM) and should allow us to explore cosmic history down to
$z\sim6$. Because the signal depends both on the cosmological model
and on astrophysics, it contains abundant information about the
Universe at early times
\citep[e.g.,][]{Furlanettoetal:2006,Barkana:2016}.

Observational efforts to date have resulted in limits on both the
global 21-cm signal and the power spectrum of 21-cm
fluctuations. Experiments that are currently taking (or analyzing)
data to detect the power spectrum from the epoch of reionization (EoR)
include the Low Frequency Array \citep[LOFAR,][]{Patil:2017}, the
Precision Array to Probe the Epoch of Reionization
\citep[PAPER,][]{Parsons:2014,Jacobs:2015,Ali:2015}, the Murchison
Wide-field Array \citep[MWA,][]{Bowman:2013,Beardsley:2016}, and the
Hydrogen Epoch of Reionization Array
\citep[HERA,][]{Pober:2014,DeBoer:2016}; meanwhile, an effort is being
made to measure the global signal at both the low-redshift (EoR) and
high-redshift (cosmic dawn) regimes, by EDGES \citep[the Experiment to
  Detect the Global EoR Signature,][]{Bowman:2010,Monsalve:2017},
SARAS \citep[the Shaped Antenna measurement of the background RAdio
  Spectrum,][]{Singh:2017}, and LEDA \citep[Large-Aperture Experiment
  to Detect the Dark Ages,][]{Bernardi:2016}. Planned experiments (or
under construction) include the New Extension in Nancay Upgrading
LOFAR \citep[NenuFAR,][]{Zarka:2012}, the Square Kilometer Array
\citep[SKA,][]{Koopmans:2015}, and the Dark Ages Radio Explorer
\citep[DARE,][]{Burns:2015}; these instruments will probe both the
global signal and its power spectrum over a wide range of epochs.

Observing the 21-cm signal will shed light on the astrophysical
processes that shaped the young Universe. One of the most important
questions that can be answered by detecting this signal is how some of
the first stars came to be. The abundance and spatial distribution of
high redshift galaxies determines the pattern of the radiative
backgrounds produced by sources of light, and thus has a strong effect
on the fluctuations in the 21-cm signal. A major parameter in this is
the minimum mass of star forming halos, $M_{\rm min}$, which is
usually set by the requirement of efficient gas cooling. Because they
are formed in highly over-dense regions, massive halos are rarer
and more highly clustered than lighter ones \citep{Barkana:2004}, and
imprint stronger fluctuations in the signal. Theoretical work shows
that star formation in a dark matter halo becomes possible only if the
halo is massive enough to radiatively cool the infalling gas
\citep[e.g.,][]{Tegmark:1997}. For instance, molecular hydrogen,
H$_2$, one of the available building blocks in the pristine
environment at high redshifts, has to be gravitationally accelerated
and shock-heated to temperatures higher than $\sim300$ K in order to
initiate radiative cooling, which leads to star formation in halos
above $M_{\rm min} \sim 10^5\,$M$_{\odot}$
\citep[e.g.,][]{Tegmark:1997,Bromm:2002,Yoshida:2003}. If molecular
hydrogen is unavailable, stars will form via cooling of atomic
hydrogen in more massive halos, above $M_{\rm
  min}\sim10^7\,$M$_{\odot}$ \citep[e.g.,][]{Barkana:2001}.

 Formation of stars through cooling of H$_2$ is sensitive to various
 feedback processes. In particular, radiation emitted by stars
 includes photons in the Lyman-Werner (LW) band (11.2-13.6 eV), which
 dissociate hydrogen molecules \citep{Haiman:1997} thus boosting the
 minimal mass of star forming halos
 \citep{Haiman:2000,Machacek:2001,Wise:2007,O'Shea:2008}.  The
 efficiency of this feedback mechanism, and thus the transition point
 of star formation from molecular cooling to atomic cooling, are
 highly uncertain \citep{Visbal:2014,Schauer:2015}. In addition to the
 LW feedback, star formation in small halos is sensitive to other
 factors, such as the relative streaming velocity between dark matter
 and gas, which suppresses star formation in a spatially inhomogeneous
 way in halos below $\sim10^6\,$M$_{\odot}$
 \citep{Tseliakhovich:2010,Dalal:2010,Tseliakhovich:2011,Visbal:2012,
   Fialkov:2014a}. Another possible way to quench star formation in
 small halos is via supernova explosions that can expel gas from light
 halos, raising $M_{\rm min}$ well above the atomic cooling threshold
 \citep[e.g.,][]{Wyithe:2013}. On the other hand, supernova explosions
 can also revive star formation in light halos by enriching gas with
 metals. Because metal-rich gas can allow a lower $M_{\rm min}$ than
 even the H$_2$ cooling channel, small halos may contribute to star
 formation at high redshifts via metal-line cooling despite the effect
 of the LW feedback. Current numerical simulations suggest that star
 formation is likely to be inefficient in small halos, but whether
 metal cooling contributes significantly to high-redshift star
 formation is still uncertain
 \citep[e.g.,][]{Jeon:2014,Wise:2014,O'Shea:2015,Cohen:2016a}. Finally,
 when the gas in the IGM is photoheated above $10^4$ K by ionizing
 photons, photoheating feedback becomes efficient and gas stops
 accreting onto halos below $10^8-10^9$M$_{\odot}$
 \citep[e.g.,][]{R86,WHK97,NS00,Sobacchi:2013,Cohen:2016a}. However,
 this process becomes significant only at relatively low redshifts,
 during the advanced stages of reionization. Future 21-cm measurements
 will be able to constrain the cooling channel and the efficiency of
 primordial star formation.

Stars and their remnants produce radiative backgrounds that strongly
affect the environment. For instance, the temperature of the IGM rises
due to the X-ray radiation produced by the first heating sources. Even
fixing the total energy emitted in X-rays, different spectral energy
distributions (SEDs) can lead to completely different heating
histories and, thus, predicted 21-cm signals \citep{Fialkov:2014b,
  Pacucci, Fialkov:2014c}. Because the mean free path is larger for
high-energy photons, hard photons travel further away from the source
before depositing their energy into the IGM. This leads to a delayed,
weaker (due to redshift losses), and more spatially uniform heating of
the universe in the case of sources with a hard SED. The temperature
of the gas directly affects the 21-cm intensity. As a result, X-ray
sources imprint their signature in the signal from neutral
hydrogen. By detecting the characteristic signature, the nature and
distribution of the first X-ray sources can be studied. At present,
the high-redshift X-ray population is poorly constrained; however,
available observations yield upper and lower limits on the X-ray
luminosity of the sources \citep{Fialkov:2016}. Upper limits come from
the unresolved soft X-ray background \citep{Dikstra:2012,
  Fialkov:2016}, while lower limits can be extracted from the upper
limits on the 21-cm power spectrum \citep{Ali:2015,Pober:2015}.

There are several candidate X-ray sources discussed in the literature,
and at present it is unclear which one dominates at high
redshift. High-mass X-ray binaries (XRBs) are currently the most
plausible dominant source \citep{Mirabel:2011}. This possibility is
supported by a detailed population synthesis simulation that indicated
that XRBs dominate over the contribution of quasars at $z \ga 6-8$
\citep{Fragos:2013}. Although the majority of X-ray photons emitted by
XRBs are hard, with the X-ray SED peaking around $1-3$ keV, the SED
used in previous models has often been a soft power law
\citep[e.g.,][]{Furlanetto:2006}, which might be more appropriate for
describing emission by gas heated by supernova explosions within
galaxies. The third possible source of X-ray radiation is a population
of mini-quasars (MQ), central black holes in early star-forming halos
\citep{Madau:2004}. The properties of these objects, including their
contribution to heating and reionization, are highly uncertain, since
their masses and the characteristics of their host galaxies are very
different from those of the central black holes observed today in much
more massive halos. Because it takes time for a supermassive black
hole to grow, mini-quasars tend to be significant only in relatively
massive halos and, therefore, in our models their heating becomes
important at lower redshifts than the other two options. The
bolometric luminosity of X-ray sources, $L_X$, is another free
parameter. Both population synthesis simulations \citep{Fragos:2013}
and observations \citep{Brorby:2016} suggest that in the case of XRBs,
$L_X$ is larger in metal-poor high-redshift galaxies compared to their
metal-rich low-redshift counterparts. The same studies also indicate
that the luminosity is proportional to the star formation rate (SFR),
allowing us to define a $L_X$-SFR relation. However, the exact
relation between $L_X$ and the SFR depends on the nature of the
sources, and it is highly uncertain, especially at high redshifts
where there are few observational constraints.
 
In addition to X-ray radiation, the first luminous objects emitted
ultraviolet (UV) radiation, which ionized the neutral hydrogen in the
IGM.  Thanks to the effect of ionized gas on the CMB, the value of
this parameter is relatively well constrained (compared to the other
parameters that we consider), with a total scattering optical depth of
$\tau = 0.055 \pm 0.009$ \citep{Planck:2016b}.  The main sources of
ionizing radiation are believed to be stars; however, reionization by
quasars may also be possible \citep{Madau:2015}.

In previous work \citep{Cohen:2016b}, we explored the space of
astrophysical parameters, varying the parameters that play crucial
roles in driving the 21-cm signal, namely: the minimal mass of
star-forming halos and the star formation efficiency (which reflect
the dominant cooling channel and the efficiency of internal feedback),
the parameters of X-ray heating (the SED and the bolometric luminosity
of X-ray sources), and the total reionization optical depth and
maximum mean free path of ionizing photons. In that work, we used our
simulations to predict the global 21-cm signal (i.e., the mean
spectrum over the relevant frequency range), for 193 different
combinations of the astrophysical parameters. We showed that the
expected signal fills a large parameter space, but with a fixed
general shape of the global 21-cm signal. Using the 193 models we
identified relations between features of the spectrum and the
astrophysical parameters. Since we showed these relations to hold over
a very wide range of possible astrophysics parameters, these relations
can be used to directly link future measurements of the global signal
to astrophysical quantities in a mostly model-independent way. This
approach is novel in that it covers a substantially wider
astrophysical parameter space than other work
\citep[e.g.,][]{Greig:2015,Hassan:2016,Shimabukuro:2017}. For example,
such works often focus on the late stages of reionization assuming
that cosmic heating has saturated and thus the parameters of the X-ray
sources do not matter. They also usually assume that small halos,
those below the atomic cooling threshold, do not contribute
significant star formation. Such simplifying assumptions are often
made because algorithms such as the Monte Carlo Markov Chain analysis
tool \citep{Greig:2015} need to run the simulation many times before
they succeed to fill up the required parameter space. However, given
the current lack of observational constraints, the space of
possibilities is still extremely wide. For example, the relative
timing between reionization and the heating era is uncertain. It has
been recently recognized that heating occurs late in many scenarios,
and the cosmic gas can still be colder than the CMB during the early
stages of reionization \citep{Fialkov:2014b, Madau:2016, Mirocha:2016,
  Cohen:2016b}.

This paper is a follow up to \citet{Cohen:2016b}. Here we use the same
compilation of models to map out the space of possible 21-cm power
spectra. We explore features of the 21-cm power spectrum and aim to
classify the main observable properties of the evolution in redshift
of the power spectrum. Another goal of ours is to establish relations
between these features and the astrophysical parameters. This paper is
organized as follows: In Section~\ref{Sec:Methods} we discuss the
general properties of the 21-cm power spectrum
(Section~\ref{Sec:MethodsPS}), and then outline and discuss the
astrophysical parameter space (Section~\ref{Sec:Param}). In
Section~\ref{Sec:results} we illustrate our predictions in detail for
a particular choice of astrophysical parameters, while in
Section~\ref{sec:entire} we generalize our results to the entire
parameter space, focusing on correlations between the redshift
evolution of the power spectrum (and that of its slope) and the
properties of early galaxies. In Section~\ref{Sec:consist} we show
that the timing of cosmic milestones (such as the redshift at which
the IGM was heated to the temperature of the CMB) can be extracted
from the evolution of the slope, as well as from the spectral shape of
the global 21-cm signal. In the same Section we also provide
consistency relations that could be used to verify experimental
results once the measurements of both the global signal and the power
spectrum are available. Finally, we summarize and conclude in
Section~\ref{Sec:sum}.

\section{Simulated 21-cm Signal}
\label{Sec:Methods}

Our goal is to explore the high-redshift astrophysical parameter space
and create a mock 21-cm signal for a large number of parameter
sets. To this end we use a flexible and fast semi-numerical method
first introduced by \citet{Visbal:2012}, inspired by 21cmfast
\citep{Mesinger:2011}. The framework follows the evolution of the
density and velocity fields in time in a large cosmological volume (a
$384^3$ Mpc$^3$ box) with coarse resolution (3 Mpc), and extensively
uses sub-grid models to implement physics on smaller scales
\citep{Fialkov:2012, Visbal:2012, Fialkov:2013, Fialkov:2014c,
  Fialkov:2014b, Fialkov:2014d, Cohen:2016a}. The star formation rate
in each cell, at each redshift and in each halo mass bin is computed
using the extended Press-Shechter formalism
\citep{Barkana:2004}. Assuming population II star formation
\citep{Barkana:2005b} and properly accounting for time delay effects,
the simulation calculates various radiative backgrounds created by
stars and their remnants, including the Ly$\alpha$ background which is
needed to source the Wouthuysen-Field (WF) coupling \citep{Wouthuysen,
  Field}, LW radiation responsible for radiative feedback, X-rays that
heat the gas, and ionizing UV radiation. This simulation takes into
account the effect of relative streaming velocity between dark and
baryonic matter \citep{Tseliakhovich:2010, Visbal:2012}, and the
photoheating feedback \citep{Cohen:2016a}. While the code is inspired by 
 21cmfast \citep{Mesinger:2011}, it goes beyond it with more
accurate X-ray heating (including the effect of local reionization),
\Lya\ fluctuations (approximately including the effect of multiple
scattering), and photoheating feedback, plus the possibility of having
substantial star formation in halos below the atomic cooling
threshold, in which case spatially-inhomogeneous processes such as the
streaming velocity and LW feedback play a key role (and are included
in our 21-cm code but not in others).

\subsection{The 21-cm power spectrum}
\label{Sec:MethodsPS}

An output of the simulation is the inhomogeneous 21-cm signal
calculated for every cell in the redshift range $6-50$. The
brightness temperature observed against the CMB is
\begin{equation}
\begin{split}
\label{eq:signal}
T_{\rm b}=26.8\, x_{\rm HI}\left( \frac{1+z}{10}\right)
^{1/2}\left(1+\delta \right) \frac{x_{\rm tot}}{1+x_{\rm
    tot}}\left[1-\frac{T_{\rm CMB}}{T_{\rm gas}} \right] \rm mK\ ,
\end{split}
\end{equation}
where $x_{\rm HI}$ is the neutral hydrogen fraction, $\delta$ is the
matter density contrast, $x_{\rm tot}=x_c+x_\alpha$ is the sum of the
coupling coefficients which includes collisional coupling ($x_c$) and
the WF coupling due to Ly$\alpha$ photons ($x_\alpha$), $T_{\rm CMB}$
is the CMB temperature and $T_{\rm gas}$ is the (kinetic) gas
temperature. There are two other effects that we include in our
simulations but omit from Eq.~(\ref{eq:signal}) in order to simplify
the discussion here. One is the effect of peculiar velocities
\citep{Bharadwaj:2004,Barkana:2005a}; while they produce line-of-sight
anisotropy that can be used for model-independent inferences of early
cosmic history \citep{2015PRL}, in this paper we focus on the more
easily measured spherically-averaged 21-cm power spectrum. The other
effect is that of low-temperature corrections to Ly-a scattering
\citep{ChSh06,Chen,Hirata,PritF,Barkana:2016}.
  
Eq.~(\ref{eq:signal}) contains four different terms that can source
fluctuations in the total brightness temperature:
\begin{enumerate}
\item Fluctuations in the matter density affect the signal via the
  matter density contrast, $\delta$, defined as
  $\delta=\rho/\bar{\rho}-1$, where $\rho$ is the local density and
  $\bar{\rho}$ is the mean density; they also produce peculiar
  velocity fluctuations, determined by the density fluctuations
  through the continuity equation. Thus, we usually show the sum of
  the two contributions and denote it $\delta+v$; the variance of this
  sum is 28/15 times that of $\delta$ alone (but the cross-correlation
  with other 21-cm fluctuations is more complicated).
\item The term $x_{\rm tot}/(1+x_{\rm tot})$ depends on the total
  coupling coefficient and fluctuates due to inhomogeneous collisions
  at the highest redshifts and the non-uniform production of
  Ly$\alpha$ photons at later times.
\item The term $1-T_{CMB}/T_{\rm gas}$ varies due to inhomogeneous
  heating.
\item The inhomogeneous process of reionization is encoded in the
  neutral fraction, $x_{\rm HI}$.
\end{enumerate}
As with the definition of $\delta$, we define contrasts for each one
of the other three terms, $\delta_{\rm coup}$, $\delta_{\rm heat}$ and
$\delta_{\rm ion}$, respectively. To first order, different sources of
fluctuations are additive (although they may have different signs),
and at each redshift the contrast in brightness temperature is
approximately
\begin{equation}
\label{eq:deltas}
\delta_{T_b} \approx (\delta+v) + \delta_{\rm coup} + \delta_{\rm
  heat} +\delta_{\rm ion}\ .
\end{equation}
In reality there are also non-linear terms, and cross-correlations
contribute to the power spectrum, but it is useful to look at the
separate contributions of the sources of fluctuations in order to
understand which of them dominate at any given time.

It is often more convenient to discuss fluctuations in Fourier (rather
than real) space, in terms of the comoving wavenumber, $k$, which is
inversely proportional to the comoving scale. The total power spectrum
$P_{T_b}(k)$ is defined by:
\begin{equation}
\label{eq:Pk}
\left\langle\tilde{\delta}_{T_b}({\bf k})
\tilde{\delta}^{*}_{T_b}({\bf k^\prime}) \right\rangle=\left(
2\pi\right)^3 \delta_D({\bf k}-{\bf k^\prime})P_{T_b}(k)\ ,
\end{equation}
where $\tilde{\delta}_{T_b}(\textbf{k})$ is the Fourier transform of
$\delta_{T_b}$, $\textbf{k}$ is the comoving wavevector, $\delta_D$
is the Dirac delta function, and angle brackets denote the ensemble
(or spatial) average. Finally, we use the convention of expressing the
power spectrum in terms of the variance in mK$^2$ units:
\begin{equation}
\label{eq:Del}
\Delta^2=\left\langle T_b\right\rangle ^2\frac{k^3P_{T_b}(k)}{2\pi^2}\ ,
\end{equation}
where the expression $k^3P_{T_b}(k)/2\pi^2$ is dimensionless. Because
the universe is homogeneous and isotropic on large cosmological
scales, the fluctuations tend to decrease with increasing scale
(decreasing wavenumber).

For a typical set of astrophysical parameters, fluctuations in the
21-cm signal on large cosmological scales, e.g., $k\sim$0.1
Mpc$^{-1}$, and in the redshift range $5\lesssim z\lesssim 35$ exhibit
three distinct peaks
\citep[e.g.,][]{Barkana:2005b,Pritchard:2007,Pritchard:2008,Fialkov:2014c}.
Prior to significant star formation, the 21-cm brightness temperature
is driven by interatomic collisions and interactions with the CMB
photons. While collisions dominate at the highest redshifts, driving
the 21-cm brightness temperature to the kinetic temperature of the
IGM, thermal equilibrium with the CMB takes over once the universe
expands enough to render collisions inefficient.  When the first stars
form in rare peaks of the density field and create an inhomogeneous
Ly$\alpha$ background, WF coupling becomes efficient. Fluctuations are
induced by the Ly-$\alpha$ peak around $z\sim25$ (when $x_{\rm tot}
\sim x_\alpha \sim 1$) and disappear once the coupling saturates
(i.e., $x_{\rm tot} \sim x_\alpha\gg1$). At the same time, the X-ray
background builds up, leading to an increase in the temperature of the
gas. This non-uniform heating creates fluctuations with a peak power
at redshift $\sim15$. When heating becomes saturated ($T_{\rm gas}\gg
T_{\rm CMB}$) the signal no longer depends on the gas
temperature. Fluctuations at low redshift ($z\sim 10$) are dominated
by patchy reionization. However, as we show below, this overall,
standard picture is not universal, and the power spectrum as a
function of redshift can have various numbers of peaks (between one
and three) depending on the scale and on the particular choice of
astrophysical parameters.

\subsection{The parameter space}
\label{Sec:Param}

Existing observational evidence and theoretical arguments place very
weak constraints on the astrophysical properties of the first luminous
objects, which translates into a large uncertainty in the predicted
21-cm signal. For instance, \citet{Cohen:2016b} showed that the
current astrophysical parameter space yields global 21-cm spectra with
the depth of the absorption trough feature anywhere in the
$-250$~mK~$\lesssim T_b\lesssim -25$~mK range. Here we explore the
implications of this large parameter space for the 21-cm power
spectrum.

We first review the relevant parameters \citep[same as in][where
  complete details are given]{Cohen:2016b} and quote the range within
which they are allowed to vary:
\begin{itemize}
\item The \textbf{star formation efficiency} is the fraction of gas in
  dark matter halos that is converted into stars, $f_*$. In general,
  this fraction depends on halo mass and the cooling channel through
  which stars form. The efficiency measured in numerical simulations
  shows a large scatter, especially for the light halos that dominate
  the early universe \citep{Wise:2014,O'Shea:2015,Xu:2016}. Our
  parameter $f_*$ is actually the star formation efficiency in large
  halos (above the atomic cooling threshold), where for smaller halos
  we consider two different dependencies of $f_*$ on the halo mass:
  (i) a constant down to the minimum mass and (ii) a gradual low-mass
  cutoff. We vary the star formation efficiency from 0.5\% to 50\%,
  with 5\% being our fiducial value.
\item The \textbf{minimum halo mass for star formation}, $M_{\rm
  min}$, discussed earlier in the Introduction, can be expressed in
  terms of the circular velocity, $V_c$. Here we entertain a few
  possibilities: star formation in molecular cooling halos (i.e., down
  to $V_c=4.2$ km s$^{-1}$, affected by LW feedback and the streaming
  velocity), atomic cooling halos ($V_c=16.5$ km s$^{-1}$), via metal
  line cooling ($V_c=4.2$ km s$^{-1}$ without LW feedback but with the
  streaming velocity), or, finally, in massive or super-massive halos
  (e.g., due to strong supernovae feedback; a minimum $V_c=35.5$ or
  76.5 km s$^{-1}$, respectively). We note that since both cooling and
  internal feedback depend on the depth of the potential, which is
  measured by $V_c$, it is more physical to assume a fixed $V_c$ with
  redshift rather than a fixed $M_{\rm min}$.
	\item In order to accommodate the variety of X-ray sources we
          use two quite different \textbf{X-ray SEDs} that bracket a
          large range: (i) the commonly used soft power-law spectrum
          \citep{Furlanetto:2006}, and (ii) a hard spectrum that
          corresponds to XRBs \citep{Fragos:2013,Fialkov:2014b}.  In
          addition, we consider mini-quasars \citep[sources expected
            to have a hard SED similar to that of XRBs,
            e.g.,][]{Tanaka:2012}. The evolution in time of a
          population of mini-quasars is different from X-ray binaries,
          i.e., with a much later build-up of the X-ray intensity,
          since the X-ray luminosity has an extra dependence on the
          halo mass (assuming a similar relation between black hole
          and halo mass as observed at low redshift).
	\item The {\bf X-ray efficiency}, $f_X$, accounts for the
          uncertain normalization of the $L_X$-SFR relation. We define
          $L_X/SFR = 3\times 10^{40} f_X$ [erg s$^{-1}$ M$^{-1}_\odot$
            yr] for the cases of XRBs or the soft SED
          \citep{Fialkov:2014b}.  For the mini-quasars the $L_X$-SFR
          relationship depends on both the redshift and halo mass. Our
          fiducial value is $f_X=1$, but we vary it between zero and a
          few hundred. A negligible amount of X-rays is not yet ruled
          out for some models, while the upper limit (which is
          model-dependent) is determined by saturating the unresolved
          soft X-ray background observed by {\it Chandra} in the
          $0.5-2$ keV band \citep{Lehmer:2012, Fialkov:2016}.
	\item Finally, we vary the \textbf{CMB optical depth},
          $\tau$. Our fiducial value is $\tau=0.066$, which
          corresponds to the most recent measurement of the optical
          depth when we began this project \citep{Planck:2015}. Given
          the rather large uncertainty in the measured value, we also
          consider higher values of $\tau$, including values of
          $\tau>0.09$ that now seem unlikely. To get the desired
          optical depth we vary the ionizing efficiency for each case
          as described in \citet{Cohen:2016b}.
\end{itemize}
As we will see below, the shape and redshift evolution of the power
spectrum varies greatly among various sets of plausible astrophysical
parameters. For instance, the difference between two otherwise
identical models but with a hard or soft X-ray SED, as discussed in
the next Section, is dramatic \citep[see also][]{Fialkov:2014b,
  Fialkov:2014c}. In order to fully explore the effect of these
parameters on the expected 21-cm signal we run our simulation for 193
different combinations of the five parameters described above (the
full list of cases appears in Appendix~\ref{appA}). Except for a few
cases, all the considered parameters are well within the limits
established by recent 21-cm power spectrum data
\citep{Ali:2015,Pober:2015} and CMB data \citep{Planck:2015}, and do
not saturate the unresolved soft X-ray background \citep{Lehmer:2012}.

\section{Case Study}
\label{Sec:results}

In order to establish some intuition we begin with a relatively simple
case, and examine in detail the predictions for a particular case
(which we refer to as our {\it standard}), which assumes the atomic
cooling minimum mass, $f_* = 0.05$, $f_X = 1$, XRBs, and $\tau =
0.066$ (\#53 in Appendix~\ref{appA}). We demonstrate how much varying
the X-ray SED affects predictions by comparing this model to case \#55
which has a soft X-ray SED but otherwise identical parameters.

\begin{figure*}
	\centering	
	\begin{subfigure}[b]{0.4\textwidth}
	\begin{center} \hspace{0.25in}	\textbf{Standard}\par\medskip \end{center}
	\includegraphics[width=3.1in]{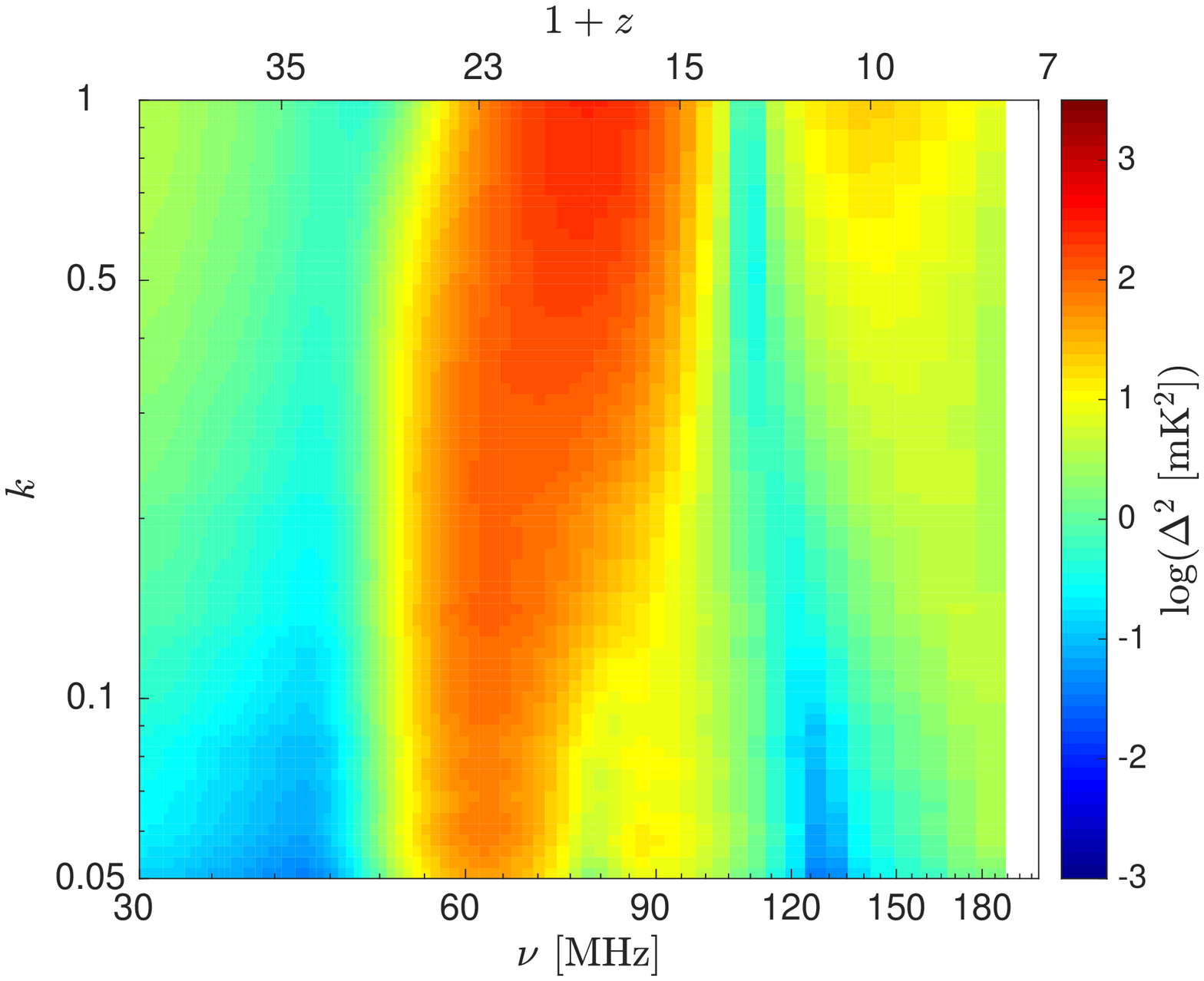}
	\vspace{0.05in}
\end{subfigure}
	\hspace{0.5in}
\begin{subfigure}[b]{0.4\textwidth}
	\begin{center} \hspace{0.25in} \textbf{Soft SED}\par\medskip \end{center}
	\includegraphics[width=3.1in]{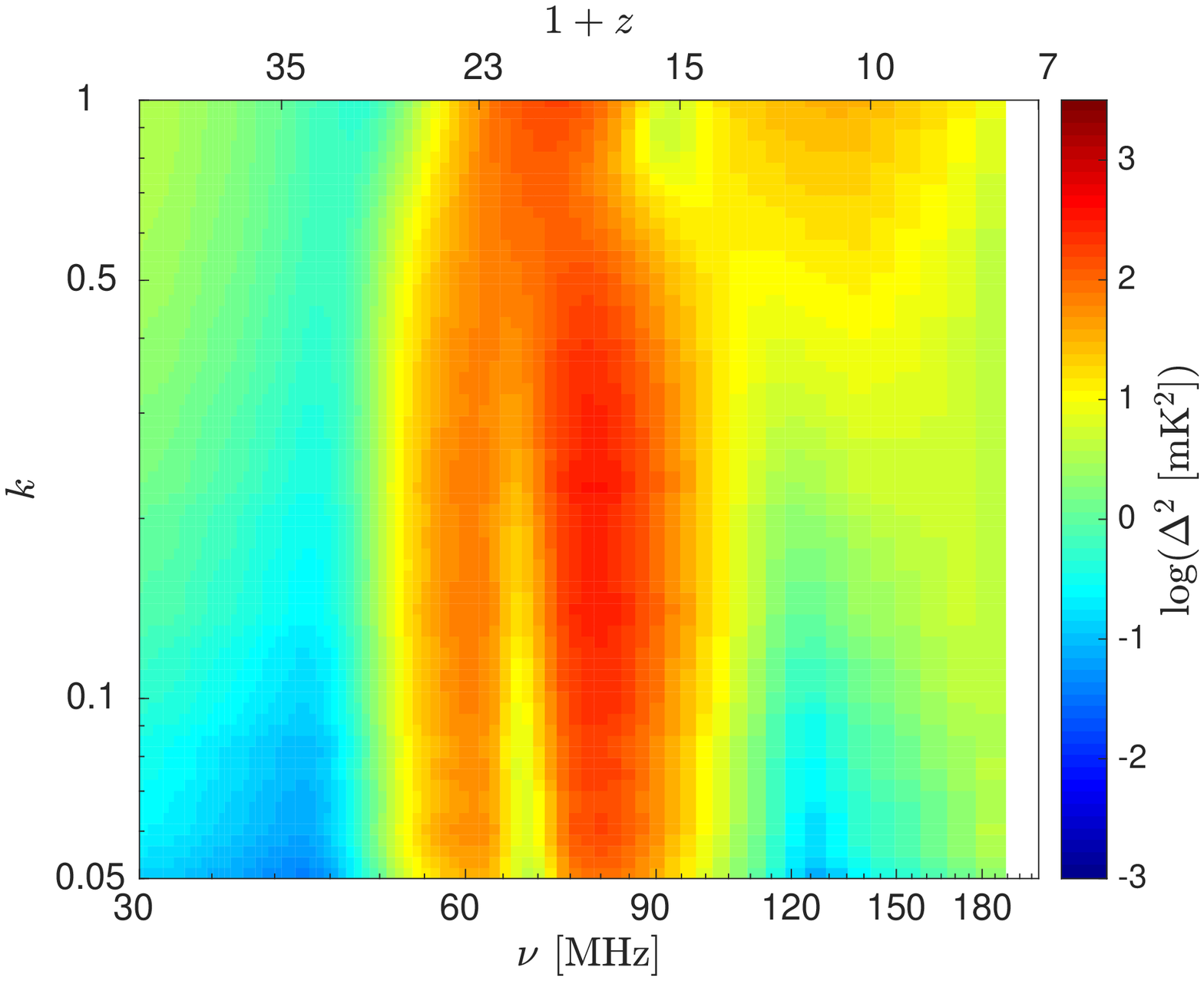}
	\vspace{0.05in}
	\end{subfigure}
	\begin{subfigure}[b]{0.4\textwidth}
	\includegraphics[width=3.1in]{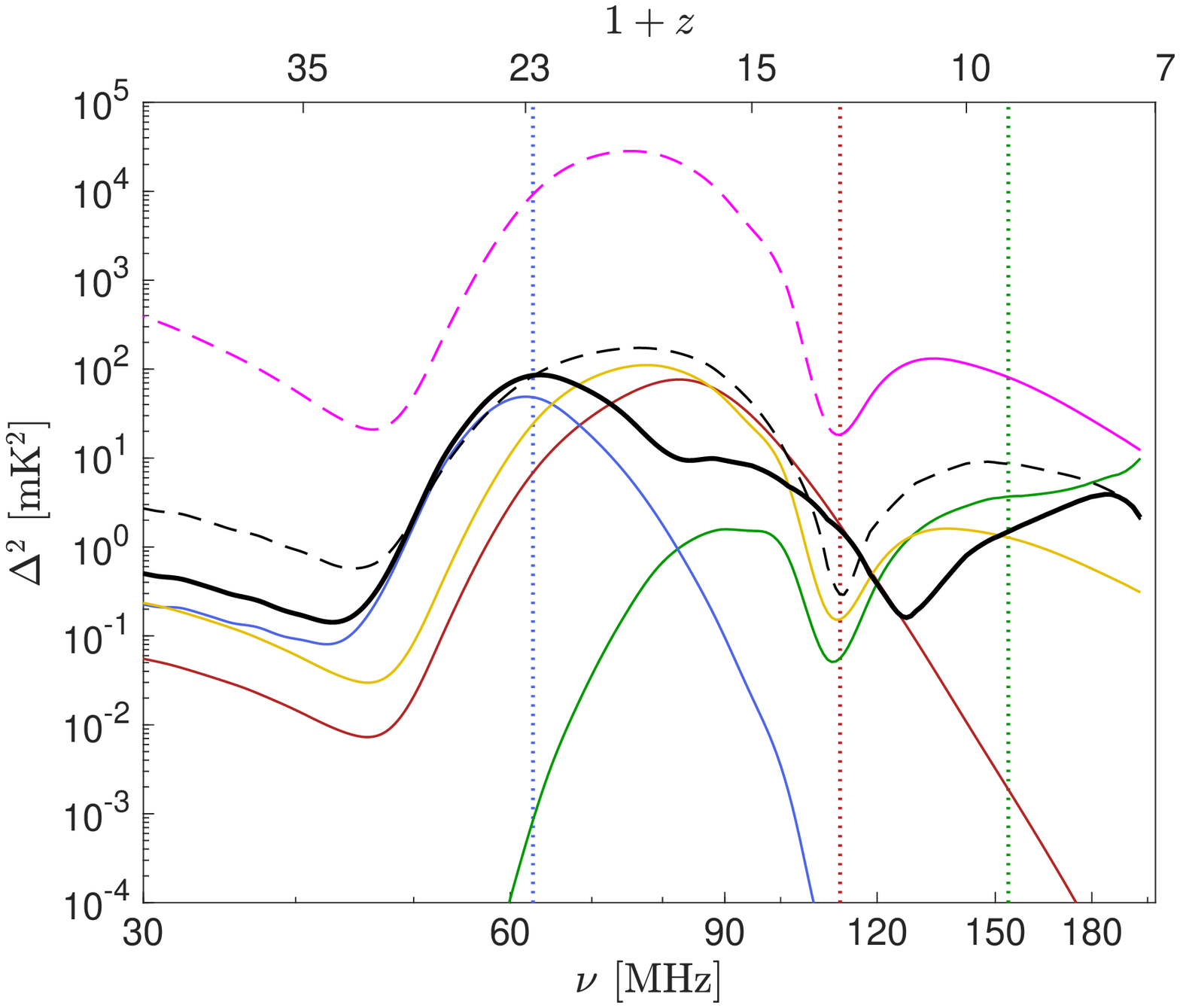}
	\vspace{0.05in}
\end{subfigure}
	\hspace{0.5in}
	\begin{subfigure}[b]{0.4\textwidth}
	\includegraphics[width=3.1in]{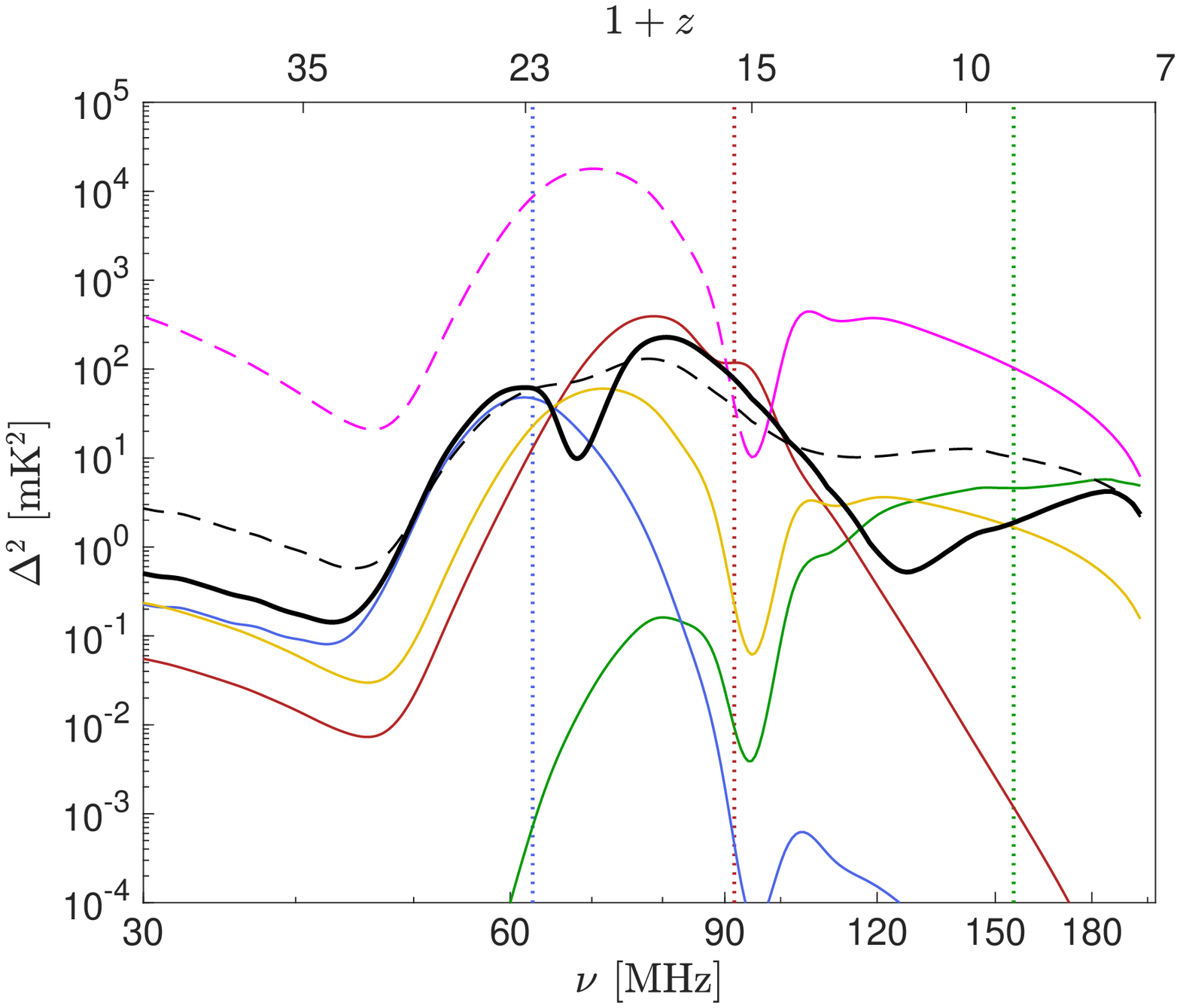}
	\vspace{0.05in}
\end{subfigure}
\caption{\label{firstPS} 21-cm power spectra for the model \#53 (our
  standard case, left column) and model \#55 (right
  column). \textbf{Top panels:} Two-dimensional map of the power
  spectra as a function of both wavenumber (vertical axis) and
  frequency/redshift (lower/upper horizontal axis). Colors correspond
  to $\log(\Delta^2)$ as indicated on the color bar.  \textbf{Bottom
    panels:} Evolution of the total power spectrum at a fixed
  wavenumber is shown for $k=0.1$ Mpc$^{-1}$ (solid black) and $k=0.5$
  Mpc$^{-1}$ (dashed black) as a function of frequency/redshift
  (lower/upper horizontal axis). For $k=0.1$ Mpc$^{-1}$ we also show
  the separate power spectra of various components: Ly$\alpha$ (blue),
  density plus velocity ($\delta+v$; yellow), temperature/heating
  (dark red) and ionization (green); these have all been expressed
  like $\Delta^2$ [Eq.~\ref{eq:Del}], in mK$^2$ units. For comparison,
  the global 21-cm signal, $T_{b}^2$, is also shown (magenta), as
  dashed when the signal is negative and solid when it is
  positive. Vertical lines mark several important milestones, namely
  Ly$\alpha$ coupling (blue), the cosmic heating transition (dark red)
  and the midpoint of reionization (green).}
\label{fig:Cases1}
\end{figure*}

The evolution of the power spectra is inherently a function of two
dimensions, as the power spectrum depends on both frequency/redshift
and scale. We show this full dependence in 2-dimensional color plots
(top panels of Fig.~\ref{fig:Cases1}).  At different stages of cosmic
history the total power in the 21-cm fluctuations can vary anywhere
between 0.001 and 1000 mK$^2$, with strong fluctuations shown in red
and weak fluctuations in blue. For our standard case, WF coupling
turns on at around $z\sim 30$ when the first significant population of
stars appears, creating an inhomogeneous Ly$\alpha$ background which
imprints a broad peak in the 21-cm spectrum at $z \sim 20$. The effect
of inhomogeneous heating is visible later at $z\sim 15$, while the
signature of reionization dominates at $z \sim 10$.

Predictions of the model \#55 are the same except for the effect of
X-rays, which imprints a strong signature in the intermediate redshift
range and on large scales \citep{Fialkov:2014c}. Consider first the
small-scale regime ($k \ga 0.5$ Mpc$^{-1}$). On these scales, which
are below the typical mean free path of X-ray photons for both the
hard and the soft SED, heating fluctuations are washed out (although
more so in the hard SED case, where there is a clear drop in the power
spectrum at the heating transition), and we only see two high power
regions caused by (i) inhomogeneous Ly$\alpha$ radiation at the
high-redshift end and (ii) ionizing fluctuations at the low-redshift
end. In both cases, the signature of reionization first appears at the
small-scale end and propagates to larger scales at lower redshifts, as
the ionized bubbles grow. The qualitative difference between the two
maps is on large scales ($k < 0.5$~Mpc$^{-1}$) that exceed the mean
free path of a typical X-ray photon in the case of the soft SED, but
are still below the mean free path of photons in the hard SED case.
(Note that at $z\sim 20$ the mean free path is longer than 100~Mpc for
photons with energies above $\sim 1$~keV). As a result, X-ray heating
imprints a strong peak in the former case, compared to a
barely-noticeable peak (and only at $k=0.1$~Mpc$^{-1}$ and below) in
the latter case.

A brief note on scales: a half-wavelength equal to our pixel size
(corresponding to the Nyquist critical frequency) is $k \sim
1$~Mpc$^{-1}$. At the other end, a half-wavelength equal to the size
of our box corresponds to $k \sim 0.008$~Mpc$^{-1}$, but even $k$
values that are a few times that suffer from being averaged over only
a small number of samples (as we have verified by comparison to larger
simulation boxes). Thus, in order to avoid edge effects at both ends,
we only use the range $k = 0.05 - 1$~Mpc$^{-1}$, and focus on the
specific values $k=0.1$~Mpc$^{-1}$ (which we refer to as representing
large scales) and $k=0.5$~Mpc$^{-1}$ (small scales).

To follow the evolution of fluctuations with redshift more closely, it
is useful to examine the behavior of the total power spectrum at a
fixed comoving scale, which we show in the bottom panels of
Fig.~\ref{fig:Cases1} for $k=0.1$ Mpc$^{-1}$ (solid black,
representing large scales) and $k=0.5$ Mpc$^{-1}$ (dashed black,
representing small scales). In agreement with the color plot, on large
scales the case with a soft SED shows three separate peaks dominated
by Ly$\alpha$, X-ray and ionization fluctuations, respectively; while
the other (hard SED) case has only two peaks, dominated by Ly$\alpha$
and by ionization fluctuations. In the latter case X-rays contribute
only a knee around $z\sim 15$. In the small-scale regime the
signatures of a soft and hard SED are more similar: the gas
temperature is nearly uniform on such scales in both cases, and there
is no heating peak. In both of these cases there are two peaks in the
absolute value of the global 21-cm temperature, one in absorption
(when the Ly$\alpha$ coupling saturates and X-ray heating first
becomes significant), and one in emission (after heating saturates);
they are separated by a minimum near the cosmic heating transition
(though $k=0.5$ Mpc$^{-1}$ is a borderline value in the soft SED case,
a scale large enough that the temperature fluctuations are significant
and nearly eliminate this minimum). This naturally tends to produce
two redshift peaks in the 21-cm power spectrum as well (at a given
$k$), although extra astrophysics (such as a rise and fall of various
sources of fluctuations) can move the peaks or create additional ones.

It is interesting to examine in detail what is the leading source of
fluctuations at every epoch. To explore this aspect we directly
extract from our simulation and separately show on the same plot
(bottom panels of Fig.~\ref{fig:Cases1}) power spectra (normalized as
in Eq.~(\ref{eq:Del})) of each of the terms $\delta+v$, $\delta_{\rm
  coup}$, $ \delta_{\rm heat}$ and $\delta_{\rm ion}$. The total power
spectrum is then a sum of individual contributions of power spectra
from each type of fluctuations plus cross-correlations among the
various sources. Note that unlike the power spectra of each component
(which are positive by definition), the cross-correlation components
of the power spectrum can be negative. In particular
\citep{Fialkov:2014c}, density fluctuations correlate positively with
Ly$\alpha$ fluctuations and anti-correlate with ionization
fluctuations; density fluctuations anti-correlate with heating
fluctuations when the gas is colder than the CMB, and correlate
positively with heating fluctuations when the gas is hotter than the
CMB. In the plots we only show the power spectra of each of the
sources of fluctuations separately. This allows us to see which
fluctuation source dominates at each redshift. For the specific choice
of astrophysical parameters shown, the coupling term (blue curve)
dominates at high redshifts, around the Ly$\alpha$ coupling transition
(defined as the redshift at which $x_{\rm tot} = 1$; vertical blue
dotted line. Note that our plots also include the dark ages ($z>30$)
but we do not focus on them here). This contribution starts to fade
when Ly$\alpha$ coupling saturates ($x_{\rm tot} \gg 1$). Next, the
build-up of the X-ray radiation background leads to a rise in
fluctuations in the $1-T_{CMB}/T_{gas}$ term (dark red curve). All
contributions except for heating experience a minimum around the
cosmic heating transition (defined as the redshift at which the mean
IGM temperature equals the CMB temperature; vertical dark red dotted
line. Note that this can be offset by a $\Delta z \sim 0.5$ compared
to the redshift at which the mean $T_{b}=0$.). Heating fluctuations
disappear once the IGM temperature is well above the CMB
temperature. Finally, the ionization fluctuations dominate (solid
green line) around the midpoint of reionization (defined as when the
cosmic mean mass-weighted $x_{\rm HI}=0.5$; vertical green line).
 
However, this is not the full story. The contribution of density (plus
velocity) fluctuations, which is often considered to be sub-dominant
relative to the radiative contributions, is significant throughout
cosmic history from the dark ages to the end of reionization. As is
evident from the plots, at some moments in cosmic history,
particularly when a dominant radiative contribution fades away,
density fluctuations can dominate the total power of the 21-cm
signal. This usually occurs in-between peaks, but as we go towards
smaller scales (including $k=0.5$~Mpc$^{-1}$), it becomes more common
for density fluctuations to be the dominant factor at a peak of
the power spectrum.

It is easy to understand the redshift evolution of the density
term. Because the density power spectrum evolves slowly with redshift
in the linear regime, $P_\delta\propto(1+z)^{-2}$, the redshift
evolution of the density (plus velocity) contribution
($\Delta_\delta^2 \propto T_b^2\times P_\delta$, yellow curve) is
largely driven by $T_b^2$ (shown with a magenta line on the plot, for
comparison). Note that the difference between $T_b^2$ and the density
contribution decreases at low redshifts because of the growth of
the density contrast.

\begin{figure*}
	\centering
	\includegraphics[width=3.2in]{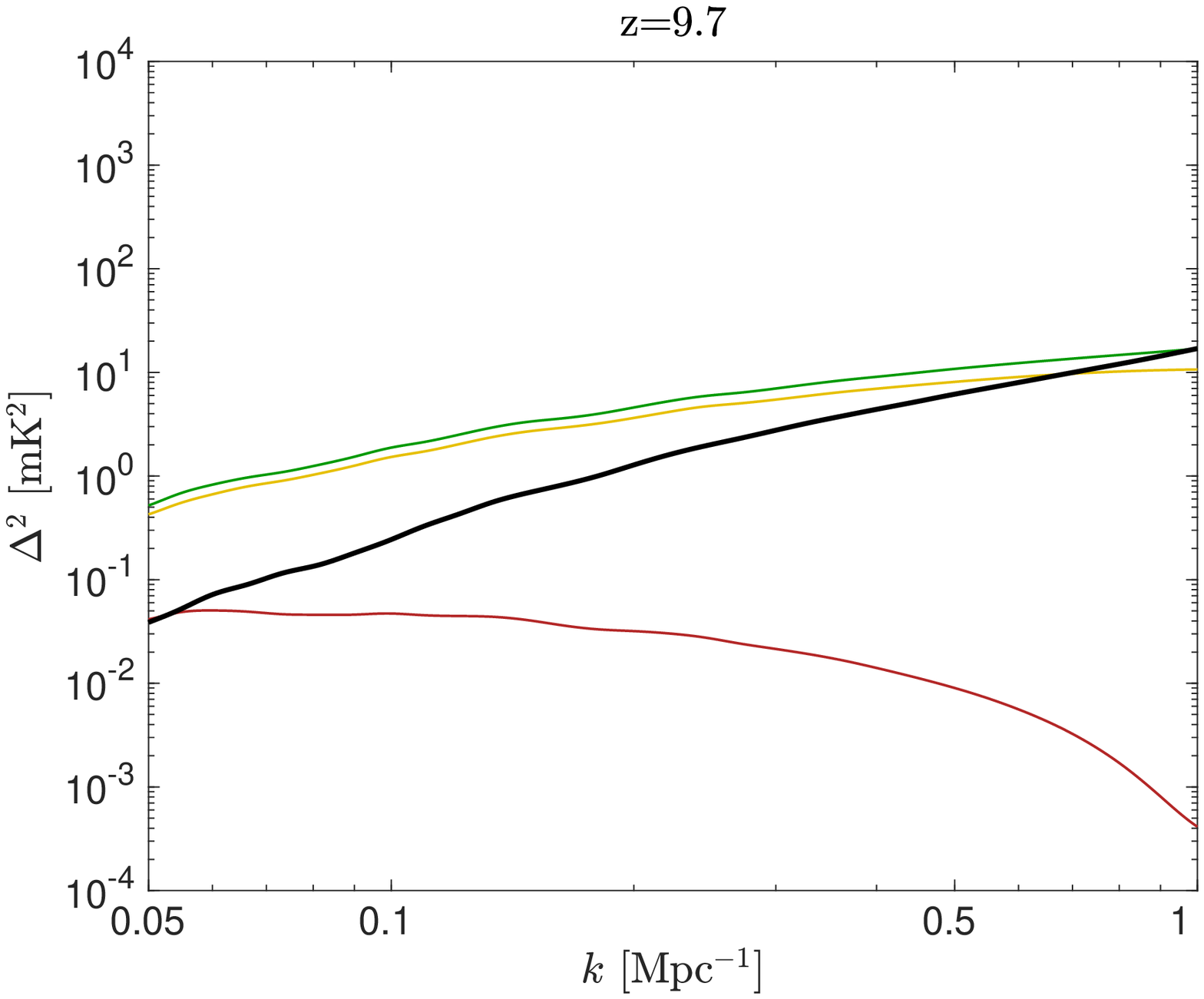}
	\includegraphics[width=3.2in]{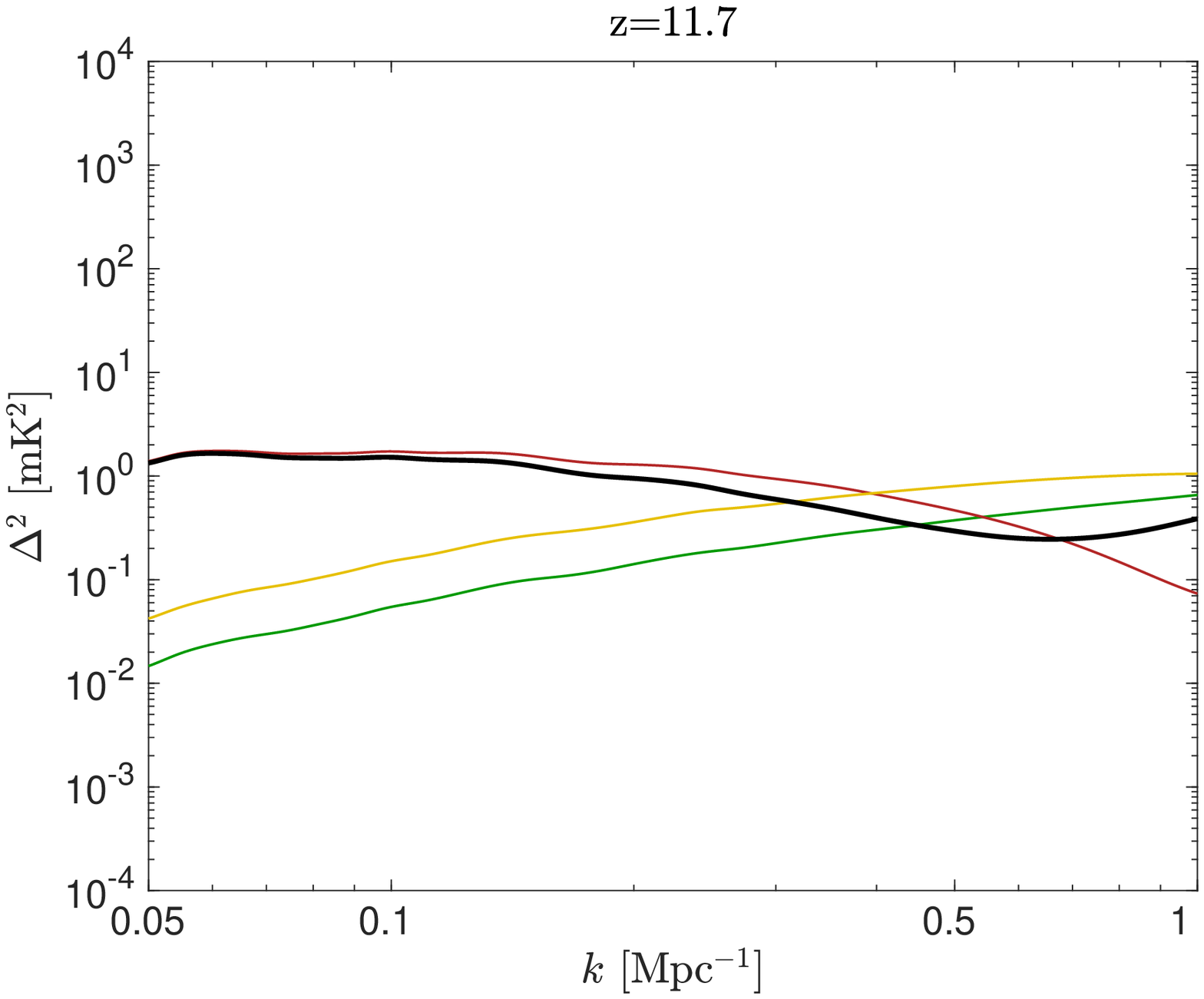}
	\includegraphics[width=3.2in]{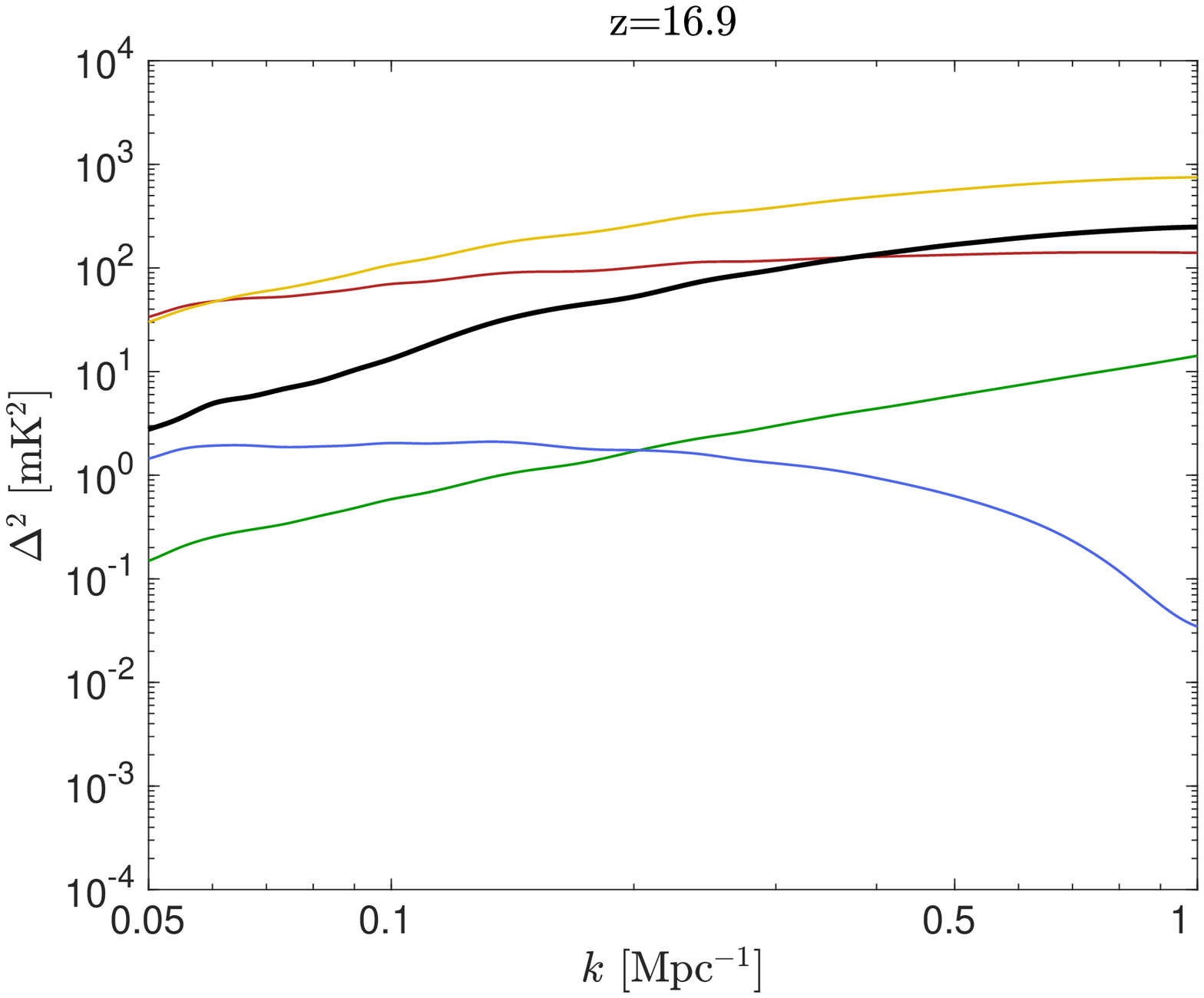}
	\includegraphics[width=3.2in]{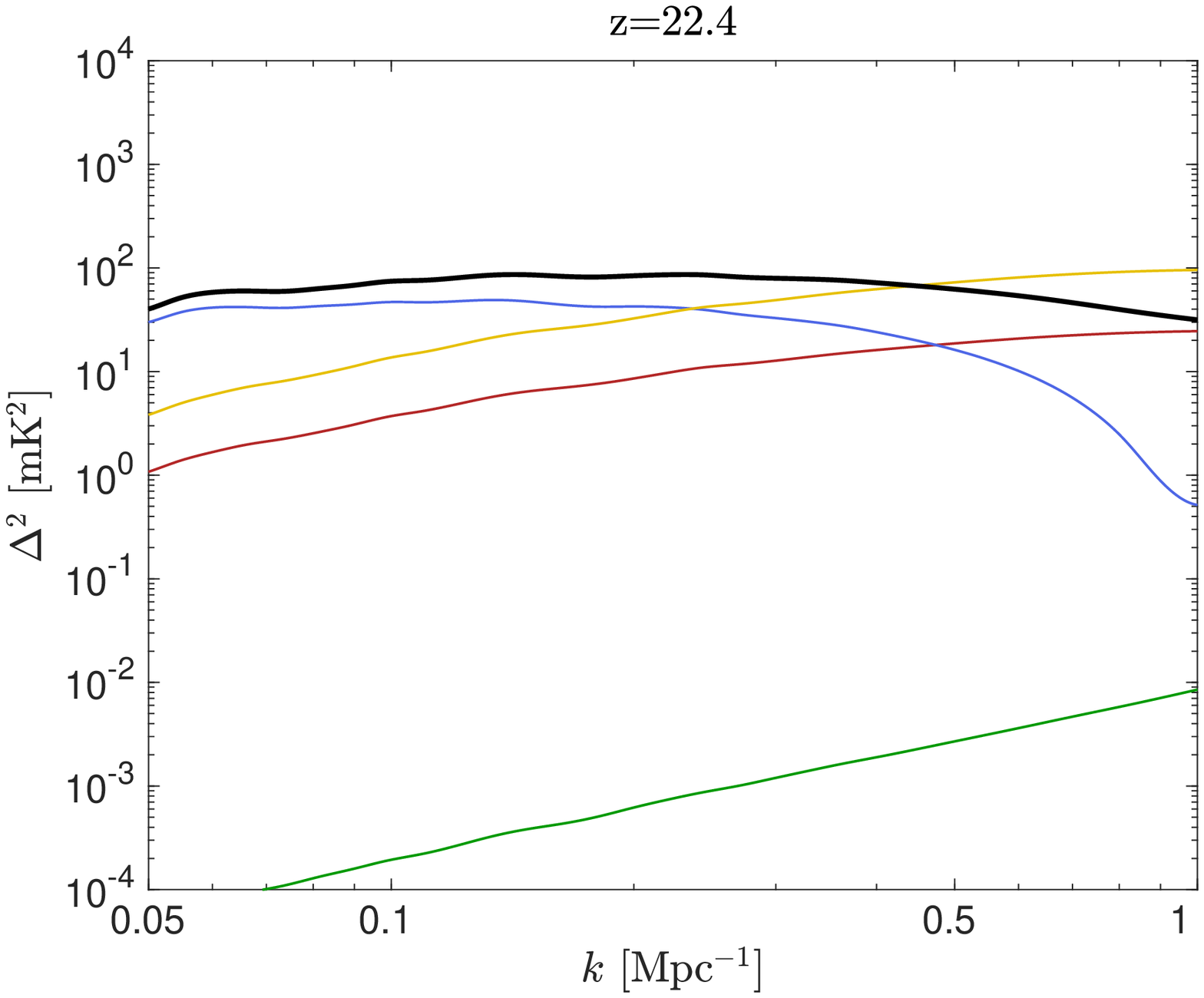}
	\caption{The 21-cm power spectrum (black) in our standard case
          as a function of $k$ at redshifts $z=9.7$, 11.7, 16.9 and
          22.4 (see panel titles) corresponding to the extrema of the
          large-scale slope as seen in Fig.~\ref{fig:Slopes1}. Also
          shown are the power spectra of the components $\delta+v$
          (yellow), $\delta_\alpha$ (blue), $ \delta_{\rm heat}$ (dark
          red) and $\delta_{\rm ion}$ (green). At $z=9.7$ and 16.9 the
          total power spectrum shows a near-cancellation of different
          components with opposite signs.}
	\label{fig:StandardPeaks}
\end{figure*}

Since the power spectrum is a function of two variables ($k$ and $z$),
we next consider the shape of the power spectrum (see Fig. \ref{fig:StandardPeaks}). Specifically, the
spectral index of the power spectrum is commonly defined as $ \beta
\equiv \frac{d\ln\Delta^2}{d\ln k}$. Naturally, if the spectral index
is positive, fluctuations have more power on smaller scales (larger
$k$) and if it is negative, there is more power on larger scales
(smaller $k$). If the spectrum is a pure power law, it has a shape
$\Delta^2 \propto k^{\beta}$ with constant $\beta$. In the 21-cm case,
there are several sources of fluctuations that drive the signal
($\delta+v$, $\delta_{\rm coup}$, $ \delta_{\rm heat}$ and
$\delta_{\rm ion}$), each having its own spectral index that can be
either positive or negative. In particular, in the hierarchical
picture of structure formation density fluctuations have more power on
small scales and, therefore, $\beta_\delta>0$. The coupling term has a
more complex temporal evolution. During the dark ages $\delta_{\rm
  coup}$ is dominated by the collisional term which follows $\delta$
on all scales and thus we expect a positive spectral index $\beta_{\rm
  coup} = \beta_{\delta}$. However, as soon as Ly$\alpha$ coupling
becomes dominant, $\beta_{\rm coup}$ should drop (and may even become
negative) on scales below the mean free path (which is of order tens
of comoving Mpc for Ly$\alpha$ photons). This is since fluctuations
are washed out on such scales. However, on scales much larger than the
mean free path, fluctuations continue to mostly follow the density and
maintain a positive spectral index. Realistically, the transition
between large and small (or negative) spectral indexes happens
smoothly versus redshift. Similarly, heating fluctuations are first
driven by the density and have $\beta_{\rm heat} = \beta_{\delta}$,
while at later times (once X-ray heating dominates) the power spectrum
of $\delta_{\rm heat}$ flattens. Finally, in the inside-out picture of
reionization \citep{Barkana:2004}, fluctuations in the ionizing field
are roughly a convolution of the density field with a bubble of a
typical size at each redshift. We expect the spectral index to be
close to $\beta_\delta$ on scales larger than the average size of an
ionized bubble (which grows with time), and to drop on smaller
scales. The total power spectrum is a combination of the four terms
and is driven by one or sometimes a few terms at a time. Therefore,
its spectral index evolves with redshift and is scale-dependent, as
can be seen in Fig.~\ref{fig:StandardPeaks} where the power spectrum
of our standard case (which has a hard SED) is shown as a function of
$k$ at several redshifts.

\begin{figure*}
	\centering
	\begin{subfigure}[b]{0.4\textwidth}
		\begin{center} \hspace{0.5in}	\textbf{Standard}\par\medskip \end{center}
	\includegraphics[width=3.1in]{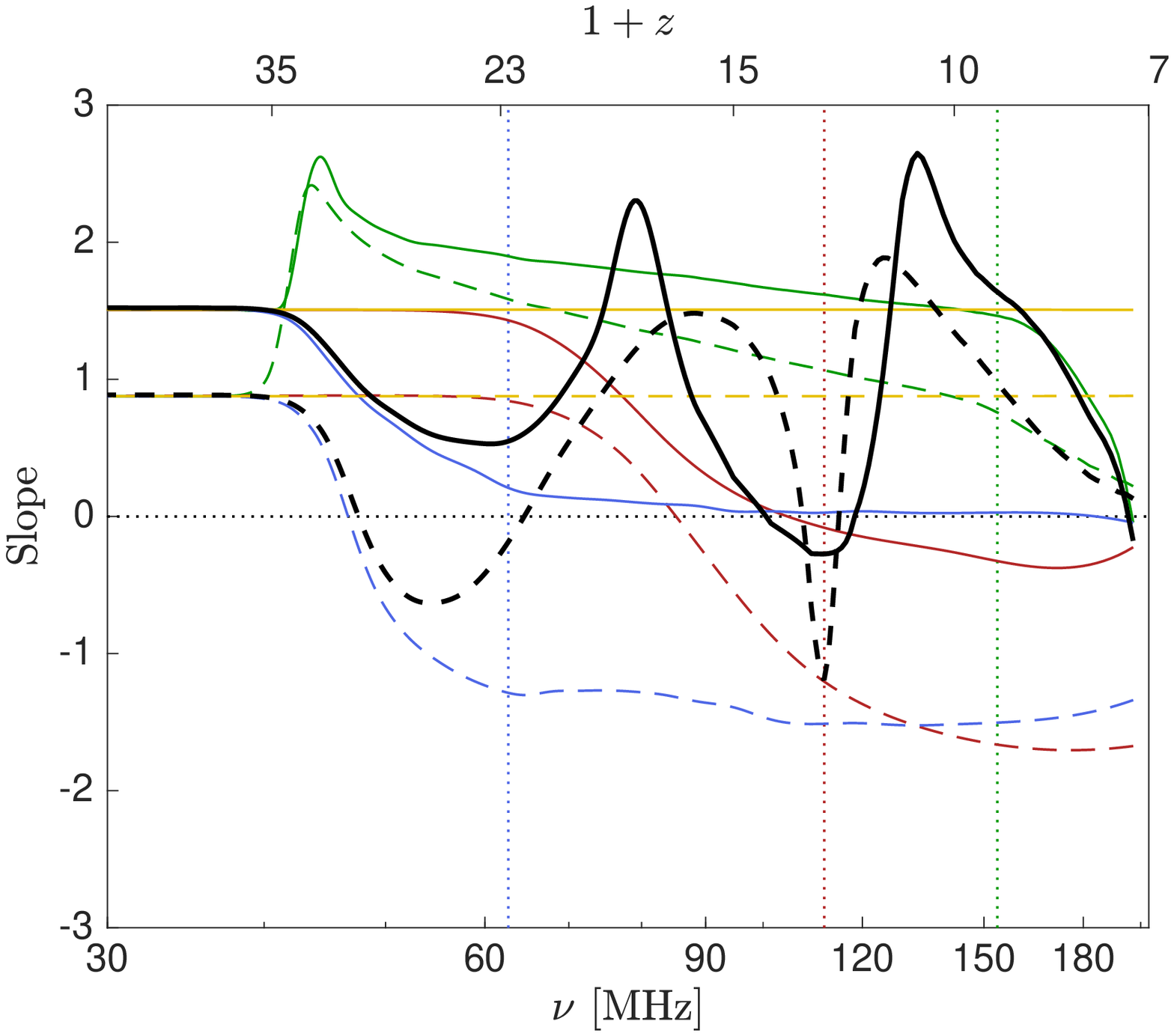}
\end{subfigure}
	\hspace{0.5in}
	\begin{subfigure}[b]{0.4\textwidth}
		\begin{center} \hspace{0.5in}	\textbf{Soft SED}\par\medskip \end{center}
	\includegraphics[width=3.1in]{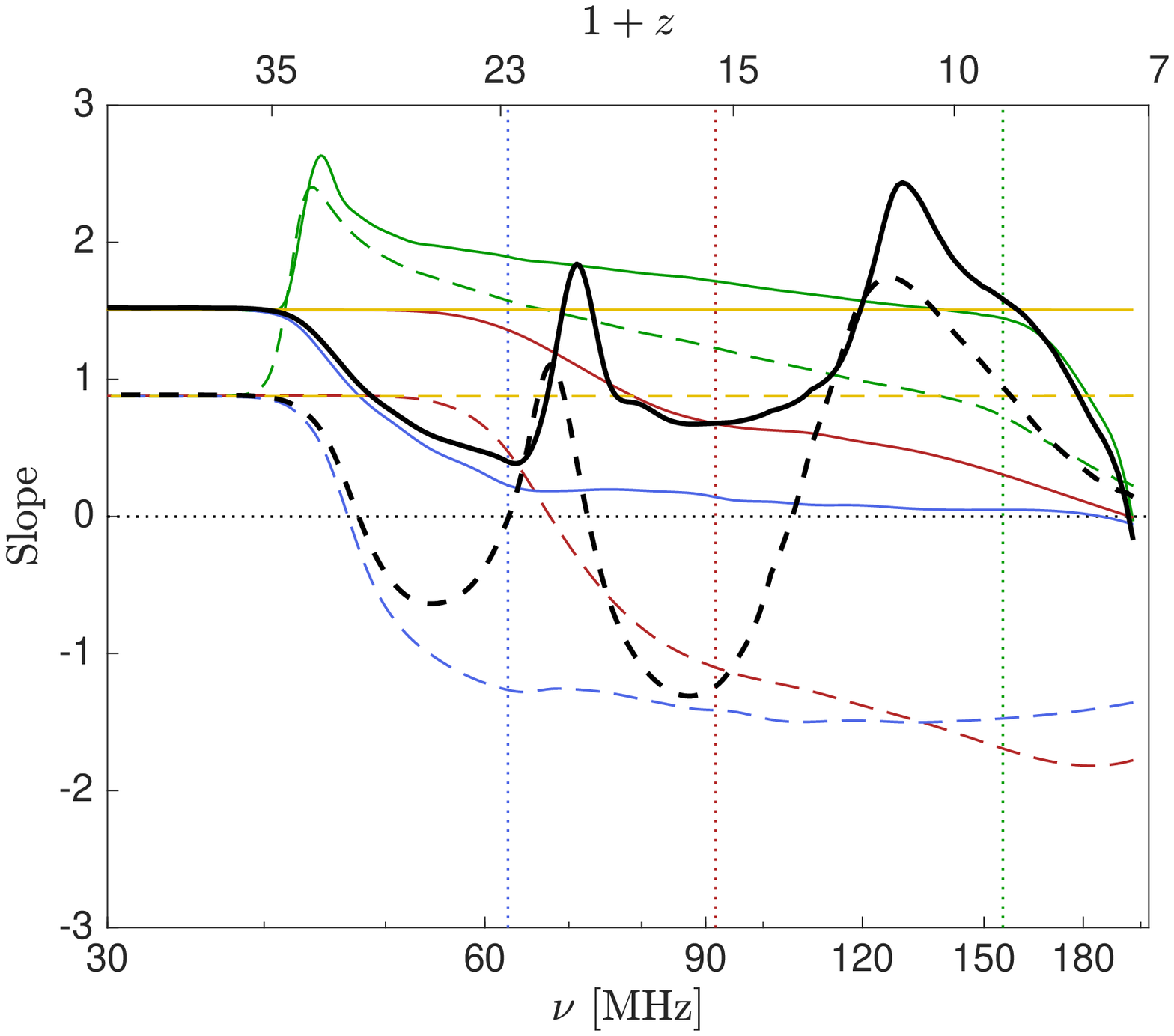}
\end{subfigure}
\caption{\label{firstSl} Slopes are shown for the standard case (left)
  and the case with soft SED (right) at large scales (between $k=0.05$
  and 0.2 Mpc$^{-1}$, solid) and small scales (between $k=0.2$ and 0.6
  Mpc$^{-1}$, dashed) as a function of frequency/redshift (lower/upper
  horizontal axis). We show $B_{\rm tot}$ (black), $B_{\delta}$
  (yellow), $B_{\rm coup}$ (blue), $B_{\rm heat}$ (red) and $B_{\rm
    ion}$ (green).}
\label{fig:Slopes1}
\end{figure*}

Rather than using the scale-dependent spectral index, we define a
simpler measure that captures the essence of the $k$ dependence,
namely, the average slope of the power spectrum between two scales
(still in log-log):
\begin{equation}
B = \frac{\ln\Delta^2(k_2)-\ln\Delta^2(k_1)}{\ln k_2-\ln k_1}\ .
\end{equation}
 In Fig.~\ref{fig:Slopes1} we show the evolution of the slope with
 redshift, both in the small-scale regime (calculated between
 $k_1=0.2$ Mpc$^{-1}$ and $k_2=0.6$ Mpc$^{-1}$, corresponding to
 $\sim$10-30 comoving Mpc) and in the large-scale regime (calculated
 between $k_1=0.05$ Mpc$^{-1}$ and $k_2=0.2$ Mpc$^{-1}$, corresponding
 to $\sim$30-125 comoving Mpc). In addition to the slope of the total
 power spectrum, $B_{\rm tot}$, we show the slopes for each of the
 four sources of fluctuations, $B_{\delta}$, $B_{\rm coup}$, $B_{\rm
   heat}$, and $B_{\rm ion}$.

The overall evolution of the spectral slope, as outlined above, can be
seen in detail in Fig.~\ref{fig:Slopes1}. First of all, $B_\delta$ is
positive and constant with redshift (because in the linear regime
within $\Lambda$CDM cosmology, the growth of density fluctuations is
scale-independent); $B_\delta \sim 3$ where the matter power spectrum
turns over (i.e., at $k_{\rm eq}\sim 0.06$ Mpc$^{-1}$), and decreases
with $k$. Both $B_{\rm coup}$ and $B_{\rm heat}$ start out equal to
$B_{\delta}$, while at later times (when they become dominated by
astrophysical radiation) they drop to zero or below (on scales below
the effective horizon of the radiation). The evolution of $B_{\rm
  ion}$ is interesting, as it has a particularly high peak value at
very high redshifts. This is because the non-linear amplification
(i.e., the fact that the ionization fraction is either 0 or 1)
increases the power on small scales, in particular causing large
Poisson fluctuations initially \citep{Barkana:2008}, for which we
would expect a slope around 3; this, though, mostly occurs long before
there is enough reionization to significant affect the total 21-cm
power spectrum. At lower redshifts, the ionized bubbles grow
and wash out small-scale fluctuations in the ionization.

The slope of the total 21-cm power spectrum is driven by one or
several components at a time and its evolution can be easily
understood (as we illustrate with the large-scale slope of our
standard case). During the cosmic dark ages, all fluctuations are
driven by the density, and so $B_{\rm tot} = B_\delta$. When the first
significant population of stars creates a Ly$\alpha$ background, the
power spectrum flattens and its slope decreases as $B_{\rm tot}$
follows $B_{\rm coup}$. In fact, $B_{\rm tot}$ of the large-scale
slope has a minimum just when the contribution of Ly$\alpha$ to the
power spectrum, shown in Fig.~\ref{fig:Cases1}, is maximal ($z = 22.4$
in the standard case). Then, when $x_\alpha$ approaches saturation,
the contribution of Ly$\alpha$ fluctuations becomes negligible
compared to density fluctuations. As a result, the total power
spectrum steepens again and the slope increases. $B_{\rm tot}$ peaks
when $\delta_{\rm heat}$ is comparable to $\delta+v$, causing
cancellation on large scales, but not on small scales (where the
heating power spectrum is flatter); see Fig.~\ref{fig:StandardPeaks}
at $z = 16.9$. At the heating transition ($z = 11.7$), heating
fluctuations dominate on large scales (Fig.~\ref{fig:StandardPeaks});
therefore, $B_{\rm tot}$ reaches a minimum as it tracks $B_{\rm
  heat}$. After the heating transition is completed and the IGM
becomes much hotter than the CMB, the impact of heating on the 21-cm
signal gradually declines. As a result, for a short time the density
fluctuations become the main source of the 21-cm power again. At the
same time, ionizing fluctuations become increasingly important and
eventually win, surpassing the contribution of density at $z = 10.8$.
The slope peaks at $z = 9.7$ when ionization fluctuations become large
enough to cancel out the sum of density (plus velocity) and heating
fluctuations, with heating being significant only on very large scales
and thus the small-scale fluctuations do not cancel out
(Fig.~\ref{fig:StandardPeaks}). After that, ionization fluctuations
dominate the 21-cm fluctuations. Around the midpoint of reionization
(vertical green line), there is an inflection point in $B_{\rm tot}$,
as the main source of fluctuations changes from ionized bubbles within
a neutral IGM to remaining neutral regions within a mostly ionized
IGM. $B_{\rm tot}$ tracks $B_{\rm ion}$ during the later stages of
reionization.

The general behavior of the small-scale slope (dashed black line) is
similar to that of the large scale slope. However, the numerical value
of this slope is generally lower because the density power spectrum
has a lower slope on small scales. Also, both Ly$\alpha$ and X-rays
are types of long-range radiation and have lower slopes on small
scales than the density. As a result, the two minima of the slope are
significantly lower (-0.63 and -1.19 for small scales compared to 0.52
and -0.27 for the large scales).

To illustrate the effect of different astrophysical parameters on the
shape of the power spectrum and its slope we show more examples in
Appendix~\ref{appB} (Figs.~\ref{fig:Cases3}, \ref{fig:Cases2} and
\ref{fig:Cases4}) in which we vary $f_*$, $f_X$, and $V_c$ from their
values in our standard case. Given the existing uncertainty in the
parameters, the evolution of the power spectrum and its slope vary
quite a lot among plausible parameter sets, and the change induced by
varying one parameter is not a small variation around the standard
case. Depending on the parameter set, there are differences in both
the location and amplitude of features in the power spectrum and its
slope. Moreover, as we see from the plots, the dominant sources of
fluctuations at various moments in cosmic history can change, altering
the history of the power spectrum and changing the number of peaks.

\section{The Entire Parameter Space}
\label{sec:entire}

\begin{figure*}
	\centering
	  	\includegraphics[width=3.2in]{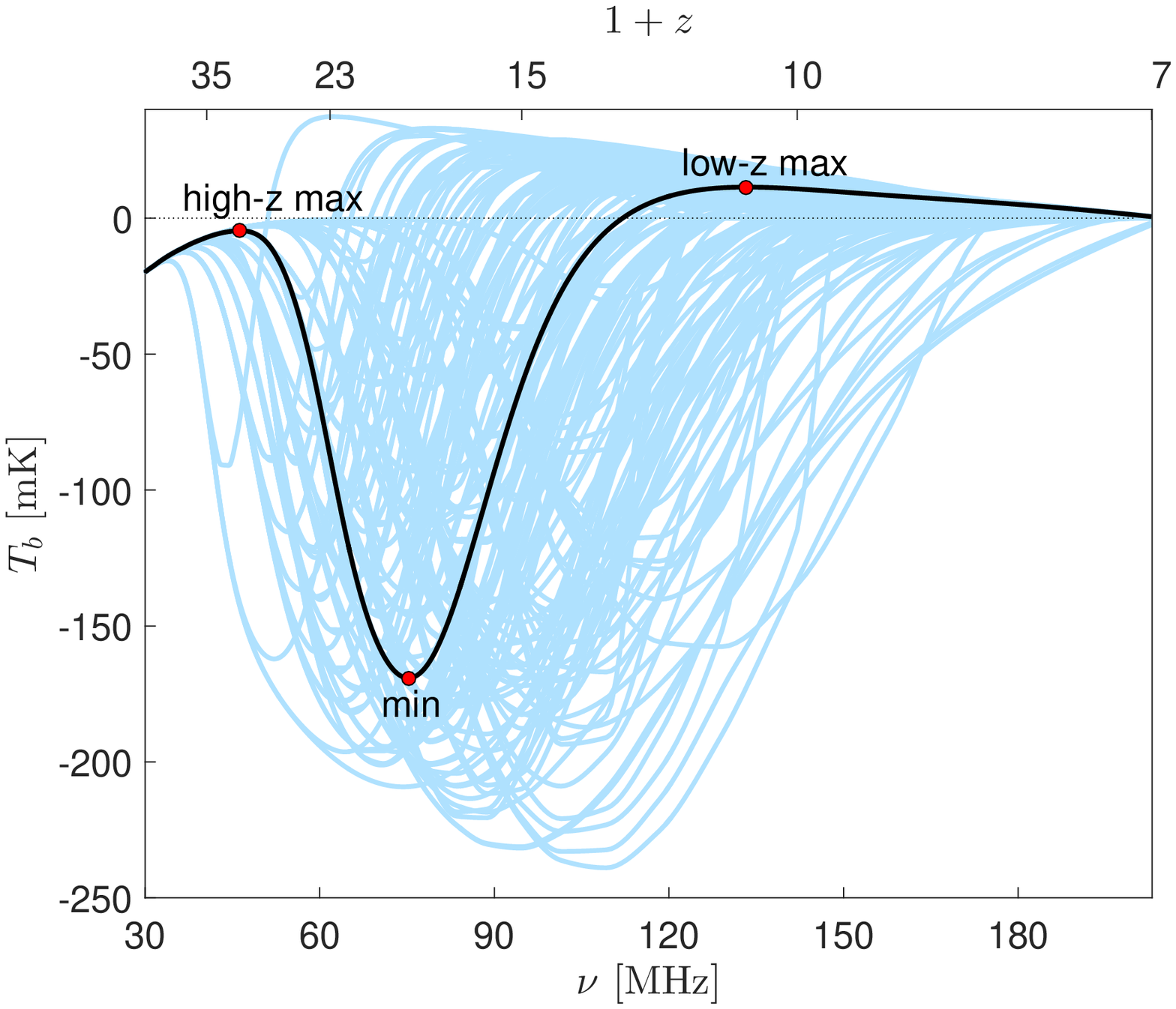}
		\includegraphics[width=3.25in]{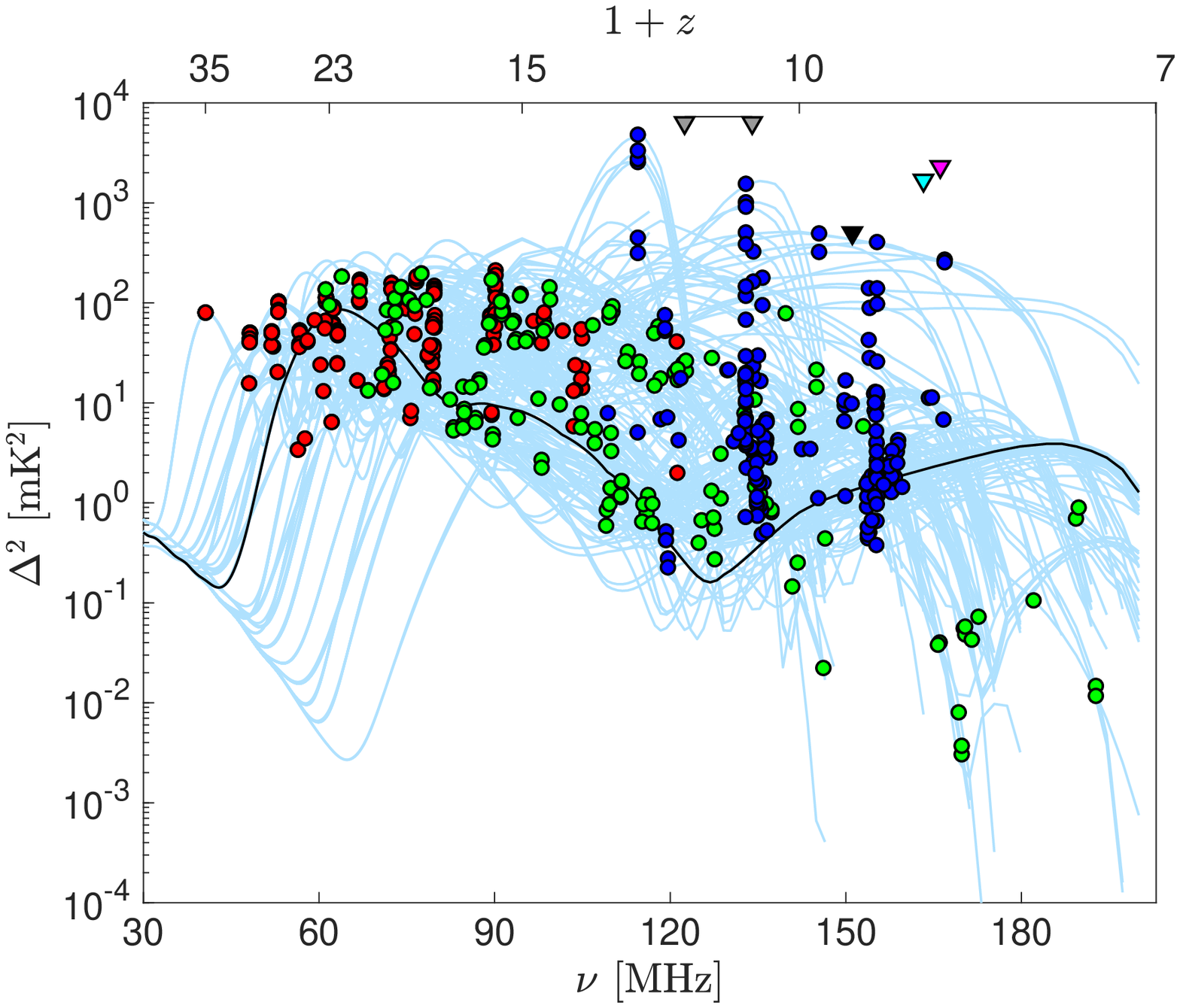}
	\caption{{\bf Left:} Parameter study of the global signal
          adopted from \citet{Cohen:2016b}. We show the 21-cm global
          signal as a function of redshift for our standard case
          (black line), with red points marking the three turning
          points (from left to right: the high-$z$ maximum, the minimum,
          and the low-$z$ maximum). Light-blue lines show the entire set
          of realizations of the 21-cm signal for the 193 different
          astrophysical models discussed in this paper. {\bf Right:}
          The corresponding complete set of realizations of the power
          spectrum at $k=0.1$ Mpc$^{-1}$ (light-blue curves) as a
          function of observed frequency/redshift (bottom/top axis).
          The standard case is shown with a black line. For each
          model, we mark the redshift of the Ly$\alpha$ coupling (red
          dot), the moment of the heating transition (green dot), and
          the midpoint of reionization (blue dot). Triangles mark
          various upper limits measured on various scales: black at
          $k=0.1$ to 0.33 Mpc$^{-1}$ \citep{Ali:2015}, cyan at
          $k=0.18$ Mpc$^{-1}$ \citep{Parsons:2014}, magenta at
          $k=0.13$ Mpc$^{-1}$ \citep{Jacobs:2015} and grey at $k=0.04$
          Mpc$^{-1}$ \citep{Patil:2017} } 
	\label{fig:AllPS}
\end{figure*}

Having considered a few specific cases in the previous section, we now
address the full set of models introduced in
Section~\ref{Sec:Param}. In \citet{Cohen:2016b} we used these models
to analyze the features of the global 21-cm spectrum (left panel of
Fig.~\ref{fig:AllPS}). The shape of this signal is universal and has
three main features: (i) a high-redshift maximum at redshift $z_{\rm
  g,max}^{\rm \, hi}$ marks the onset of significant stellar
radiation, (ii) a minimum (i.e., absorption trough) located at $z_{\rm
  g,min}$ is the beginning of the X-ray heating era, and (iii) a
high-redshift maximum (emission peak) at $z_{\rm g,max}^{\rm \, lo}$
occurs when heating saturates. Here we use the same set of models to
first analyze the shapes versus redshift of the power spectrum and its
slope, and then relate the features to the astrophysical parameters
that we vary. We summarize in Table~\ref{Table:Notations} our notation
for the various features of the global 21-cm signal, the power
spectrum, and its slope.

\subsection{Shapes}

To demonstrate the span of possibilities that could be realized in
Nature, we begin by placing all the power spectra on the same plot
(right panel of Fig.~\ref{fig:AllPS}). On every curve we mark the
cosmic milestones: the redshift at which Ly$\alpha$ coupling saturates
(red dot), the moment of the heating transition (green dot), and the
midpoint of reionization (blue dot). The large scatter in the location
of these markers and the large variety of shapes express our ignorance
about the high-redshift astrophysical parameters. The markers of the
midpoint of reionization fill a relatively small range because this
transition is the most constrained, being pinned down by the {\it
  Planck} measurements. On the other hand, the timing of the
Ly$\alpha$ transition as well as the heating transition are very
unconstrained and show large scatter. The former event depends on the
cooling channel and the efficiency with which the first stars where
formed, while the latter depends on both the properties of the first
stars and of the first heating sources (which may be two significantly
different populations). Note that in some cases the gas temperature
does not reach the CMB temperature even at the end of reionization,
due to very inefficient heating or no heating at all.

\begin{figure*}
	\centering
	\includegraphics[width=3.2in]{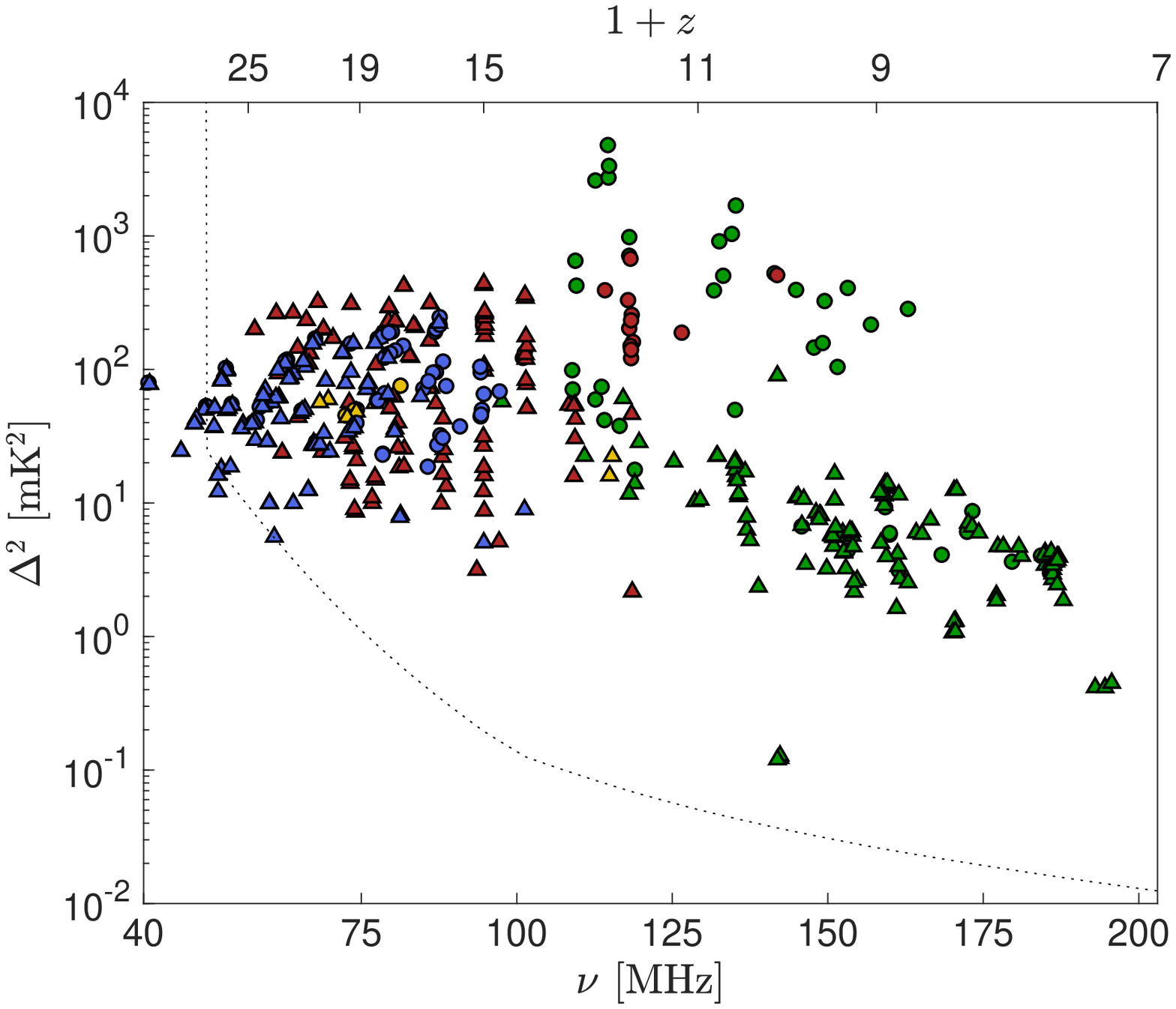}
	\hspace{0.5in}
	\includegraphics[width=3.2in]{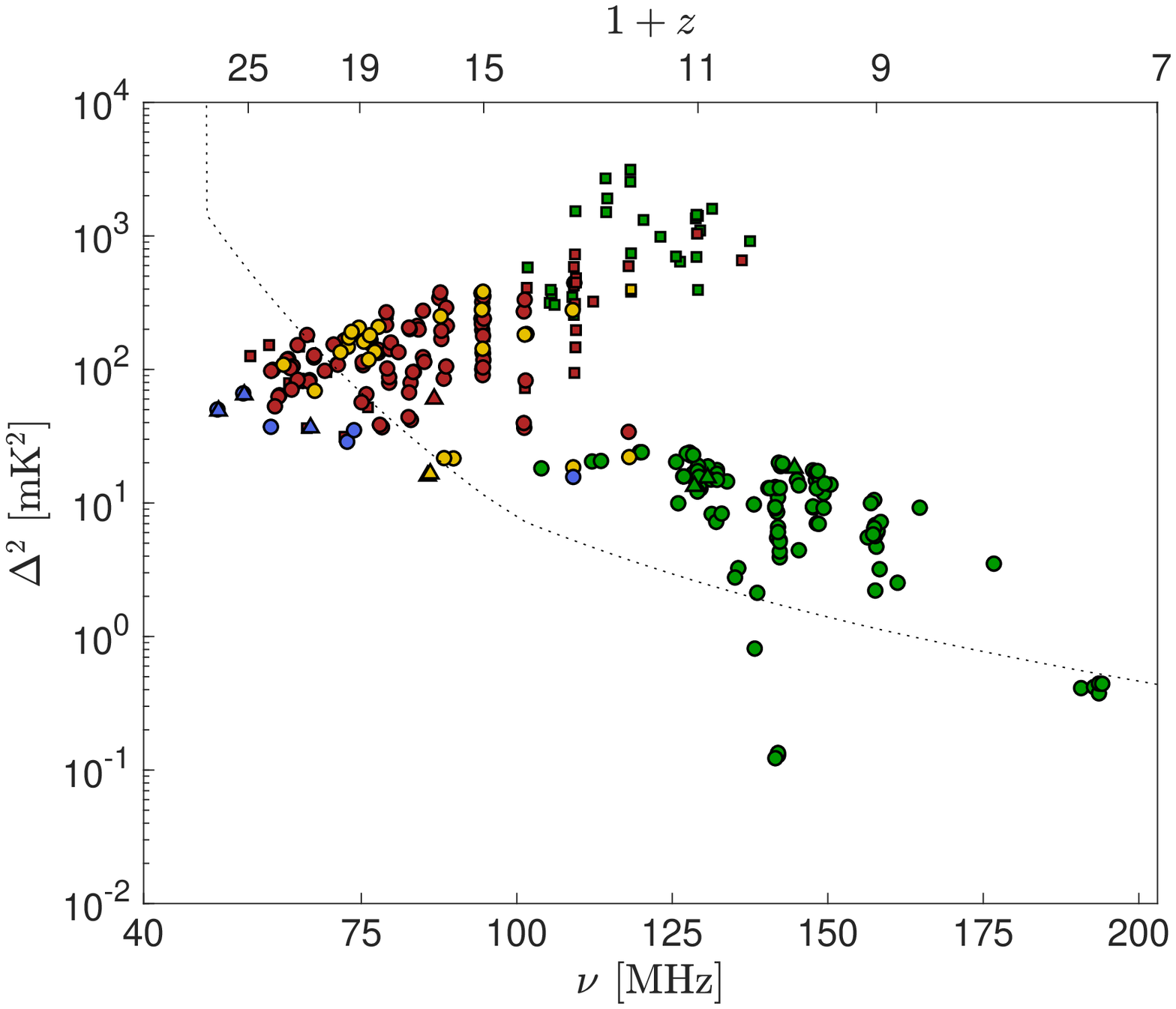}
	\caption{The power spectrum at $k=0.1$ Mpc$^{-1}$ (left) and
          $k=0.5$ Mpc$^{-1}$ (right) as a function of observed
          frequency/redshift (bottom/top axis) at various
          peaks. Marker colors indicate the dominant source of
          fluctuations for each peak: density (yellow), Ly$\alpha$
          (blue), heating (red) and ionization (green). Marker shapes
          indicate whether the corresponding case has a total (over
          all redshifts) of 1 (squares), 2 (circles), or 3 (triangles)
          peaks at that scale (Note that there are no squares in the
          left panel). The black dotted line shows the power spectrum
          of the thermal noise for the SKA1 assuming a single beam,
          integration time of 1000 hours, 10 MHz bandwidth, and bins
          $\Delta k = k$.}
	\label{fig:peaks}
\end{figure*}

Our approach to classifying different cases is by using the properties
of the peaks (which are easiest to observe), with each maximum being
tagged according to the dominant source of fluctuations. The
unconstrained astrophysical parameters introduce at least an order of
magnitude uncertainty in the maximal power produced by each type of
source. Fig.~\ref{fig:peaks} shows the maxima for all the cases at
$k=0.1$ Mpc$^{-1}$ (left) and $k=0.5$ Mpc$^{-1}$ (right),
demonstrating the scatter in peaks dominated by density (yellow),
Ly$\alpha$ (blue), heating (red) and ionization (green). Within the
uncertainty introduced by the astrophysical parameters, the peak power
is typically higher by $1-2$ orders of magnitude at high redshifts
compared to the low-redshift ionization peak. This is an important
conclusion demonstrating that the high-redshift signal might be as
accessible as its low-redshift counterpart with an SKA-like
instrument. On large scales ($k=0.1$ Mpc$^{-1}$) and prior to
reionization the peak power for all the models varies in the range
$5\lesssim \Delta^2\lesssim 500$ mK$^2$ and drops as soon as
reionization progresses (except for the cases when reionization is
cold). On smaller scales the typical peak power prior to reionization
is slightly higher and is in the range $10\lesssim \Delta^2\lesssim
1000$ mK$^2$. It is interesting to note that, whereas on large scales
all sources of fluctuations can contribute a strong peak, on scales
smaller than the typical mean free path of the radiation (including
$k=0.5$ Mpc$^{-1}$) density fluctuations play a relatively important
role, dominating the high-redshift peak (usually attributed to
heating) in a significant fraction of cases. Finally we note that the
number of peaks on small-scales ($1-3$, usually 2) is typically
smaller than on large scales ($2-3$, usually 3).

\begin{figure*}
	\centering
	\includegraphics[width=3.2in]{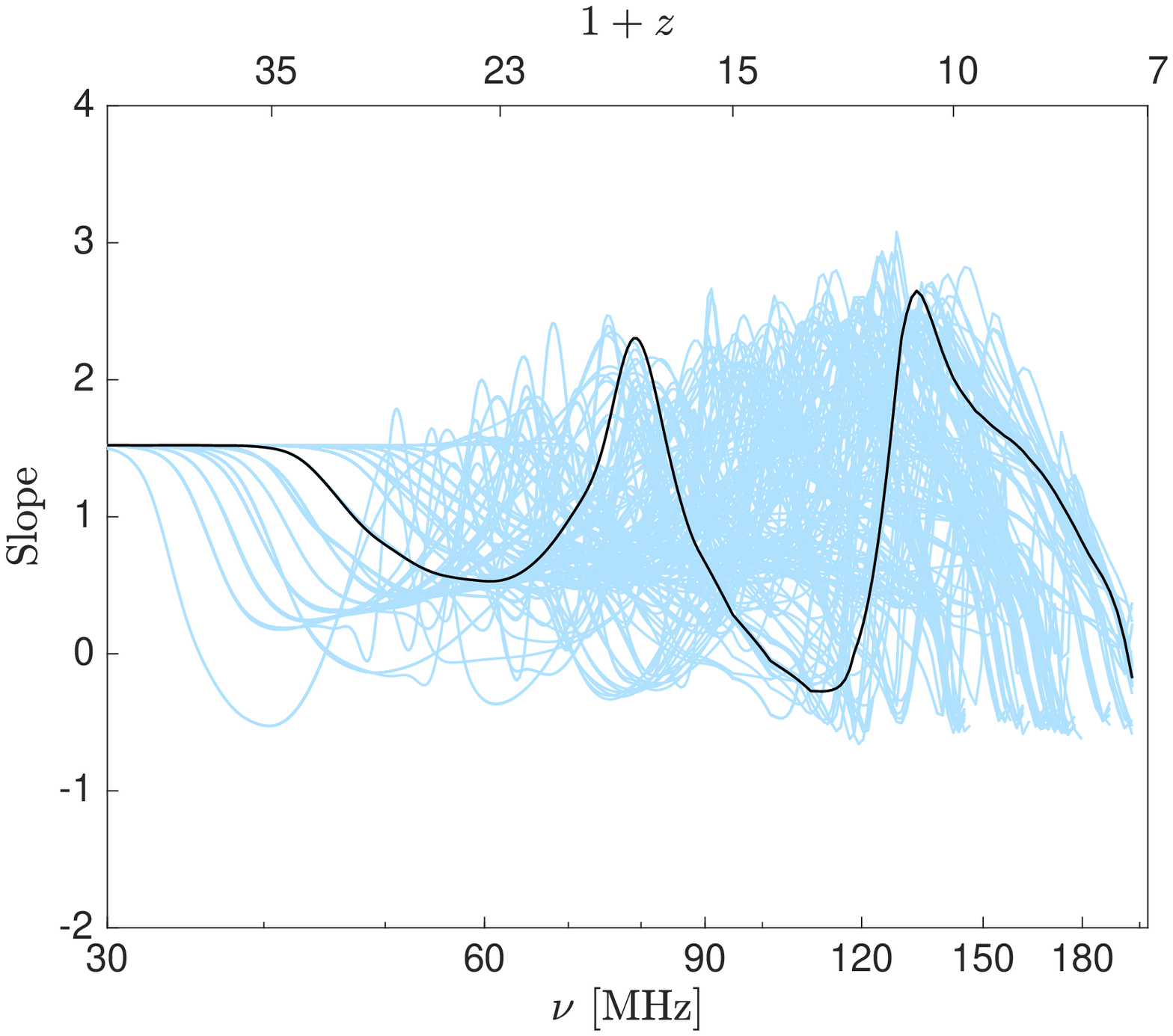}
		\hspace{0.5in}
	\includegraphics[width=3.2in]{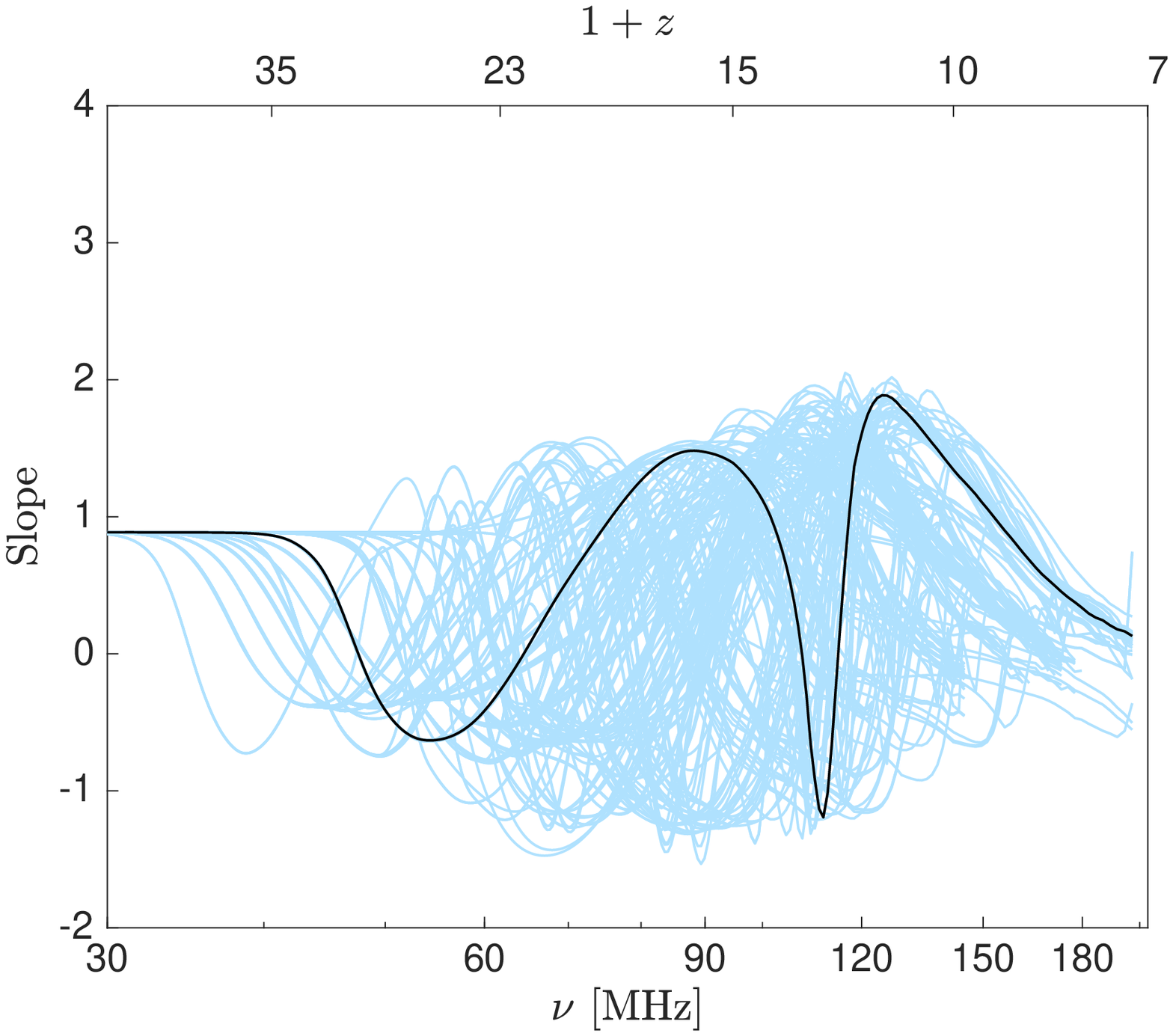}
 	\includegraphics[width=3.2in]{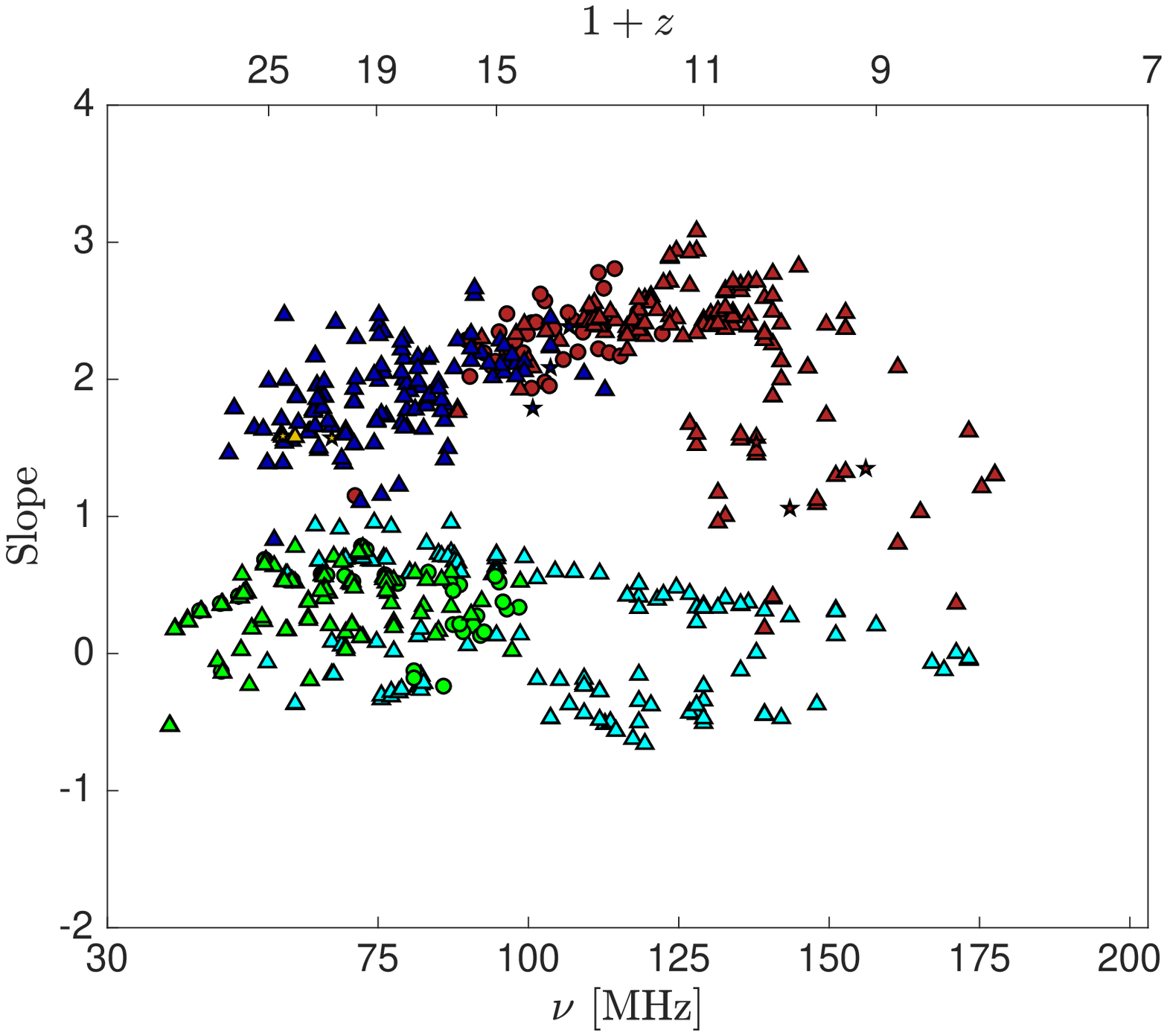}
 		\hspace{0.5in}
 	\includegraphics[width=3.2in]{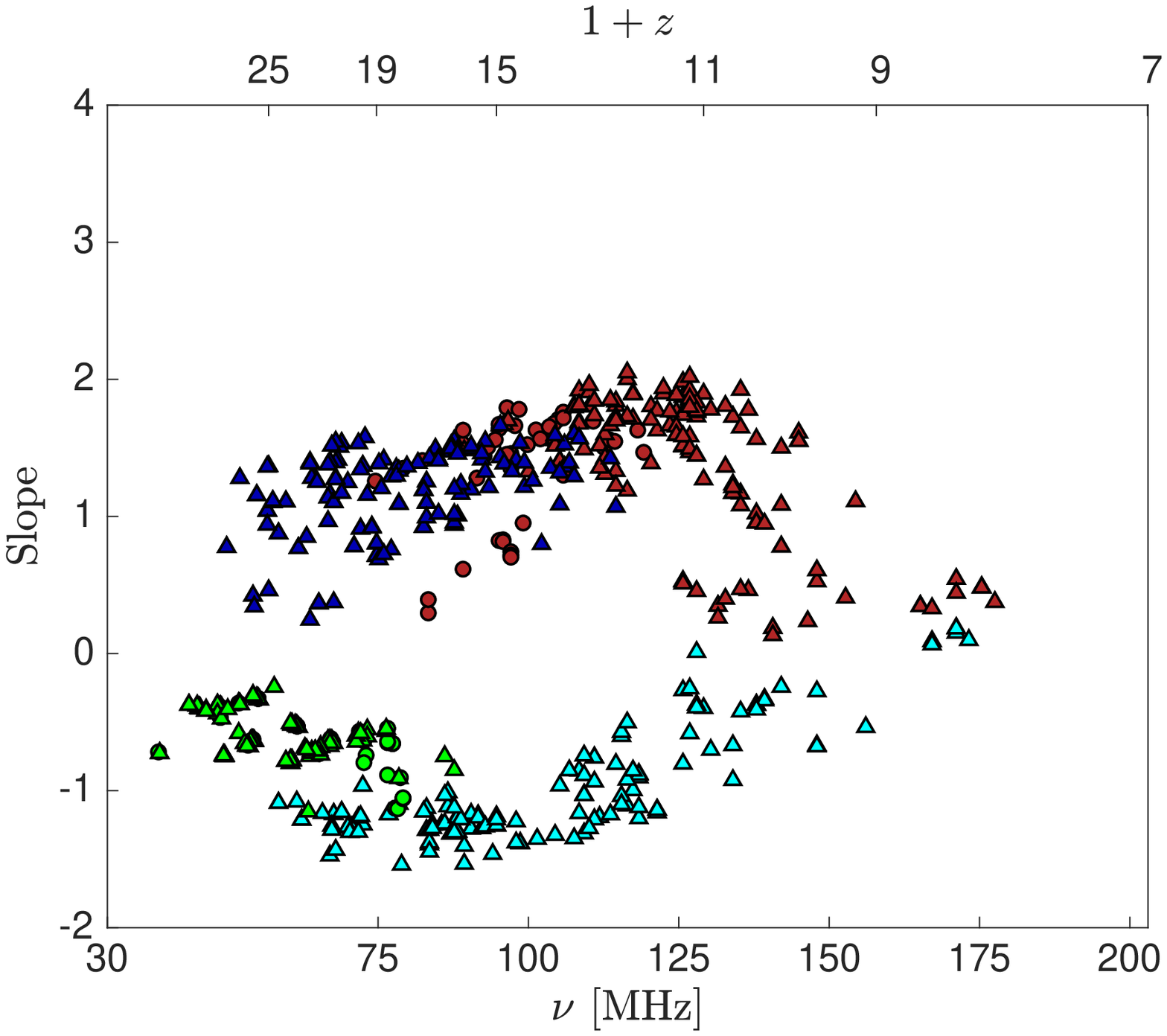}
 	\caption{{\bf Top} The slope of the 21-cm power spectrum on
          large ($k=0.05-0.2$ Mpc$^{-1}$, left) and small ($k=0.2-0.6$
          Mpc$^{-1}$, right) scales as a function of observed
          frequency/redshift (bottom/top axis) for all the cases
          discussed in this paper (light-blue). The standard case is
          shown as a black line. The full list of models appears in
          Appendix~\ref{appA}. {\bf Bottom} The extrema of the slope
          for large ($k=0.05-0.2$ Mpc$^{-1}$; left panel) and small
          ($k=0.2-0.6$ Mpc$^{-1}$; right panel) scales, presented for
          all cases. Marker shapes indicate the total number of minima
          or maxima for each case/model: 1 (circle), 2 (triangle) or 3
          (star). Marker colors indicate the type of the point: red
          (low-redshift maximum), cyan (low-redshift minimum), blue
          (high-redshift maximum), green (high-redshift minimum) and
          yellow (extra bump; see the right panel of
          Fig.~\ref{fig:Cases4}). Note that there are no stars or
          yellow points in the panel on the right.}
 	\label{fig:SlopeAllPeaks}
 \end{figure*}

As expected, the large uncertainty in the power spectrum leads to a
large uncertainty in its slope (top panels of
Fig.~\ref{fig:SlopeAllPeaks}).  It turns out that, unlike in the case
of the power spectra where the number of peaks varies for different
sets of parameters, in the vast majority of cases the slopes exhibit a
universal shape having two maxima and two minima (as was shown for the
standard case). This rule breaks down only in cases of very
inefficient heating where there is no heating transition before the
end of reionization, resulting in only one maximum and one minimum
(e.g., left panel of Fig.~\ref{fig:Cases4}). Another exception are the
cases with super-massive halos, which show a small extra peak (bump)
in the large scale slope before the high-redshift minimum (e.g., right
panel of Fig.~\ref{fig:Cases4}). The explanation of this bump is as
follows. When the radiative sources first turn on (and significantly
produce mainly Ly$\alpha$), there is a first stage where they only
reach relatively short distances (due to time retardation: at the
retarded time corresponding to a large distance, there were far fewer
sources in existence). Thus, at this stage the radiation amplifies
small-scale fluctuations, and the power-spectrum slope rises. After a
short time, the radiation reaches large scales and begins to smooth
over the small scales, and the power-spectrum slope falls. What
determines if this bump is seen in the total power-spectrum slope is
whether, during this short initial period, the radiation already has a
significant effect on the 21-cm signal. In most models, the Ly$\alpha$
flux turns on (and goes through this initial period) very early (at
$z>30$), when collisional coupling is still significant, so that the
Ly$\alpha$ coupling at this time is negligible. However, in models
with very massive halos, the Ly$\alpha$ turn-on occurs so late (at $z
< 25$) that collisional coupling is negligible, and even the earliest
Ly$\alpha$ coupling immediately dominates the 21-cm emission. Note,
thought, that since the bump is only seen when the coupling is very
weak, the overall height of the power spectrum at this point is very
low ($< 10^{-2}$~mK$^2$) and thus very hard to observe.

In total, (i) 150 out of 193 cases have four extrema in the slope, including both
high- and low-redshift maxima and minima, and (ii) the rest of the
cases have two extrema: a high-redshift minimum and low-redshift
maximum. In both type (i) and (ii) cases, a bump can additionally
appear depending on $M_{\rm min}$. Although the exact position of each
feature might vary, the overall shape with redshift of the slope is
uniform (apart from the low heating cases) and allows us to correlate
each of its features with specific cosmic events, as we discuss in the
next subsection. The bottom panels of Fig.~\ref{fig:SlopeAllPeaks}
shows the distribution of the extrema of the slopes. It is interesting
to note that on the scatter plots there is little overlap in the
redshift--slope plane between the areas populated by the various
extrema, unlike in the case of the power spectra where there is more
overlap (Fig. \ref{fig:peaks}). This property should make it easier to
correctly classify each feature in a slope versus redshift plot made
from future observed data.

To summarize (see also Table~\ref{Table:Notations}), the
classification of the power spectra and the slopes is more difficult
than that of the global 21-cm spectra because of the diversity of
features and shapes. In the case of the power spectrum we use peaks to
classify each case and label the peaks according to the leading source
of fluctuations ($z_{\rm PS,density}$, $z_{\rm PS, coup}$, $z_{\rm
  PS,heat}$, $z_{\rm PS,ion}$); while for the slope we refer to the
high and low minima and maxima ($z_{\rm dPS,min}^{\rm \, hi}$, $z_{\rm
  dPS,max}^{\rm \, hi}$, $z_{\rm dPS,min}^{\rm \, lo}$, $z_{\rm
  dPS,max}^{\rm \, lo}$). Although we do show cases with $\tau>0.09$,
we exclude them from all the quantitative results in the paper as
those are more than $3\sigma$ away from the optical depth measured by
the {\it Planck} satellite.

  \begin{table*}
  	\begin{center}
  		\begin{tabular}{|l|l|l|l}
  			\hline  & \textbf{Structure} & \textbf{Notations} & \textbf{Figure}\\ 
  			\hline \textbf{Global signal} & \vtop{\hbox{\strut  2 maxima and 1 minimum point}} & $z_{\rm g,max}^{\rm \, hi}$, $z_{\rm g,min}$, $z_{\rm g,max}^{\rm \, lo}$ & Fig.~\ref{fig:AllPS} left panel\\ 
  			\hline \textbf{Power spectrum} & \vtop{\hbox{\strut Between 2 and 4 peaks}}   & \vtop{\hbox{\strut $z_{\rm PS, coup}$, $z_{\rm PS,density}$, $z_{\rm PS,heat}$, $z_{\rm PS,ion}$,}\hbox{\strut each peak classified by its dominant component}}  & Figs.~\ref{fig:Cases1} and \ref{fig:peaks}\\ 
  			\hline \textbf{Power spectrum slope} & \vtop{\hbox{\strut 2 minima and 2 maxima or}\hbox{\strut 1 maximum and 1 minimum}} & \vtop{\hbox{\strut $z_{\rm dPS,min}^{\rm \, hi}$, $z_{\rm dPS,max}^{\rm \, hi}$, $z_{\rm dPS,min}^{\rm \, lo}$, $z_{\rm dPS,max}^{\rm \, lo}$ or}\hbox{\strut $z_{\rm dPS,min}^{\rm \, hi}$, $z_{\rm dPS,max}^{\rm \, lo}$}}  & Figs.~\ref{fig:Slopes1} and \ref{fig:SlopeAllPeaks}\\ 
  			\hline 
  		\end{tabular} 
  		\caption{Summary of features and notations that we use
                  to categorize the global signal, the power spectrum,
                  and its slope. Note that the slope also sometimes
                  has an extra small bump at very high redshift (right
                  panel of Fig.~\ref{fig:Cases4}).}
  		\label{Table:Notations}
  	\end{center}
  \end{table*}

\subsection{Extracting  Astrophysical Properties}

As we discussed in Section~\ref{Sec:results}, features of both the
power spectrum and its slope are closely related to the physical
properties of the early universe, as are features of the global 21-cm
spectrum.  \citet{Cohen:2016b} related each of the observable
features of the global signal to the mean astrophysical properties. In
particular, the location of the high-redshift maximum is closely
related to $V_c$ and $f_*$ and can be translated to the mean intensity
of the Ly$\alpha$ background at that redshift \citep[Fig. 3
  of][]{Cohen:2016b}; the depth of the absorption trough is set by
$V_c$, $f_*$, $f_X$ and the X-ray SED, and correlates with the ratio
of the Ly$\alpha$ intensity to the X-ray heating rate at the redshift
of the minimum \citep[Fig. 6 of][]{Cohen:2016b}; while the emission
peak is driven by $V_c$, $f_*$, $f_X$, the X-ray SED and $\tau$, and
can constrain the mean heating rate over the mean ionizing efficiency
\citep[Fig. 8 of][]{Cohen:2016b}. Here we use the same set of models
to establish universal relationships between the features of the power
spectrum (and its slope) and the astrophysical parameters.

\begin{figure*}
	\centering
	\includegraphics[width=3.2in]{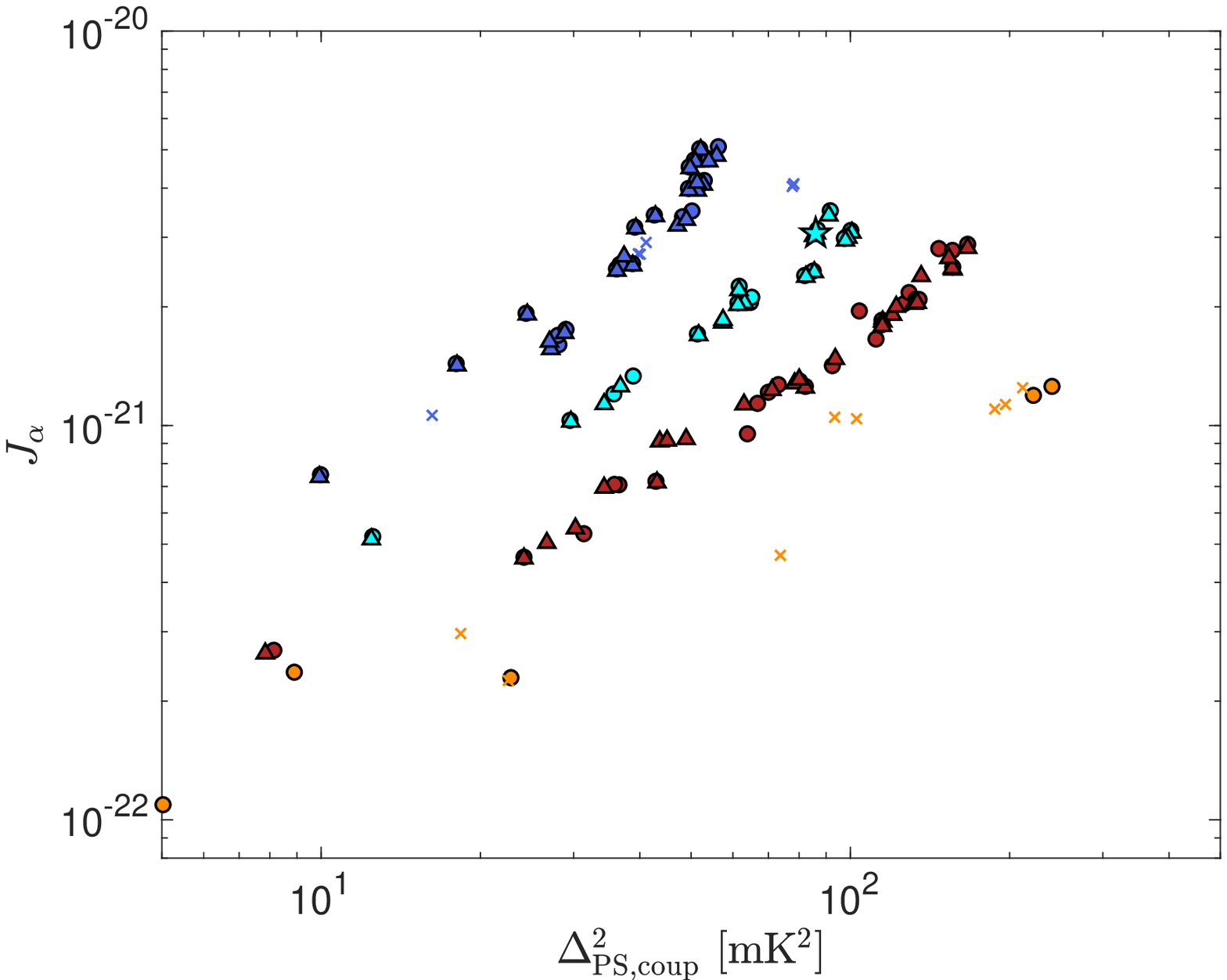}
	\hspace{0.3in}
		\includegraphics[width=3.35in]{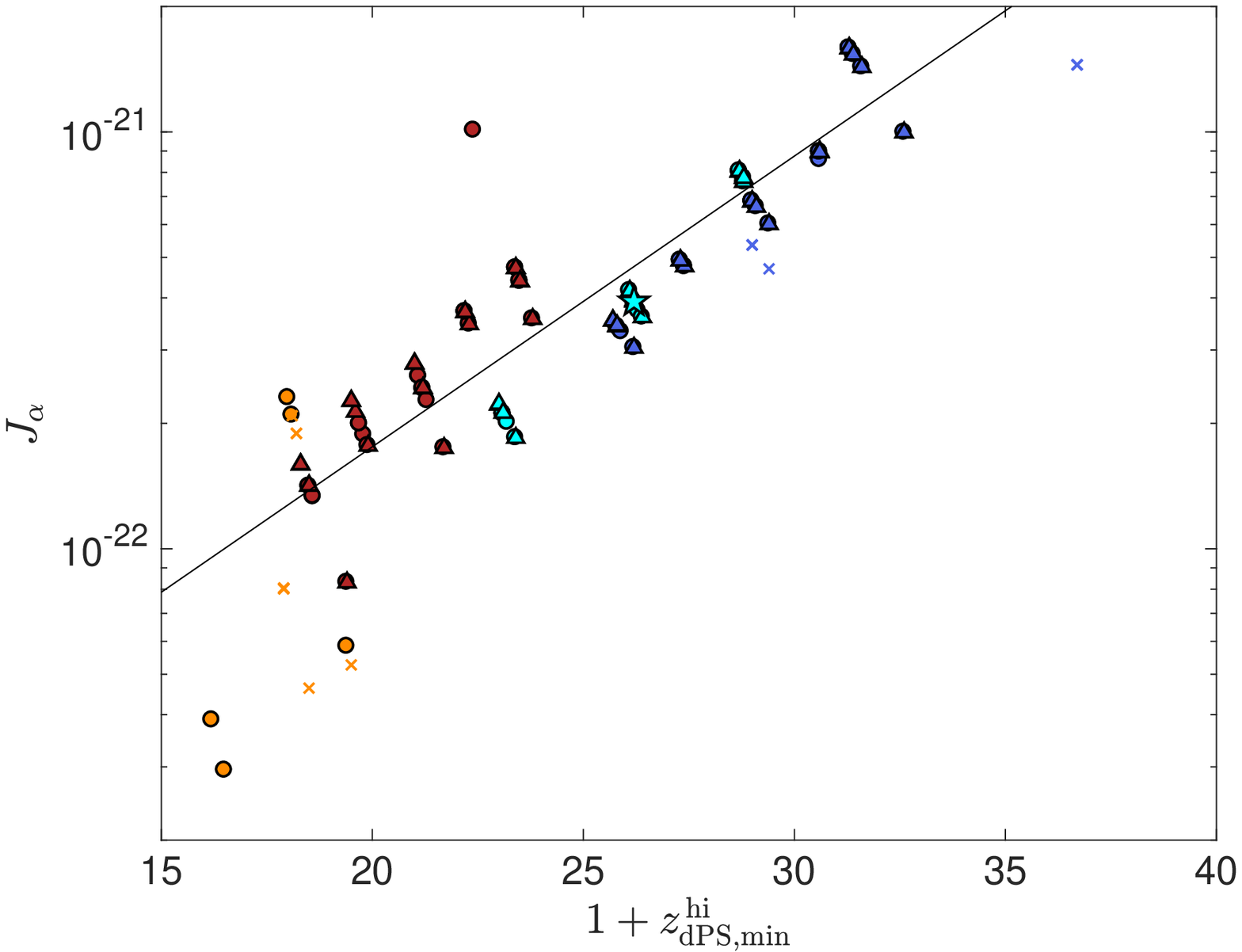}
	\caption{{\bf Left:} The Ly$\alpha$ intensity in units of erg
          s$^{-1}$ cm$^{-2}$ Hz$^{-1}$ sr$^{-1}$ as a function of the
          power spectrum at $k=0.1$ Mpc$^{-1}$ at the Ly$\alpha$
          dominant peak. Marker colors indicate the minimum circular
          velocity of star-forming halos for each case: $V_c=4.2$
          (blue), 16.5 (cyan), 35.5 (red), and 76.5 km s$^{-1}$
          (orange). Marker shapes indicate the optical depth for each
          case: $\tau=0.060 - 0.075$ (circles), $0.082 - 0.09$
          (triangles), $0.09 - 0.111$ (crosses), while the star is our
          standard case. {\bf Right} Mean Ly$\alpha$ intensity in
          units of erg s$^{-1}$ cm$^{-2}$ Hz$^{-1}$ sr$^{-1}$ as a
          function of (one plus) the redshift of the high-redshift
          minimum of the small-scale power-spectrum slope. Markers
          (both colors and shapes) are as in the left panel. A solid
          line shows the best fit, Eq.~(\ref{eq:min2S_Jalpha}).}
	\label{fig:alphamaxP_Jalpha}
\end{figure*}

We start by analyzing the high-redshift domain where the 21-cm signal
is largely driven by the Ly$\alpha$ background. We find that for most
of the considered cases, on large scales, the Ly$\alpha$ background
dominates the highest redshift peak of the power spectrum. In general,
increasing the typical mass of halos that dominate star formation
lowers the mean Ly$\alpha$ intensity at the redshift of the peak, and
raises the power-spectrum (since massive halos are more strongly
biased). However, there is a lot of overlap in the ranges (see
Fig.~\ref{fig:alphamaxP_Jalpha}), so that a given observed
$\Delta^2_{\rm PS, coup}$ cannot be used to deduce the value of
$J_\alpha$ without knowing the minimum mass of star-forming halos.
The minimum cooling mass can be estimated separately, e.g., from
measuring the high-redshift maximum of the global signal
\citep{Cohen:2016b}. 
 
The dependence of the slope on redshift has a more robust shape than
that of the power spectrum. Therefore, it is simpler to analyze and
classify its features. By analyzing the high-redshift minimum of the
slope, $z_{\rm dPS,min}^{\rm \, hi}$, we find that the mean Ly$\alpha$
intensity at that redshift can be estimated (right panel of
Fig.~\ref{fig:alphamaxP_Jalpha}). The intensity can be fitted by a
simple relation:
\begin{equation}
\label{eq:min2S_Jalpha}
\log_{10}(J_\alpha)=a(1+z_{\rm dPS,min}^{\rm \, hi})+b\ ,
\end{equation}
where $[a,b]=[0.070,-23.1]$. As is evident from the plot, there is a
correlation between a model's position in the $z_{\rm dPS,min}^{\rm \,
  hi}-J_\alpha$ plane and the minimum mass of star-forming halos.

\begin{figure}
	\centering
	\includegraphics[width=3.2in]{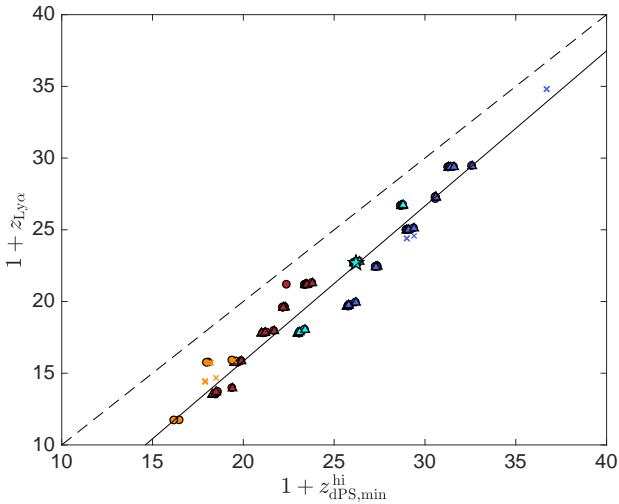}
	\caption{Redshift of the Ly$\alpha$ coupling transition (i.e.,
          $x_\alpha=1$) as a function of the redshift of the
          high-redshift minimum of the slope (at $k=0.2-0.6$
          Mpc$^{-1}$). Marker colors indicate the minimum circular
          velocity of star-forming halos for each case: $V_c=4.2$
          (blue), 16.5 (cyan), 35.5 (red), and 76.5 km s$^{-1}$
          (orange). Marker shapes indicate the optical depth for each
          case: $\tau=0.060 -0.075$ (circles), $0.082 - 0.09$
          (triangles), and $0.09 - 0.111$ (crosses), while the star is
          our standard case. We also show the line $Y=X$ (dashed) and
          the fitting formula from Eq.~(\ref{eq:min2S_LyaC}).  }
\label{fig:min2S_alpha}
\end{figure}

We find a cleaner correlation (Fig.~\ref{fig:min2S_alpha}) between the
redshift of Ly$\alpha$ coupling (i.e., $x_\alpha=1$ when averaged over
the simulated volume) and $z_{\rm dPS,min}^{\rm \, hi}$:
\begin{equation}
\label{eq:min2S_LyaC}
1+z_{\rm Ly\alpha}=a(1+z_{\rm dPS,min}^{\rm \, hi})+b\ ,
\end{equation}
where $[a,b]=[1.08,-5.80]$. We find that in all the models, the
Ly$\alpha$ coupling transition happens slightly later than the minimum
in the slope (here shown on small scales, where the correlation is
clearer).

\begin{figure*}
	\centering
	\includegraphics[width=3.2in]{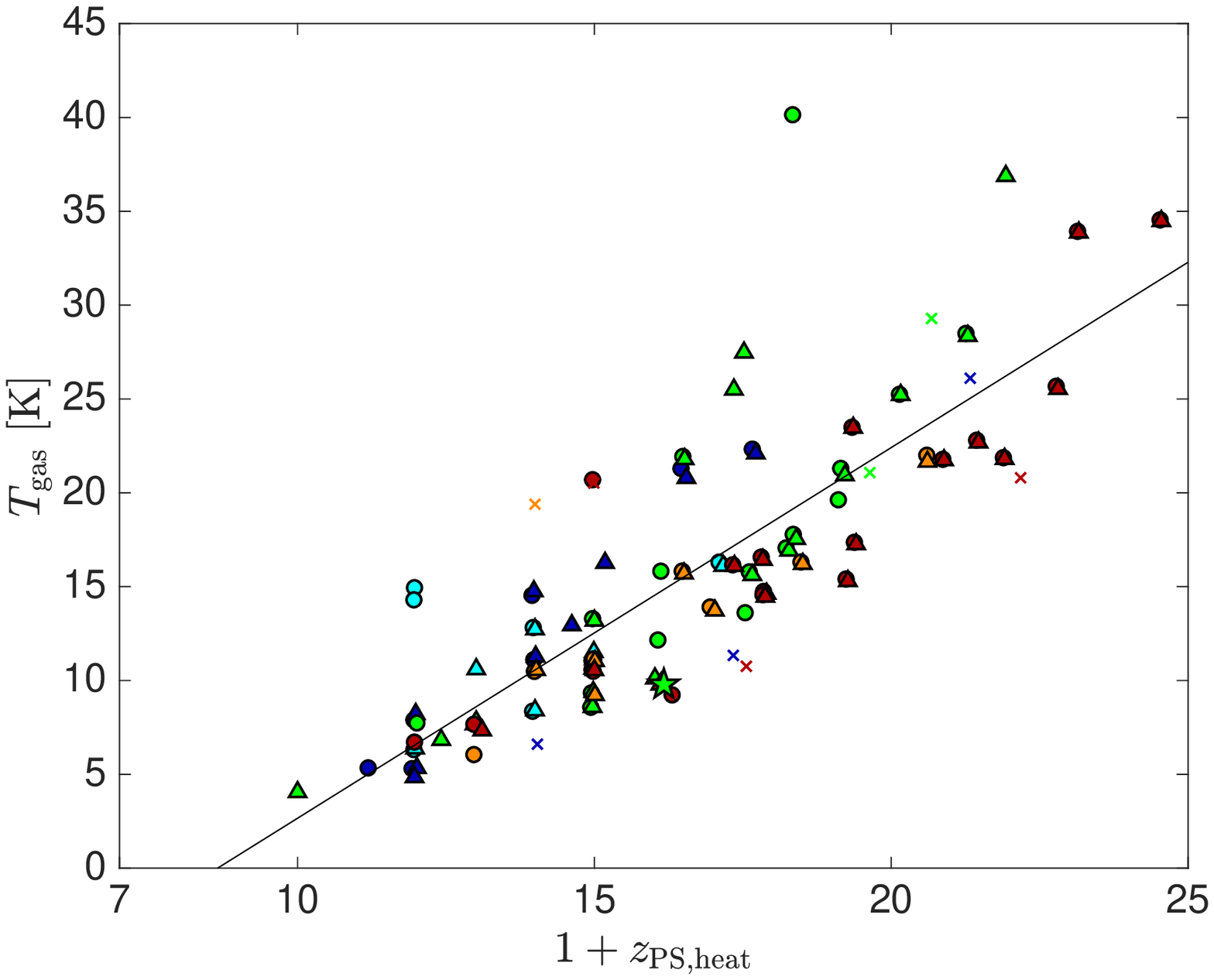}
			\hspace{0.5in}
			\includegraphics[width=3.2in]{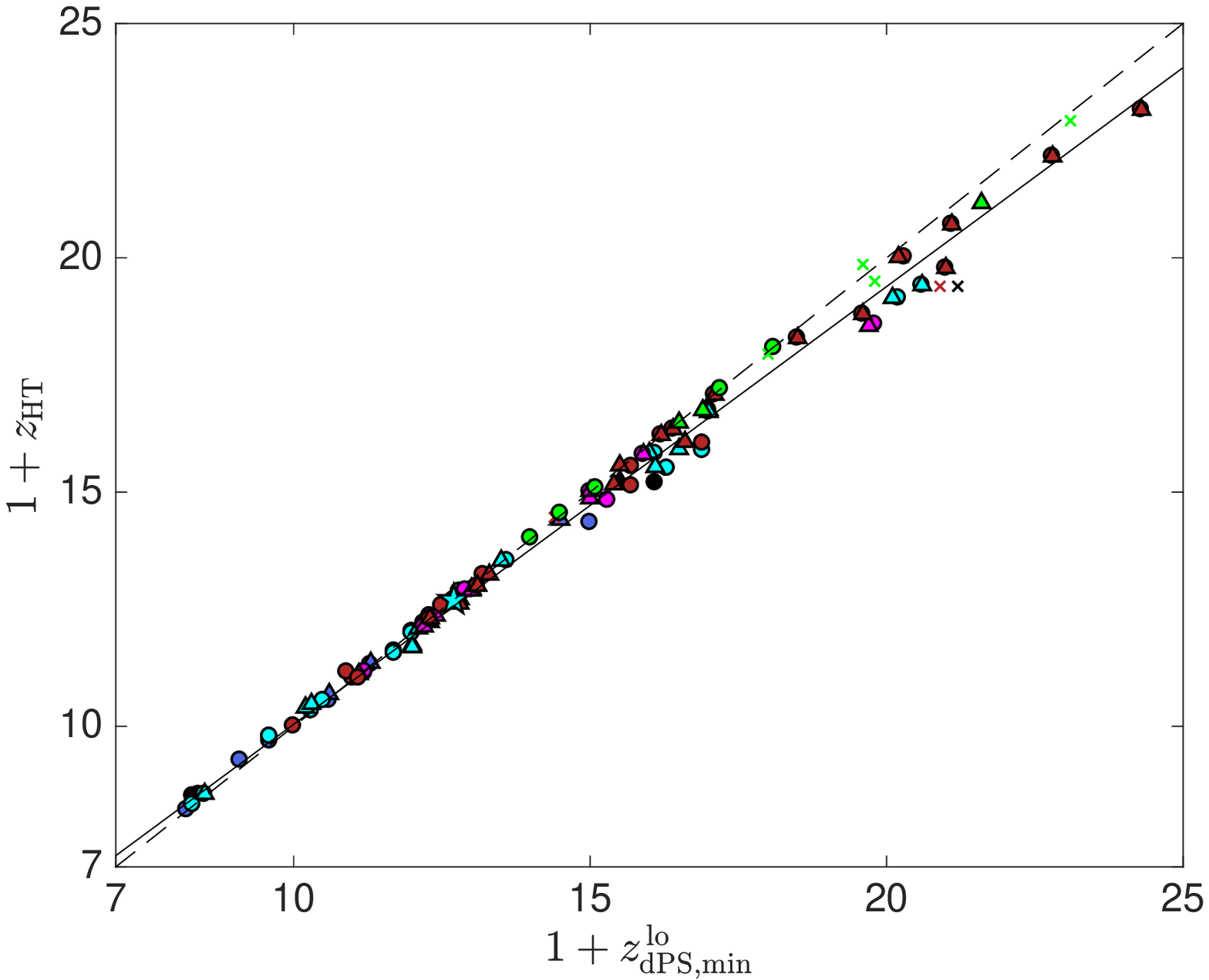}
	\caption{{\bf Left:} Mean gas temperature (at $z_{\rm
            PS,heat}$) as a function of $z_{\rm PS,heat}$ (at $k=0.1$
          Mpc$^{-1}$). The solid line shows the fitting formula of
          Eq.~(\ref{eq:XmaxZ_TK}). Colors indicate the star formation
          efficiency for each case: $f_*=0.005$ (blue), 0.016 (cyan),
          0.05 (green), 0.16 (orange) and 0.5 (red). Shapes indicate
          the optical depth for each case: $\tau=0.060 - 0.075$
          (circles), $0.082 - 0.09$ (triangles), $0.09 - 0.111$
          (crosses). {\bf Right:} Redshift of the heating transition
          ($T_{\rm CMB}=T_{\rm gas}$) as a function of the redshift of
          the low-redshift minimum of the slope (at
          $k=0.2-0.6$~Mpc$^{-1}$). The colors indicate the star
          formation efficiency for each case: $f_X=0.1$ (black), 0.32
          (blue), 1 (cyan), 3.16 (magenta), 8 or 10 (red) and upper
          limits (green). Shapes indicate the optical depth for each
          case: $\tau=0.060 -0.075$ (circles), $0.082 - 0.09$
          (triangles), $0.09 - 0.111$ (crosses). Also shown is the
          $Y=X$ line (dashed) and the fitting formula (solid) of
          Eq.~(\ref{eq:min1S_THT}). Note that this plot only includes
          models that have a heating transition before the end of
          reionization; the 39 excluded cases can be observationally
          recognized since they fall in the $\Delta^2_{\rm PS,ion} >
          150$ mK$^2$ part of Fig.~\ref{fig:ionmaxP_xHI} (except for
          one that does not have an ionization peak at
          $k=0.5$~Mpc$^{-1}$). }
	\label{fig:minVX_max}
\end{figure*}

Cosmic heating can be an extended process starting soon after the
first population of luminous sources is formed and extending
throughout the first half of reionization. X-rays emitted by the first
sources inject their energy into the IGM, driving its temperature up
and above that of the CMB. Therefore, properties of the first heating
sources are directly related to the thermal evolution of the IGM,
which affects the 21-cm signal and its fluctuations. The left panel of
Fig.~\ref{fig:minVX_max} shows how the redshift of the peak heating
fluctuations correlates with the mean temperature of the neutral gas
in the simulated volume.  In other words, with some scatter, the mean
gas temperature at high redshifts can be estimated from the position of
the heating peak in the 21-cm power spectrum, with the linear fit:
\begin{equation}
\label{eq:XmaxZ_TK}
T_{\rm gas}=a(1+z_{\rm PS,heat})+b\ ,
\end{equation}
where $[a,b]=[1.97,-17.1]$.

The effect of heating fluctuations on the 21-cm signal is most
significant around the heating transition ($T_{\rm CMB}=T_{\rm gas}$)
when the other fluctuations vanish. As heating becomes stronger it
spatially smooths the signal on small scales, decreasing the
slope. Therefore we expect the redshift of the heating transition to
correlate with the low-redshift minimum of the slope. Indeed, there is
almost a one-to-one correspondence with little scatter, as can be seen
in the right panel of Fig.~\ref{fig:minVX_max}.  The fitting formula
is
\begin{equation}
\label{eq:min1S_THT}
(1+z_{\rm HT})=a(1+z_{\rm dPS,min}^{\rm \, lo})+b\ ,
\end{equation}
where $[a,b]=[0.93,0.70]$. In other words, the low-redshift minimum
point is an excellent tracer of the redshift of the heating
transition, and can be used to establish the moment at which the IGM
was heated to the temperature of the CMB, which of course is known and
equals $2.725\times(1+z)$~K.

\begin{figure}
 	\centering
 	\includegraphics[width=3.2in]{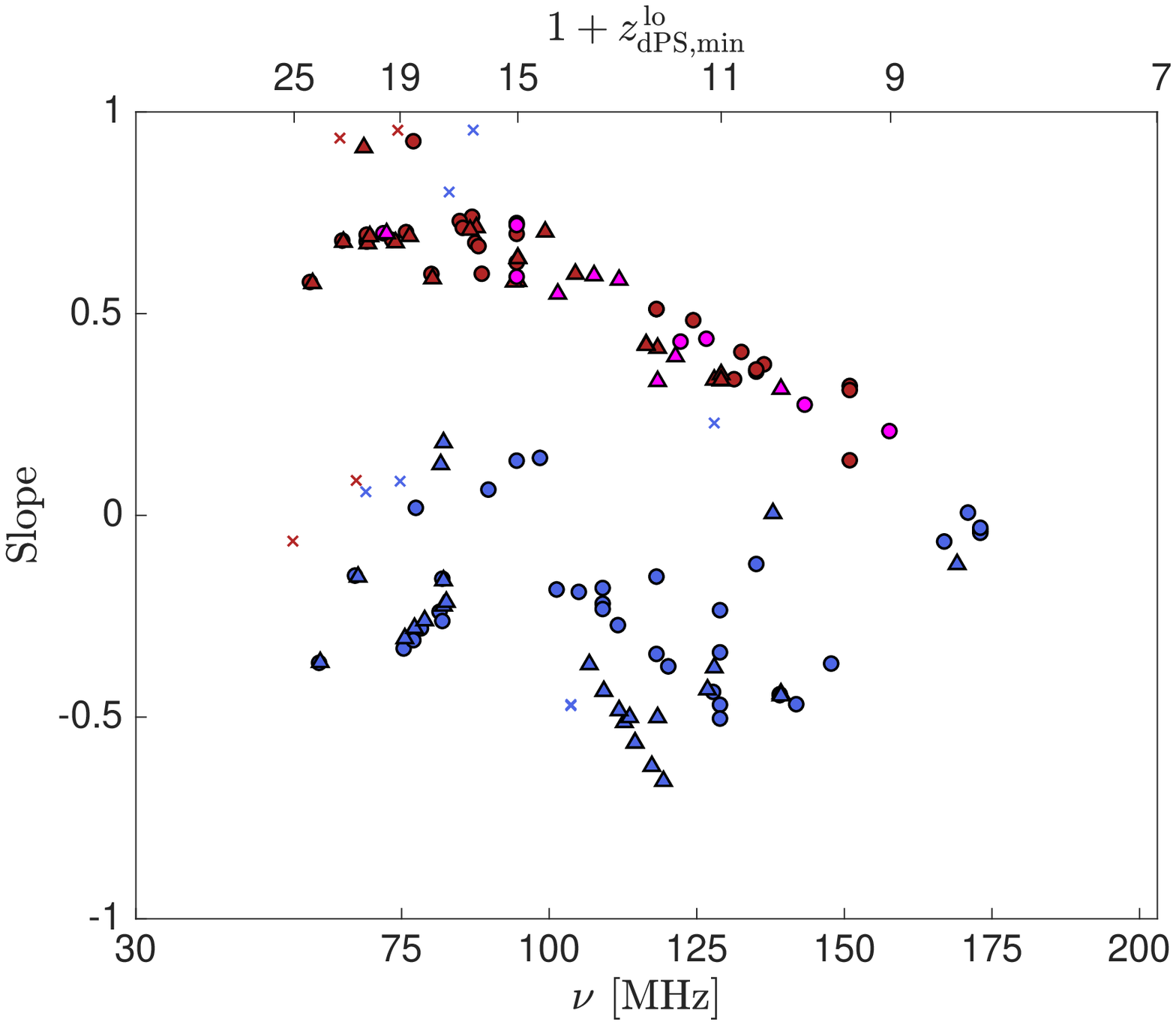}
 	\caption{Low-redshift minimum of the large-scale slope (i.e.,
          cyan points from the bottom left panel of
          Fig.~\ref{fig:SlopeAllPeaks}), with colors indicating the
          SED: soft (red), soft \& MQ (magenta) and hard (blue; i.e.,
          XRB, MQ or XRB \& MQ). Shapes indicate the optical depth for
          each case: $\tau=0.060 - 0.075$ (circles), $0.082 - 0.09$
          (triangles), $0.09 - 0.111$ (crosses). }
 	\label{fig:SED}
 \end{figure}

The depth of the low-redshift minimum of the slope depends on how
efficiently heating acts to smooth small-scale power in the 21-cm
signal. X-ray photons with high energies have a large mean free path,
thus contributing to smoother heating, and we expect the slope to be
low in the scenarios with hard SEDs. On the other hand, if the
majority of produced X-rays are soft, heating fluctuations occur on
small scales, producing a higher power-spectrum slope. This trend is
indeed seen in the simulations. Fig.~\ref{fig:SED} shows the
distribution of the low-redshift minimum point of the slope, with
colors indicating the SED of X-ray sources. Ignoring cases with an
unreasonably large $\tau$ (marked by crosses), there is the expected
clear separation between models with a hard (blue) and soft (red)
SED. Models where the X-ray population is mixed and includes both soft
sources and mini-quasars yield similar predictions to what we find in
the cases with a pure soft SED.  This is because mini-quasars become
significant only at relatively low redshifts. The prediction is clear:
if the slope is negative at the low-redshift minimum point, the
heating is due to sources emitting mainly hard X-rays; while if it is
positive, the heating is very likely dominated by soft X-rays. Thus,
measuring the power spectrum at its low-redshift minimum point will
allow us to strongly constrain the hardness of X-ray photons and shed
light on the nature of their sources. This is complementary to the
inference of the X-ray slope of the heating sources that is available
based on the anisotropy of the 21-cm power spectrum \citep{2015PRL}.

\begin{figure}
	\centering
	\includegraphics[width=3.2in]{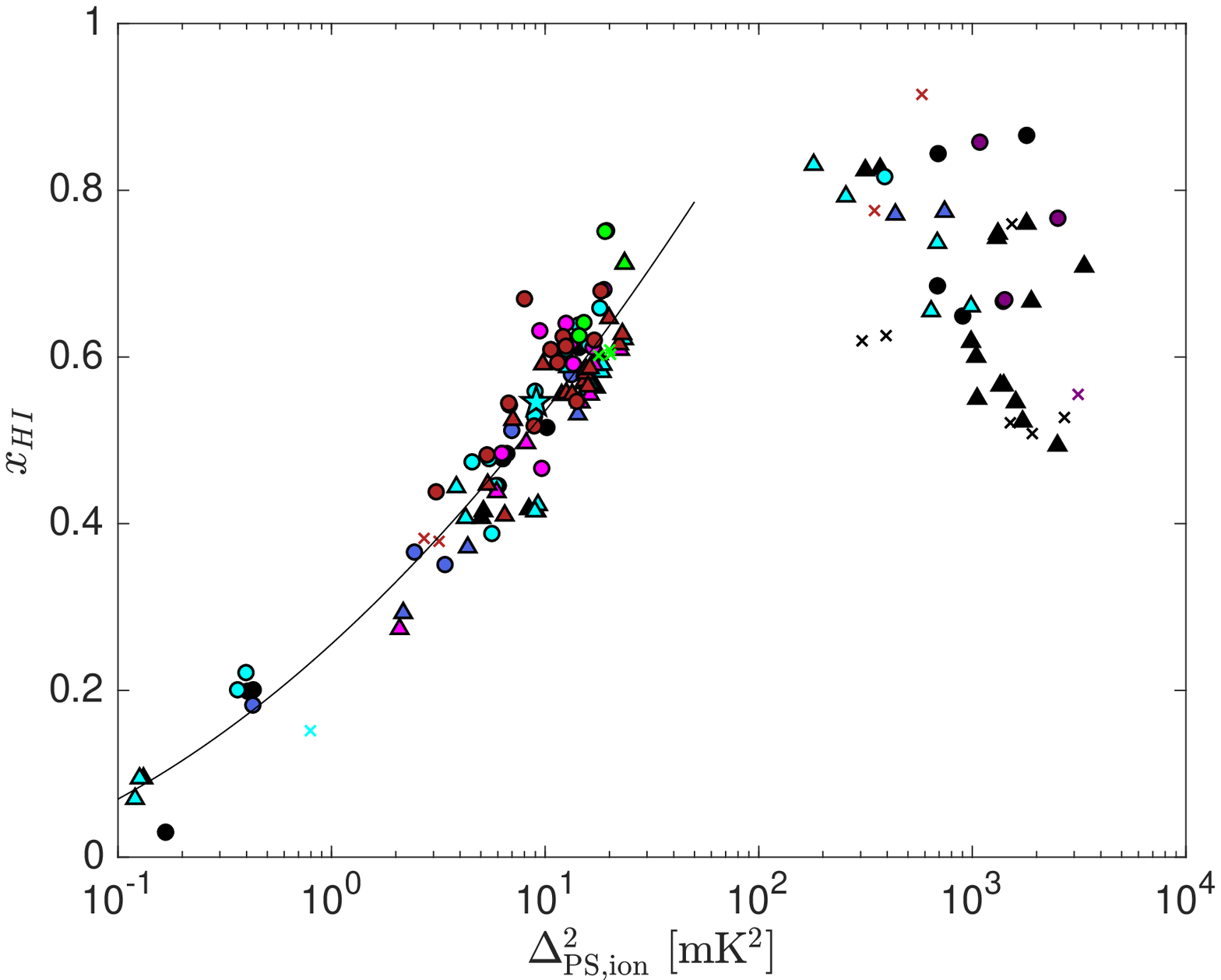}
	\caption{Neutral hydrogen fraction as a function of the power
          spectrum at $k=0.5$ Mpc$^{-1}$ at $z_{\rm PS,ion}$. The
          colors indicate the heating efficiency for each case:
          $f_X=0.1$ (black), 0.32 (blue), 1 (cyan), 3.16 (magenta), 8
          or 10 (red), lower limits (purple) and upper limits
          (green). Shapes indicate the optical depth for each case:
          $\tau=0.060 -0.075$ (circles), $0.082 - 0.09$ (triangles),
          $0.09 - 0.111$ (crosses).  Also shown is the fitting formula
          from Eq.~(\ref{eq:ionmaxP_xHI}).}
	\label{fig:ionmaxP_xHI}
\end{figure}

For a generic set of parameters the 21-cm signal during reionization
is expected to be strongly affected by both the parameters of heating
and the ionizing properties of stars. However, in cases when heating
is efficient and saturates early enough, fluctuations in the signal
are driven by the bubble structure and density perturbations. As
expected, when plotting the neutral fraction at the redshift of the
peak versus the peak power during reionization
(Fig.~\ref{fig:ionmaxP_xHI}), we find a one to one correspondence for
the models where heating is strong. In these models heating is
saturated (or almost saturated) during reionization, which results in
a relatively low ionization peak ($\Delta^2_{\rm PS,ion}\lesssim 30$
mK$^2$).  In these cases the 21-cm signal can be directly used as a
tracer of the reionization history and as a tool to reconstruct the
optical depth to reionization \citep{Barkana:2009, Liu:2016,
  Fialkov:2016b}. However, the spectrum does not trace the neutral
fraction when heating is weak. In this case the peak is very strong
($\Delta^2_{\rm PS,ion}\gtrsim 150$ mK$^2$), because the cosmic gas is
much colder than the CMB, and the peak occurs when the universe is
still mostly neutral; the peak depends strongly on the thermal history
in this case. Fig.~\ref{fig:ionmaxP_xHI} shows how an ionization peak
with a relatively low amplitude ($\Delta^2_{\rm PS,ion}\lesssim 30$
mK$^2$) can be used to estimate the neutral fraction at the redshift
$z_{\rm PS,ion}$. The fitting formula is:
\begin{equation}
\label{eq:ionmaxP_xHI}
x_{\rm HI}=a\log_{10}^2(\Delta^2_{\rm
  PS,ion})+b\log_{10}(\Delta^2_{\rm PS,ion})+c\ ,
\end{equation}
where $[a,b,c]=[0.047,0.23,0.25]$. On the other hand, most models with
low $f_X$ have high peaks which are uncorrelated with $x_{\rm HI}$.

Overall we find that it is difficult to extract useful astrophysical
information directly from the features of the power spectrum. In
addition, once a peak is detected it might not always be easy to
identify its origin, i.e., whether it is driven by fluctuations in
density, coupling, heating or ionization fraction. Out of our 193
cases, 177 have a peak at $k=0.1$ dominated by ionization
fluctuations, and it is always the lowest-redshift peak of the power
spectrum; the other 16 have a lowest-redshift peak dominated by
heating (these are the red circles in the left panel of
Fig.~\ref{fig:peaks}). The same is true at $k=0.5$ for 155 cases, 34
of which have only one peak, the ionization peak. At $k=0.1$, 184
cases have a peak dominated by Ly$\alpha$ fluctuations, and it is
always the highest-redshift peak. At $k=0.1$, 114 cases have three
power spectrum peaks; the highest-redshift one is
Ly$\alpha$-dominated, the lowest-redshift by ionization, and the
intermediate one by heating (except for two cases dominated by
$\delta+v$). We plan to further explore how to classify the peaks
directly from observations, using a larger number of simulated
cases. The slope of the power spectrum, though, appears to be a more
easily useful tool. It is easier to label its features and identify
the nature of cosmic events from its redshift evolution.

 \section{Consistency Check}
 \label{Sec:consist}

Using our bank of models we can look for relations between the
features of the power spectra (or the slope) and the features of the
global signal (high-redshift maximum, absorption trough, and emission
peak). If such relations exist they could be used as consistency
checks to verify the results of observations (e.g., comparing the
detected global signal and power spectrum and making sure that there
are not any large systematic errors). We do indeed find that the two
types of signal are correlated. In this Section we list a few selected
relations using the power spectrum slope, rather than directly the
power spectrum height, as our main tracer of fluctuations.

\begin{figure*}
	\centering	
	\begin{subfigure}[b]{0.4\textwidth}
	\includegraphics[width=3.1in]{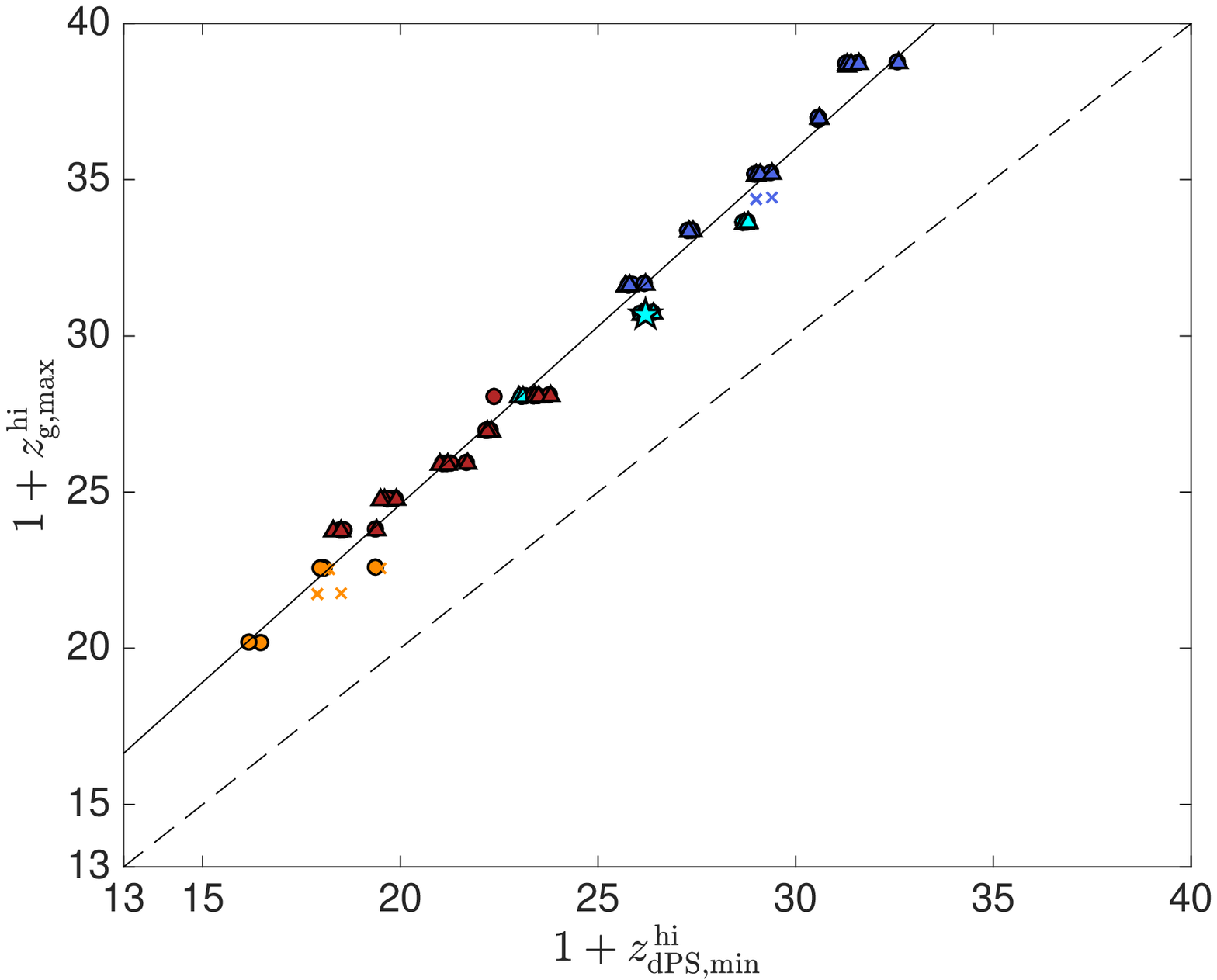}
	\vspace{0.05in}
\end{subfigure}
	\hspace{0.5in}
	\begin{subfigure}[b]{0.4\textwidth}
	\includegraphics[width=3.1in]{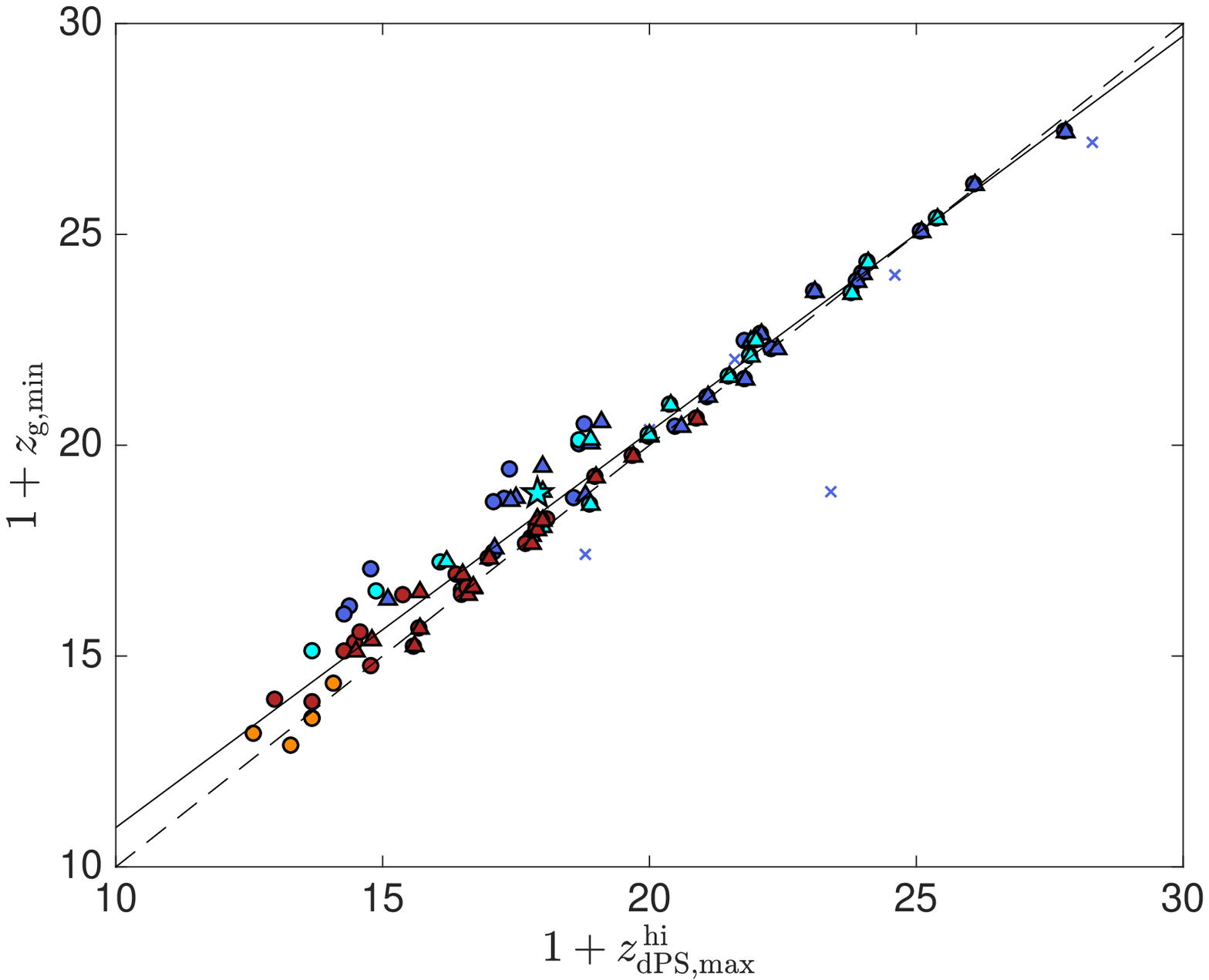}
	\vspace{0.05in}
\end{subfigure}
	\hspace{0.5in}
\begin{subfigure}[b]{0.4\textwidth}
	\includegraphics[width=3.1in]{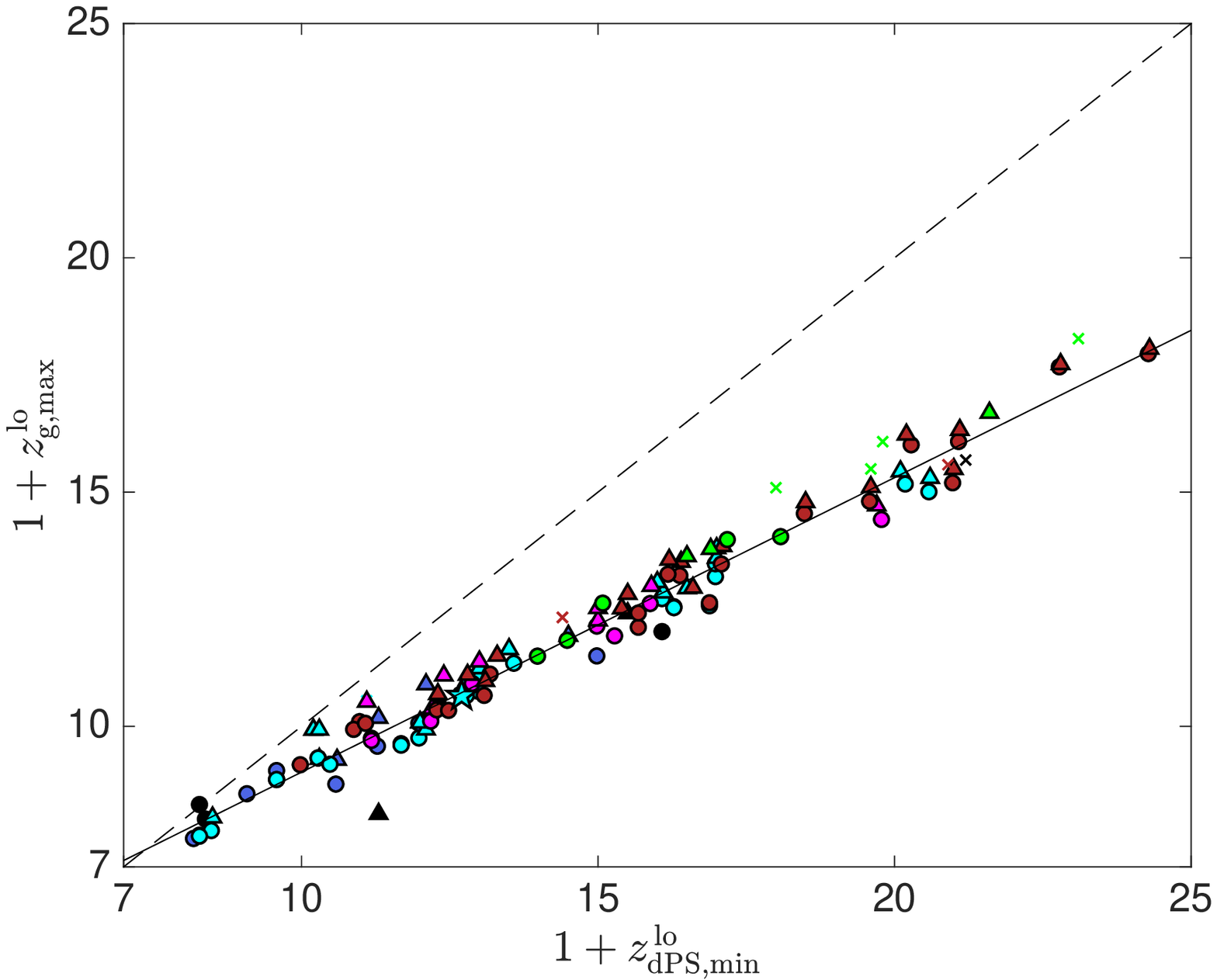}
	\vspace{0.05in}
	\end{subfigure}
\hspace{0.5in}
\begin{subfigure}[b]{0.4\textwidth}
	\includegraphics[width=3.1in]{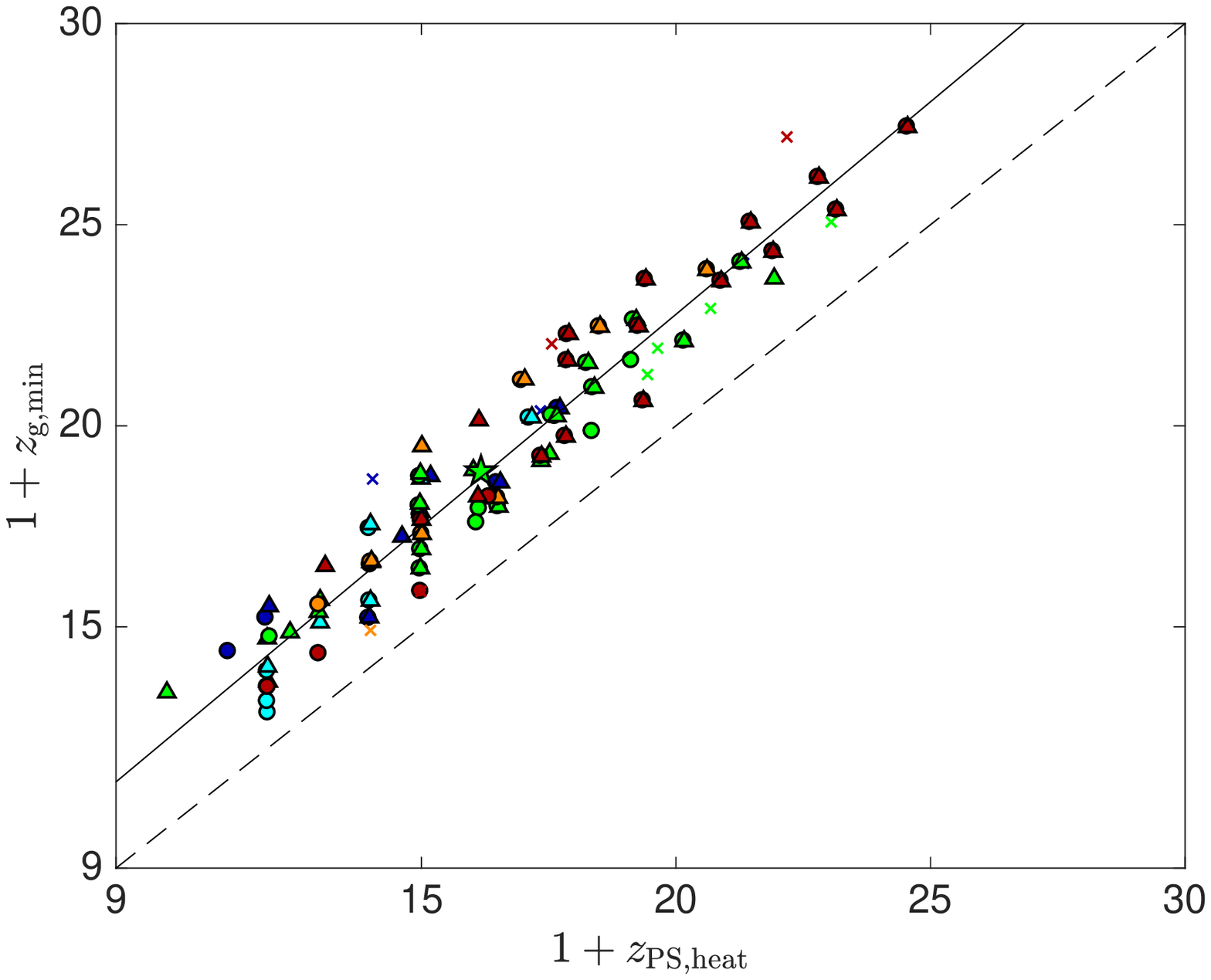}
	\vspace{0.05in}
\end{subfigure}
\caption{Consistency relations between the features of the global
  signal and the features of the power spectrum/slope. In addition to
  the distribution of data points we show the line $Y=X$ (dashed) and
  a linear fit in each case (solid line). In all panels, marker shapes
  indicate the optical depth for each case: $\tau=0.060 - 0.075$
  (circles), $0.082 - 0.09$ (triangles), and $0.09 - 0.111$ (crosses),
  while the star is our standard case. {\bf Top Left}: The redshift of
  the high-redshift maximum in the global signal as a function of the
  high-redshift minimum of the small-scale slope ($k=0.2-0.6$
  Mpc$^{-1}$). Marker colors indicate the minimum circular velocity of
  star-forming halos: $V_c=4.2$ (blue), 16.5 (cyan), 35.5 (red), and
  76.5 km s$^{-1}$ (orange).  We also show the fit,
  Eq.~(\ref{eq:min2S_GlobalfMax}). {\bf Top Right}: $1+z_{\rm g,min}$
  versus $1+z_{\rm dPS,max}^{\rm \, hi}$ (calculated at $k=0.05-0.2$
  Mpc$^{-1}$). Colors and shapes of the markers are as in the left
  panel. The fitting formula is from
  Eq.~(\ref{eq:max2L_GlobalMin}). {\bf Bottom Left:} Redshift of the
  low-redshift maximum of the global signal as a function of the
  low-redshift minimum of the slope (at $k=0.2-0.6$
  Mpc$^{-1}$). Marker colors indicate the star formation efficiency
  for each case: $f_X=0.1$ (black), 0.32 (blue), 1 (cyan), 3.16
  (magenta), 8 or 10 (red) and upper limits (green). The linear fit is
  from Eq.~(\ref{eq:min1S_GlobalMax}).  {\bf Bottom Right:} The
  redshift of the heating peak at $k=0.1$ Mpc$^{-1}$, $z_{\rm
    PS,heat}$, as a function of the redshift of the absorption trough
  in the global signal, $z_{\rm g,min}$. Also shown is the linear fit
  from Eq.~(\ref{eq:X_peak}).}
\label{fig:min2S_GlobalfMax}
\end{figure*}

The end of the cosmic dark ages, marked by the high-redshift maximum
in the global signal, is correlated with the high-redshift minimum of
the slope, with little scatter (top left panel of
Fig.~\ref{fig:min2S_GlobalfMax}). This is because both of these
redshifts are directly related to the early Ly$\alpha$ sources. The
maximum of the global signal occurs when the Ly$\alpha$ background is
first significant, while the minimum of the slope occurs somewhat
later, near the Ly$\alpha$ transition. We find the following best-fit
relation between the two:
\begin{equation}
\label{eq:min2S_GlobalfMax}
1+z_{\rm g,max}^{\rm \, hi}=a(1+z_{\rm dPS,min}^{\rm \, hi})+b\ ,
\end{equation}
where $[a,b]=[1.14,1.83]$. As we discussed in the previous Section,
the redshift of this transition is also related to the minimum mass of
halos in which stars form. Therefore, measuring either $z_{\rm
  g,max}^{\rm \, hi}$ or $z_{\rm dPS,min}^{\rm \, hi}$ will yield
constraints on the main cooling channel in the early Universe.

The high-redshift maximum of the slope depends on fluctuations in both
heating and density. Before the heating transition, these two
components have different signs (anti-correlate) and partially cancel
each other, and at the maximum point the heating fluctuations are
large enough to cancel out $\delta+v$ on large scales. We expect this
instant to be correlated with the position of the absorption trough
of the global signal, $z_{\rm g,min}$, which occurs when heating is
strong enough to oppose the adiabatic cooling due to the expanding
universe and start to heat the gas up. In fact, as we can see from the
top right panel of Fig.~\ref{fig:min2S_GlobalfMax}, the two redshifts,
$z_{\rm g,min}$ and $z_{\rm dPS,max}^{\rm \, hi}$, are nearly equal,
with relatively low scatter (especially at high redshifts). In most
cases, the minimum point of the global signal occurs first. Our best
fit is
\begin{equation}
\label{eq:max2L_GlobalMin}
1+z_{\rm g,min}=a(1+z_{\rm dPS,max}^{\rm \, hi})+b\ ,
\end{equation}
where $[a,b]=[0.94,1.55]$. We also show a relation directly with the
amplitude of the power spectrum, namely between $z_{\rm g,min}$ and
the redshift of a peak dominated by heating fluctuations. The
correlation, shown on the bottom right panel of
Fig.~\ref{fig:min2S_GlobalfMax}, can be fitted by the following
formula:
\begin{equation}
\label{eq:X_peak}
(1+z_{\rm g,min}) = a(1+z_{\rm PS,heat})+b\ ,
\end{equation}
where $[a,b]=[1.06,1.63]$.

Finally, as we know from our earlier work \citep[Figure 7
  of][]{Cohen:2016b} both the position and height of the low-redshift
maximum of the global signal depend on the average heating rate, as
this maximum occurs when heating saturates. Because the redshift of
the low-redshift minimum of the slope occurs around the heating
transition, we expect $z_{\rm g,max}^{\rm hi}$ and $z_{\rm
  dPS,min}^{\rm lo}$ to be correlated, with $z_{\rm dPS,min}^{\rm lo}$
somewhat higher. Our models yield the best-fit relation
\begin{equation}
\label{eq:min1S_GlobalMax}
(1+z_{\rm g,max}^{\rm hi})=a(1+z_{\rm dPS,min}^{\rm lo})+b\ ,
\end{equation}
where $[a,b]=[0.63,2.73]$, shown in the bottom left panel of
Fig.~\ref{fig:min2S_GlobalfMax}.

\section{Summary and Discussion}
\label{Sec:sum}

The power spectrum of the 21-cm signal contains rich information about
the early universe; however, it might be challenging to interpret the
signal in terms of astrophysical parameters. In this paper we have
used 193 different astrophysical scenarios to build relations between
the observable features of the power spectra and astrophysical
properties at high redshifts. We applied a semi-numerical simulation
to produce a realization of the 21-cm signal for each model, using the
same parameter space introduced in detail by \citet{Cohen:2016b}. The
astrophysical parameters can be sub-divided into three main
categories: properties of early star formation (minimum halo mass and
star formation efficiency), properties of the heating sources (the
X-ray spectrum and the total luminosity), and properties of
reionization (the CMB optical depth). In addition, feedback mechanisms
are taken into account when appropriate. The uncertainty in the
astrophysical parameters and feedback processes defines wide margins
within which the actual 21-cm signal can vary. We used the 193
different models to fill up the parameter space and explore possible
realizations of the signal.

First, using a few particular cases, we explored the components of the
21-cm signal. There are four possible sources of fluctuations: density
(+velocity), Ly$\alpha$ background radiation, heating of the gas in
the IGM, and reionization of the IGM. Among these sources, the
radiative sources (UV, X-rays and Ly$\alpha$) are usually stressed
while the density fluctuations are thought to have a minor
contribution at redshifts and scales of interest (accessible by the
current and the next generation of ground-based experiments such as
the SKA). However, by carefully tagging power spectra peaks and
exploring their origins, we found that density fluctuations can play
an important role in driving the signal in the intermediate redshift
range and on observable scales. This makes the 21-cm signal directly
sensitive to fundamental cosmology, though in practice it will still
be difficult to separate the density contribution from those
of the other sources of 21-cm fluctuations.

The 21-cm power spectrum can have up to three peaks (dominated by
density+velocity, Ly$\alpha$, X-rays and/or ionization). After tagging
each peak of the power spectrum according to the dominant source of
fluctuations, we related the observable properties (redshift and
amplitude of the peaks) to the astrophysical inputs in each case.
Generally, we found that the slope of the power spectrum (i.e., its
dependence on scale) has a more universal structure than the power
spectrum. The features include (from high to low redshifts): a minimum
point when the Ly$\alpha$ radiative background imprints large scale
fluctuations in the 21-cm signal, a maximum when heating cancels
density fluctuations on large scales, a minimum when heating
dominates, and finally a maximum when ionization cancels out the
density and heating fluctuations on large scales. (For cases with
inefficient heating, the high-redshift maximum and low-redshift
minimum disappear since heating fluctuations are not significant
enough.) Moreover, the magnitude of the slope gives useful information
for the astrophysical interpretation. A large slope indicates more
structure on small scales, and is characteristic of density+velocity
or moments where multiple fluctuation sources cancel each other out on
large scales. On the other hand, when a single radiative background
dominates, the power spectrum slope is low, indicating that the
structure on small scales has been washed out.

We showed results for the cleanest correlations that we found, and
provided simple fits. Our main results include:
\begin{itemize}
\item A map of the possible locations of peaks (versus redshift) of
  the power spectrum (Fig.~\ref{fig:peaks}) and of its slope (bottom
  panels of Fig.~\ref{fig:SlopeAllPeaks}).
\item The redshift of the Ly$\alpha$ coupling transition can be
  estimated from the redshift of the high-redshift minimum of the
  slope (Fig.~\ref{fig:min2S_alpha}).
\item The redshift of the heating transition (and the mean gas
  temperature of the IGM at that time) can be estimated from the
  redshift of the low-redshift minimum of the slope (right panel of
  Fig.~\ref{fig:minVX_max}).
\item The spectrum of the X-ray heating sources can be estimated from
  the properties of the low-redshift minimum of the slope
  (Fig.~\ref{fig:SED}).
\item By measuring the properties of the ionization-dominated peak of
  the power spectrum, it is possible to deduce whether heating
  occurred early or late; and if early, then the mean reionized
  fraction of the IGM at that time can also be estimated
  (Fig.~\ref{fig:ionmaxP_xHI}).
\item Significant correlations (with low scatter) are expected between
  some peaks of the power spectrum and its slope, and peaks of the
  global 21-cm signal (Fig.~\ref{fig:min2S_GlobalfMax}). This will
  give an important consistency check once these various observations
  are made.
\end{itemize}

\section{Acknowledgments}
This project/publication was made possible through the support of a
grant from the John Templeton Foundation. The opinions expressed in
this publication are those of the authors and do not necessarily
reflect the views of the John Templeton Foundation. R.B.\ and
A.C.\ also acknowledge Israel Science Foundation grant 823/09 and the
Ministry of Science and Technology, Israel. R.B.'s work has been
partly done within the Labex Institut Lagrange de Paris (ILP,
reference ANR-10-LABX-63) part of the Idex SUPER, and received
financial state aid managed by the Agence Nationale de la Recherche,
as part of the programme Investissements d'avenir under the reference
ANR-11-IDEX-0004-02. R.B.\ also acknowledges a Leverhulme Trust
Visiting Professorship. This research was supported in part by
Perimeter Institute for Theoretical Physics. Research at Perimeter
Institute is supported by the Government of Canada through the
Department of Innovation, Science and Economic Development Canada and
by the Province of Ontario through the Ministry of Research,
Innovation and Science.

\appendix
\section{Cases List}
\label{appA}

\begin{table*}
	\begin{center}
		\begin{tabular}{lllllllllll}
		\hline \#  &                       & $f_*$ & $V_c$ [km/s]& $f_X$ & SED        & $\tau$ & LW  & Low-mass cutoff                & $\zeta$ & $R_{\rm mfp}$ [Mpc] \\
			\hline 1   & Filler                & 0.005 & 4.2   & 0.1   & Hard       & 0.066  & On  & Eq.~(4) from C2016b  & 20   & 70   \\
			2   & Filler                & 0.005 & 4.2   & 0.1   & Hard       & 0.082  & On  & Eq.~(4) from C2016b  & 32    & 70  \\
			3   & Filler                & 0.005 & 4.2   & 0.1   & Soft       & 0.066  & On  & Eq.~(4) from C2016b  & 20    & 70  \\
			4   & Filler                & 0.005 & 4.2   & 0.1   & Soft       & 0.082  & On  & Eq.~(4) from C2016b  & 32    & 70  \\
			5   & Filler                & 0.005 & 4.2   & 1     & Hard       & 0.066  & On  & Eq.~(4) from C2016b  & 20    & 70  \\
			6   & Filler                & 0.005 & 4.2   & 1     & Hard       & 0.082  & On  & Eq.~(4) from C2016b  & 32    & 70  \\
			7   & Filler                & 0.005 & 4.2   & 1     & Soft       & 0.066  & On  & Eq.~(4) from C2016b  & 20    & 70  \\
			8   & Filler                & 0.005 & 4.2   & 1     & Soft       & 0.082  & On  & Eq.~(4) from C2016b  & 32    & 70  \\
			9   & Filler                & 0.005 & 4.2   & 8     & Hard       & 0.066  & On  & Eq.~(4) from C2016b  & 20    & 70  \\
			10  & Filler                & 0.005 & 4.2   & 8     & Hard       & 0.082  & On  & Eq.~(4) from C2016b  & 32    & 70  \\
			11  & Filler                & 0.005 & 4.2   & 8     & Soft       & 0.066  & On  & Eq.~(4) from C2016b  & 19    & 70  \\
			12  & Filler                & 0.005 & 4.2   & 8     & Soft       & 0.082  & On  & Eq.~(4) from C2016b  & 31    & 70  \\
			13  & Filler                & 0.05  & 4.2   & 0.1   & Hard       & 0.066  & On  & Eq.~(4) from C2016b  & 20    & 70  \\
			14  & Filler                & 0.05  & 4.2   & 0.1   & Hard       & 0.082  & On  & Eq.~(4) from C2016b  & 36    & 70  \\
			15  & Filler                & 0.05  & 4.2   & 0.1   & Soft       & 0.066  & On  & Eq.~(4) from C2016b  & 20    & 70  \\
			16  & Filler                & 0.05  & 4.2   & 0.1   & Soft       & 0.082  & On  & Eq.~(4) from C2016b  & 36    & 70  \\
			17  & Filler                & 0.05  & 4.2   & 1     & Hard       & 0.066  & On  & Eq.~(4) from C2016b  & 20    & 70  \\
			18  & Filler                & 0.05  & 4.2   & 1     & Hard       & 0.082  & On  & Eq.~(4) from C2016b  & 36    & 70  \\
			19  & Filler                & 0.05  & 4.2   & 1     & Soft       & 0.066  & On  & Eq.~(4) from C2016b  & 19    & 70  \\
			20  & Filler                & 0.05  & 4.2   & 1     & Soft       & 0.082  & On  & Eq.~(4) from C2016b  & 35    & 70  \\
			21  & Filler                & 0.05  & 4.2   & 8     & Hard       & 0.066  & On  & Eq.~(4) from C2016b  & 18    & 70  \\
			22  & Filler                & 0.05  & 4.2   & 8     & Hard       & 0.082  & On  & Eq.~(4) from C2016b  & 34    & 70  \\
			23  & Filler                & 0.05  & 4.2   & 8     & Soft       & 0.066  & On  & Eq.~(4) from C2016b  & 15    & 70  \\
			24  & Filler                & 0.05  & 4.2   & 8     & Soft       & 0.082  & On  & Eq.~(4) from C2016b  & 31    & 70  \\
			25  & Filler                & 0.5   & 4.2   & 0.1   & Hard       & 0.066  & On  & Eq.~(4) from C2016b  & 26    & 70  \\
			26  & Filler                & 0.5   & 4.2   & 0.1   & Hard       & 0.082  & On  & Eq.~(4) from C2016b  & 51    & 70  \\
			27  & Filler                & 0.5   & 4.2   & 0.1   & Soft       & 0.066  & On  & Eq.~(4) from C2016b  & 25    & 70  \\
			28  & Filler                & 0.5   & 4.2   & 0.1   & Soft       & 0.082  & On  & Eq.~(4) from C2016b  & 51    & 70  \\
			29  & Filler                & 0.5   & 4.2   & 1     & Hard       & 0.066  & On  & Eq.~(4) from C2016b  & 24    & 70  \\
			30  & Filler                & 0.5   & 4.2   & 1     & Hard       & 0.082  & On  & Eq.~(4) from C2016b  & 50    & 70  \\
			31  & Filler                & 0.5   & 4.2   & 1     & Soft       & 0.066  & On  & Eq.~(4) from C2016b  & 20    & 70  \\
			32  & Filler                & 0.5   & 4.2   & 1     & Soft       & 0.082  & On  & Eq.~(4) from C2016b  & 44    & 70  \\
			33  & Filler                & 0.5   & 4.2   & 8     & Hard       & 0.066  & On  & Eq.~(4) from C2016b  & 16    & 70  \\
			34  & Filler                & 0.5   & 4.2   & 8     & Hard       & 0.082  & On  & Eq.~(4) from C2016b  & 39    & 70  \\
			35  & Filler                & 0.5   & 4.2   & 8     & Soft       & 0.066  & On  & Eq.~(4) from C2016b  & 1.5   & 70  \\
			36  & Filler                & 0.5   & 4.2   & 8     & Soft       & 0.082  & On  & Eq.~(4) from C2016b  & 19    & 70  \\
			37  & Filler                & 0.005 & 16.5  & 0.1   & Hard       & 0.066  & -   & -                              & 20    & 70  \\
			38  & Filler                & 0.005 & 16.5  & 0.1   & Hard       & 0.082  & -   & -                              & 37    & 70  \\
			39  & Filler                & 0.005 & 16.5  & 0.1   & Soft       & 0.066  & -   & -                              & 20    & 70  \\
			40  & Filler                & 0.005 & 16.5  & 0.1   & Soft       & 0.082  & -   & -                              & 37    & 70  \\
			41  & Filler                & 0.005 & 16.5  & 1     & Hard       & 0.066  & -   & -                              & 20    & 70  \\
			42  & Filler                & 0.005 & 16.5  & 1     & Hard       & 0.082  & -   & -                              & 37    & 70  \\
			43  & Filler                & 0.005 & 16.5  & 1     & Soft       & 0.066  & -   & -                              & 20    & 70  \\
			44  & Filler                & 0.005 & 16.5  & 1     & Soft       & 0.082  & -   & -                              & 36    & 70  \\
			45  & Filler                & 0.005 & 16.5  & 8     & Hard       & 0.066  & -   & -                              & 20    & 70  \\
			46  & Filler                & 0.005 & 16.5  & 8     & Hard       & 0.082  & -   & -                              & 36    & 70  \\
			47  & Filler                & 0.005 & 16.5  & 8     & Soft       & 0.066  & -   & -                              & 19    & 70  \\
			48  & Filler                & 0.005 & 16.5  & 8     & Soft       & 0.082  & -   & -                              & 36    & 70  \\
			49  & Filler                & 0.05  & 16.5  & 0.1   & Hard       & 0.066  & -   & -                              & 20    & 70  \\
			50  & Filler                & 0.05  & 16.5  & 0.1   & Hard       & 0.082  & -   & -                              & 37    & 70  \\
			51  & Filler                & 0.05  & 16.5  & 0.1   & Soft       & 0.066  & -   & -                              & 20    & 70  \\
			52  & Filler                & 0.05  & 16.5  & 0.1   & Soft       & 0.082  & -   & -                              & 36    & 70  \\
			53  & Standard              & 0.05  & 16.5  & 1     & Hard       & 0.066  & -   & -                              & 20     & 70 \\
			54  & Filler                & 0.05  & 16.5  & 1     & Hard       & 0.082  & -   & -                              & 36    & 70  \\
			55  & Filler                & 0.05  & 16.5  & 1     & Soft       & 0.066  & -   & -                              & 19    & 70  \\
			56  & Filler                & 0.05  & 16.5  & 1     & Soft       & 0.082  & -   & -                              & 36    & 70  \\
			57  & Filler                & 0.05  & 16.5  & 8     & Hard       & 0.066  & -   & -                              & 19    & 70  \\
			58  & Filler                & 0.05  & 16.5  & 8     & Hard       & 0.082  & -   & -                              & 35    & 70  \\
			59  & Filler                & 0.05  & 16.5  & 8     & Soft       & 0.066  & -   & -                              & 15    & 70  \\
			60  & Filler                & 0.05  & 16.5  & 8     & Soft       & 0.082  & -   & -                              & 31   & 70   \\
			61  & Filler                & 0.5   & 16.5  & 0.1   & Hard       & 0.066  & -   & -                              & 20   & 70   \\
		\end{tabular}
	\end{center}
\end{table*}
\begin{table*}
	\begin{center}
		\begin{tabular}{lllllllllll}
			\hline \#  &                       & $f_*$ & $V_c$ [km/s]& $f_X$ & SED        & $\tau$ & LW  & Low-mass cutoff                & $\zeta$ & $R_{\rm mfp}$ [Mpc] \\
			\hline
			62  & Filler                & 0.5   & 16.5  & 0.1   & Hard       & 0.082  & -   & -                              & 36    & 70  \\
			63  & Filler                & 0.5   & 16.5  & 0.1   & Soft       & 0.066  & -   & -                              & 19    & 70  \\
			64  & Filler                & 0.5   & 16.5  & 0.1   & Soft       & 0.082  & -   & -                              & 36    & 70  \\
			65  & Filler                & 0.5   & 16.5  & 1     & Hard       & 0.066  & -   & -                              & 18    & 70  \\
			66  & Filler                & 0.5   & 16.5  & 1     & Hard       & 0.082  & -   & -                              & 35    & 70  \\
			67  & Filler                & 0.5   & 16.5  & 1     & Soft       & 0.066  & -   & -                              & 14    & 70  \\
			68  & Filler                & 0.5   & 16.5  & 1     & Soft       & 0.082  & -   & -                              & 30    & 70  \\
			69  & Filler                & 0.5   & 16.5  & 8     & Hard       & 0.066  & -   & -                              & 11.5  & 70  \\
			70  & Filler                & 0.5   & 16.5  & 8     & Hard       & 0.082  & -   & -                              & 27    & 70  \\
			71  & Filler                & 0.5   & 16.5  & 8     & Soft       & 0.066  & -   & -                              & 0.01  & 70  \\
			72  & Filler                & 0.5   & 16.5  & 8     & Soft       & 0.082  & -   & -                              & 9     & 70  \\
			73  & Filler                & 0.005 & 35.5  & 0.1   & Hard       & 0.082  & -   & -                              & 130   & 70  \\
			74  & Filler                & 0.005 & 35.5  & 0.1   & Soft       & 0.066  & -   & -                              & 53    & 70  \\
			75  & Filler                & 0.005 & 35.5  & 0.1   & Soft       & 0.082  & -   & -                              & 130   & 70  \\
			76  & Filler                & 0.005 & 35.5  & 1     & Hard       & 0.066  & -   & -                              & 53    & 70  \\
			77  & Filler                & 0.005 & 35.5  & 1     & Hard       & 0.082  & -   & -                              & 130   & 70  \\
			78  & Filler                & 0.005 & 35.5  & 1     & Soft       & 0.066  & -   & -                              & 53    & 70  \\
			79  & Filler                & 0.005 & 35.5  & 1     & Soft       & 0.082  & -   & -                              & 130   & 70  \\
			80  & Filler                & 0.005 & 35.5  & 8     & Hard       & 0.066  & -   & -                              & 53    & 70  \\
			81  & Filler                & 0.005 & 35.5  & 8     & Hard       & 0.082  & -   & -                              & 130   & 70  \\
			82  & Filler                & 0.005 & 35.5  & 8     & Soft       & 0.066  & -   & -                              & 53    & 70  \\
			83  & Filler                & 0.005 & 35.5  & 8     & Soft       & 0.082  & -   & -                              & 129   & 70  \\
			84  & Filler                & 0.05  & 35.5  & 0.1   & Hard       & 0.066  & -   & -                              & 53    & 70  \\
			85  & Filler                & 0.05  & 35.5  & 0.1   & Hard       & 0.082  & -   & -                              & 130   & 70  \\
			86  & Filler                & 0.05  & 35.5  & 0.1   & Soft       & 0.066  & -   & -                              & 53    & 70  \\
			87  & Filler                & 0.05  & 35.5  & 0.1   & Soft       & 0.082  & -   & -                              & 130   & 70  \\
			88  & Filler                & 0.05  & 35.5  & 1     & Hard       & 0.066  & -   & -                              & 53    & 70  \\
			89  & Filler                & 0.05  & 35.5  & 1     & Hard       & 0.082  & -   & -                              & 130   & 70  \\
			90  & Filler                & 0.05  & 35.5  & 1     & Soft       & 0.066  & -   & -                              & 52    & 70  \\
			91  & Filler                & 0.05  & 35.5  & 1     & Soft       & 0.082  & -   & -                              & 129   & 70  \\
			92  & Filler                & 0.05  & 35.5  & 8     & Hard       & 0.066  & -   & -                              & 52    & 70  \\
			93  & Filler                & 0.05  & 35.5  & 8     & Hard       & 0.082  & -   & -                              & 129   & 70  \\
			94  & Filler                & 0.05  & 35.5  & 8     & Soft       & 0.066  & -   & -                              & 48    & 70  \\
			95  & Filler                & 0.05  & 35.5  & 8     & Soft       & 0.082  & -   & -                              & 124   & 70  \\
			96  & Filler                & 0.5   & 35.5  & 0.1   & Hard       & 0.066  & -   & -                              & 53    & 70  \\
			97  & Filler                & 0.5   & 35.5  & 0.1   & Hard       & 0.082  & -   & -                              & 130   & 70  \\
			98  & Filler                & 0.5   & 35.5  & 0.1   & Soft       & 0.066  & -   & -                              & 52    & 70  \\
			99 & Filler                & 0.5   & 35.5  & 0.1   & Soft       & 0.082  & -   & -                              & 129   & 70  \\
			100 & Filler                & 0.5   & 35.5  & 1     & Hard       & 0.066  & -   & -                              & 52    & 70  \\
			101 & Filler                & 0.5   & 35.5  & 1     & Hard       & 0.082  & -   & -                              & 129   & 70  \\
			102 & Filler                & 0.5   & 35.5  & 1     & Soft       & 0.066  & -   & -                              & 47    & 70  \\
			103 & Filler                & 0.5   & 35.5  & 1     & Soft       & 0.082  & -   & -                              & 122   & 70  \\
			104 & Filler                & 0.5   & 35.5  & 8     & Hard       & 0.066  & -   & -                              & 43    & 70  \\
			105 & Filler                & 0.5   & 35.5  & 8     & Hard       & 0.082  & -   & -                              & 119   & 70  \\
			106 & Filler                & 0.5   & 35.5  & 8     & Soft       & 0.066  & -   & -                              & 22    & 70  \\
			107 & Filler                & 0.5   & 35.5  & 8     & Soft       & 0.082  & -   & -                              & 91    & 70  \\
			108 & Large                 & 0.5   & 4.2   & 10    & MQ         & 0.098  & Off & Eq.~(3) from C2016b & 27    & 70  \\
			109 & Large                 & 0.5   & 4.2   & 0.1   & Soft       & 0.098  & Off & Eq.~(3) from C2016b & 26    & 70  \\
			110 & Large                 & 0.5   & 4.2   & 0.1   & MQ         & 0.098  & Off & Eq.~(3) from C2016b & 28    & 70  \\
			111 & Large                 & 0.005 & 4.2   & 10    & Soft       & 0.098  & Off & Eq.~(3) from C2016b & 26    & 70  \\
			112 & Large                 & 0.005 & 4.2   & 10    & MQ         & 0.098  & Off & Eq.~(3) from C2016b & 27    & 70  \\
			113 & Large                 & 0.005 & 4.2   & 0.1   & Soft       & 0.098  & Off & Eq.~(3) from C2016b & 28    & 70  \\
			114 & Large                 & 0.005 & 4.2   & 0.1   & MQ         & 0.098  & Off & Eq.~(3) from C2016b & 28    & 70  \\
			115 & Large                 & 0.5   & 76.5  & 10    & Soft       & 0.066  & -   & -                              & 387   & 70  \\
			116 & Large                 & 0.5   & 76.5  & 10    & Soft       & 0.098  & -   & -                              & 6000  & 70  \\
			117 & Large                 & 0.5   & 76.5  & 10    & MQ         & 0.066  & -   & -                              & 450   & 70  \\
			118 & Large                 & 0.5   & 76.5  & 10    & MQ         & 0.098  & -   & -                              & 6060  & 70  \\
			119 & Large                 & 0.5   & 76.5  & 0.1   & Soft       & 0.066  & -   & -                              & 455   & 70  \\
			120 & Large                 & 0.5   & 76.5  & 0.1   & Soft       & 0.098  & -   & -                              & 6060  & 70  \\
			121 & Large                 & 0.5   & 76.5  & 0.1   & MQ         & 0.098  & -   & -                              & 6060  & 70  \\
			122 & Large                 & 0.016 & 76.5  & 10    & Soft       & 0.066  & -   & -                              & 455   & 70  \\
		\end{tabular}
	\end{center}
\end{table*}
\begin{table*}
	\begin{center}
		
		\begin{tabular}{lllllllllll}
		\hline \#  &                       & $f_*$ & $V_c$ [km/s]& $f_X$ & SED        & $\tau$ & LW  & Low-mass cutoff                & $\zeta$ & $R_{\rm mfp}$ [Mpc] \\
			\hline
			123 & Large                 & 0.16 & 76.5  & 10    & Soft       & 0.098  & -   & -                              & 6060  & 70  \\
			124 & Large                 & 0.016 & 76.5  & 10    & MQ         & 0.066  & -   & -                              & 455   & 70  \\
			125 & Large                 & 0.16 & 76.5  & 10    & MQ         & 0.098  & -   & -                              & 6060  & 70  \\
			126 & Large                 & 0.16 & 76.5  & 0.1   & Soft       & 0.098  & -   & -                              & 6060  & 70  \\
			127 & Large                 & 0.16 & 76.5  & 0.1   & MQ         & 0.098  & -   & -                              & 6060  & 70  \\
			128 & Small                 & 0.16 & 4.2   & 3.16  & Hard \& MQ & 0.066  & On  & Eq.~(4) from C2016b  & 21    & 70  \\
			129 & Small                 & 0.16 & 4.2   & 3.16  & Hard \& MQ & 0.082  & On  & Eq.~(4) from C2016b  & 41    & 70  \\
			130 & Small                 & 0.16 & 4.2   & 3.16  & Soft \& MQ & 0.066  & On  & Eq.~(4) from C2016b  & 18    & 70  \\
			131 & Small                 & 0.16 & 4.2   & 3.16  & Soft \& MQ & 0.082  & On  & Eq.~(4) from C2016b  & 38    & 70  \\
			132 & Small                 & 0.16 & 4.2   & 0.32  & Hard \& MQ & 0.066  & On  & Eq.~(4) from C2016b  & 21    & 70  \\
			133 & Small                 & 0.16 & 4.2   & 0.32  & Hard \& MQ & 0.082  & On  & Eq.~(4) from C2016b  & 42    & 70  \\
			134 & Small                 & 0.16 & 4.2   & 0.32  & Soft \& MQ & 0.066  & On  & Eq.~(4) from C2016b  & 21    & 70  \\
			135 & Small                 & 0.16 & 4.2   & 0.32  & Soft \& MQ & 0.082  & On  & Eq.~(4) from C2016b  & 42    & 70  \\
			136 & Small                 & 0.016 & 4.2   & 3.16  & Hard \& MQ & 0.066  & On  & Eq.~(4) from C2016b  & 20    & 70  \\
			137 & Small                 & 0.016 & 4.2   & 3.16  & Hard \& MQ & 0.082  & On  & Eq.~(4) from C2016b  & 33    & 70  \\
			138 & Small                 & 0.016 & 4.2   & 3.16  & Soft \& MQ & 0.066  & On  & Eq.~(4) from C2016b  & 19    & 70  \\
			139 & Small                 & 0.016 & 4.2   & 3.16  & Soft \& MQ & 0.082  & On  & Eq.~(4) from C2016b  & 33    & 70  \\
			140 & Small                 & 0.016 & 4.2   & 0.32  & Hard \& MQ & 0.066  & On  & Eq.~(4) from C2016b  & 20    & 70  \\
			141 & Small                 & 0.016 & 4.2   & 0.32  & Hard \& MQ & 0.082  & On  & Eq.~(4) from C2016b  & 33    & 70  \\
			142 & Small                 & 0.016 & 4.2   & 0.32  & Soft \& MQ & 0.066  & On  & Eq.~(4) from C2016b  & 20    & 70  \\
			143 & Small                 & 0.016 & 4.2   & 0.32  & Soft \& MQ & 0.082  & On  & Eq.~(4) from C2016b  & 33    & 70  \\
			144 & Small                 & 0.16 & 35.5  & 3.16  & Hard \& MQ & 0.066  & -   & -                              & 52     & 70 \\
			145 & Small                 & 0.16 & 35.5  & 3.16  & Hard \& MQ & 0.082  & -   & -                              & 129   & 70  \\
			146 & Small                 & 0.16 & 35.5  & 3.16  & Soft \& MQ & 0.066  & -   & -                              & 49     & 70 \\
			147 & Small                 & 0.16 & 35.5  & 3.16  & Soft \& MQ & 0.082  & -   & -                              & 126   & 70  \\
			148 & Small                 & 0.16 & 35.5  & 0.32  & Hard \& MQ & 0.066  & -   & -                              & 53     & 70 \\
			149 & Small                 & 0.16 & 35.5  & 0.32  & Hard \& MQ & 0.082  & -   & -                              & 130    & 70 \\
			150 & Small                 & 0.16 & 35.5  & 0.32  & Soft \& MQ & 0.066  & -   & -                              & 53     & 70 \\
			151 & Small                 & 0.16 & 35.5  & 0.32  & Soft \& MQ & 0.082  & -   & -                              & 130   & 70  \\
			152 & Small                 & 0.016 & 35.5  & 3.16  & Hard \& MQ & 0.066  & -   & -                              & 53    & 70  \\
			153 & Small                 & 0.016 & 35.5  & 3.16  & Hard \& MQ & 0.082  & -   & -                              & 130   & 70  \\
			154 & Small                 & 0.016 & 35.5  & 3.16  & Soft \& MQ & 0.066  & -   & -                              & 53    & 70  \\
			155 & Small                 & 0.016 & 35.5  & 3.16  & Soft \& MQ & 0.082  & -   & -                              & 130    & 70 \\
			156 & Small                 & 0.016 & 35.5  & 0.32  & Hard \& MQ & 0.066  & -   & -                              & 53     & 70 \\
			157 & Small                 & 0.016 & 35.5  & 0.32  & Hard \& MQ & 0.082  & -   & -                              & 130    & 70 \\
			158 & Small                 & 0.016 & 35.5  & 0.32  & Soft \& MQ & 0.066  & -   & -                              & 53     & 70 \\
			159 & Small                 & 0.016 & 35.5  & 0.32  & Soft \& MQ & 0.082  & -   & -                              & 130   & 70  \\
			160 & F2017 & 0.05  & 16.5  & 1     & MQ         & 0.0738   & -   & -                              & 24     & 70 \\
			161 & F2017 & 0.05  & 16.5  & 1     & MQ         & 0.0956  & -   & -                              & 57    & 70  \\
			162 & F2017 & 0.05  & 16.5  & 10.8  & Hard       & 0.0756   & -   & -                              & 24     & 70 \\
			163 & F2017 & 0.05  & 16.5  & 29.5  & Soft       & 0.0859   & -   & -                              & 24    & 70  \\
			164 & F2017 & 0.05  & 16.5  & 11.4  & MQ         & 0.0747   & -   & -                              & 24    & 70  \\
			165 & F2017 & 0.05  & 16.5  & 44.4  & Hard       & 0.0990  & -   & -                              & 57   & 70   \\
			166 & F2017 & 0.05  & 16.5  & 102   & Soft       & 0.1111  & -   & -                              & 57  & 70    \\
			167 & F2017 & 0.05  & 16.5  & 74.4  & MQ         & 0.0977  & -   & -                              & 57   & 70   \\
			168 & F2017 & 0.05  & 16.5  & 0.01  & Hard       & 0.0739   & -   & -                              & 24   & 70   \\
			169 & F2017 & 0.05  & 16.5  & 0.0023     & Soft       & 0.0746   & -   & -                              & 24   & 70   \\
			170 & F2017 & 0.05  & 16.5  & 0     & Hard       & 0.0957  & -   & -                              & 57  & 70    \\
			171 & F2017 & 0.05  & 35.5  & 1     & MQ         & 0.0597   & -   & -                              & 32   & 70   \\
			172 & F2017 & 0.05  & 35.5  & 1     & MQ         & 0.0831  & -   & -                              & 112 & 70    \\
			173 & F2017 & 0.05  & 35.5  & 14.7  & Hard       & 0.0609   & -   & -                              & 32  & 70    \\
			174 & F2017 & 0.05  & 35.5  & 41.4  & Soft       & 0.0688   & -   & -                              & 32  & 70    \\
			175 & F2017 & 0.05  & 35.5  & 12.1  & MQ         & 0.0606   & -   & -                              & 32  & 70    \\
			176 & F2017 & 0.05  & 35.5  & 79.2  & Hard       & 0.0850  & -   & -                              & 112  & 70   \\
			177 & F2017 & 0.05  & 35.5  & 188   & Soft       & 0.0934  & -   & -                              & 112 & 70    \\
			178 & F2017 & 0.05  & 35.5  & 87.9  & MQ         & 0.0847  & -   & -                              & 112  & 70   \\
			179 & F2017 & 0.05  & 35.5  & 0     & Hard       & 0.0831  & -   & -                              & 112  & 70   \\
			180 & F2017 & 0.05  & 35.5  & 0.036  & Hard       & 0.0597   & -   & -                              & 32  & 70    \\
			181 & F2017 & 0.05  & 35.5  & 0.0095  & Soft       & 0.0601   & -   & -                              & 32 & 70 \\
			182 & $R_{\rm mfp}$ & 0.005  & 35.5  & 0.1  & Hard       & 0.082   & -   & -                              & 125  & 20 \\
			183 & $R_{\rm mfp}$ & 0.005  & 35.5  & 0.1  & Soft       & 0.082   & -   & -                              & 125  & 20 \\
			184 & $R_{\rm mfp}$ & 0.005  & 35.5  & 1  & Hard       & 0.082   & -   & -                              & 125  & 20 \\
			185 & $R_{\rm mfp}$ & 0.05  & 35.5  & 0.1  & Hard       & 0.082   & -   & -                              & 125  & 20 \\
			186 & $R_{\rm mfp}$ & 0.5  & 76.5  & 0.1  & Hard       & 0.066   & -   & -                              & 389  & 20 \\
			187 & $R_{\rm mfp}$ & 0.5  & 76.5  & 0.1  & Hard       & 0.082   & -   & -                              & 1411  & 20 \\
			188 & $R_{\rm mfp}$ & 0.005  & 35.5  & 0.1  & Hard       & 0.082   & -   & -                              & 202  & 5 \\
			189 & $R_{\rm mfp}$ & 0.005  & 35.5  & 0.1  & Soft       & 0.082   & -   & -                              & 202  & 5 \\
			190 & $R_{\rm mfp}$ & 0.005  & 35.5  & 1  & Hard       & 0.082   & -   & -                              & 202  & 5 \\
			191 & $R_{\rm mfp}$ & 0.05  & 35.5  & 0.1  & Hard       & 0.082   & -   & -                              & 202  & 5 \\
			192 & $R_{\rm mfp}$ & 0.5  & 76.5  & 0.1  & Hard       & 0.066   & -   & -                              & 1172  & 5 \\
			193 & $R_{\rm mfp}$ & 0.5  & 76.5  & 0.1  & Hard       & 0.082   & -   & -                              & 6125  & 5 \\
			\hline   
		\end{tabular}
		\caption{List of the parameter sets used in this
                  paper. C2016b means \citet{Cohen:2016b} and F2017
                  means \citet{Fialkov:2016}.}
		\label{Table:List}
	\end{center}
\end{table*}

\section{Additional Figures}
\label{appB}

To demonstrate the variety of possible realizations of the power
spectra and corresponding slopes we show figures analogues to
Fig.~\ref{fig:Cases1} and \ref{fig:Slopes1} but for various sets of
astrophysical parameters.

\begin{figure*}
	
	\centering
	
	\begin{subfigure}[b]{0.4\textwidth}
		\begin{center} \hspace{0.25in}	\textbf{Strong X-rays}\par\medskip \end{center}
		\includegraphics[width=3.1in]{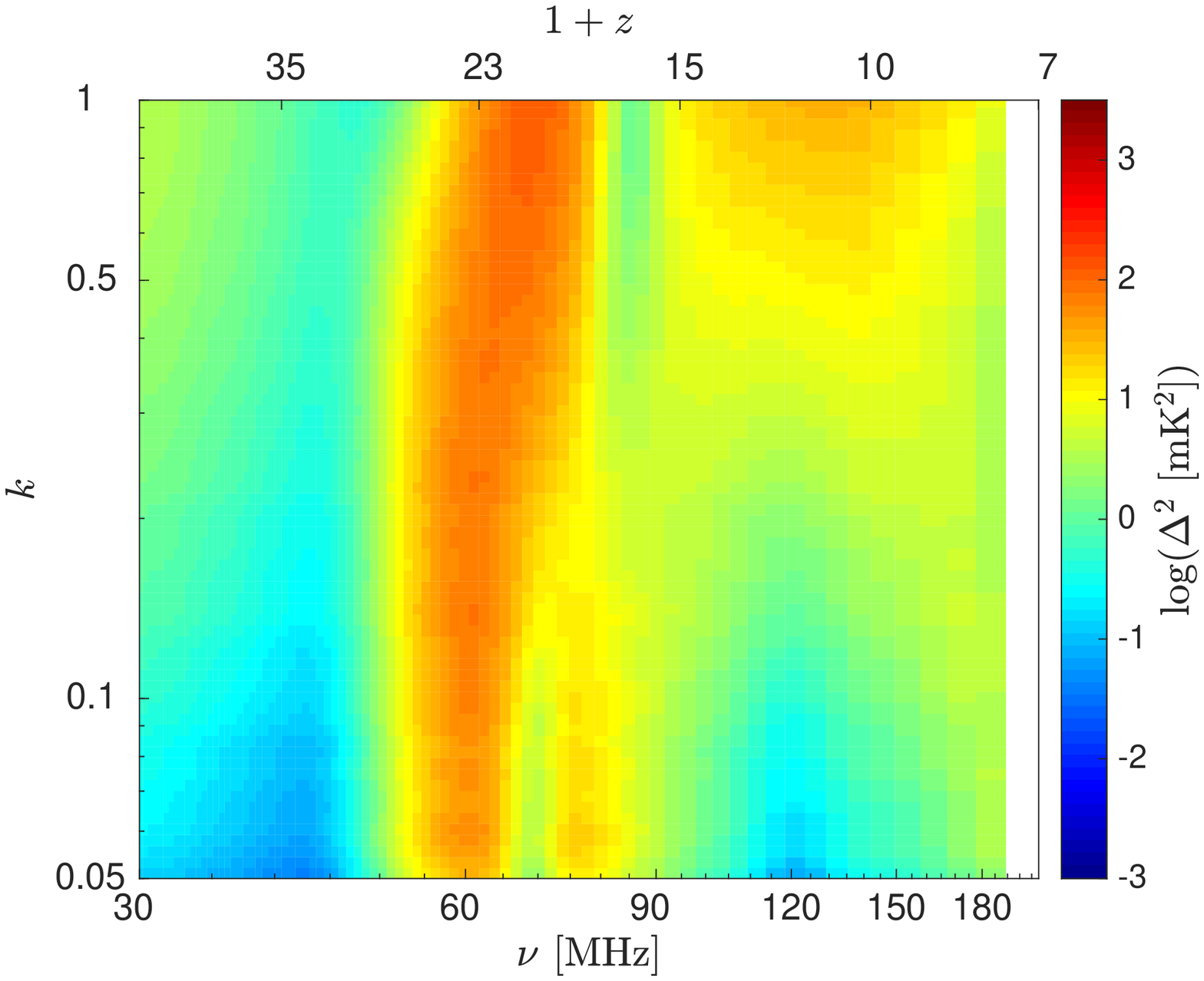}
		\vspace{0.1in}
	\end{subfigure}
	\hspace{0.5in}
	\begin{subfigure}[b]{0.4\textwidth}
		\begin{center} \hspace{0.25in} \textbf{Massive halos}\par\medskip \end{center}
		\includegraphics[width=3.1in]{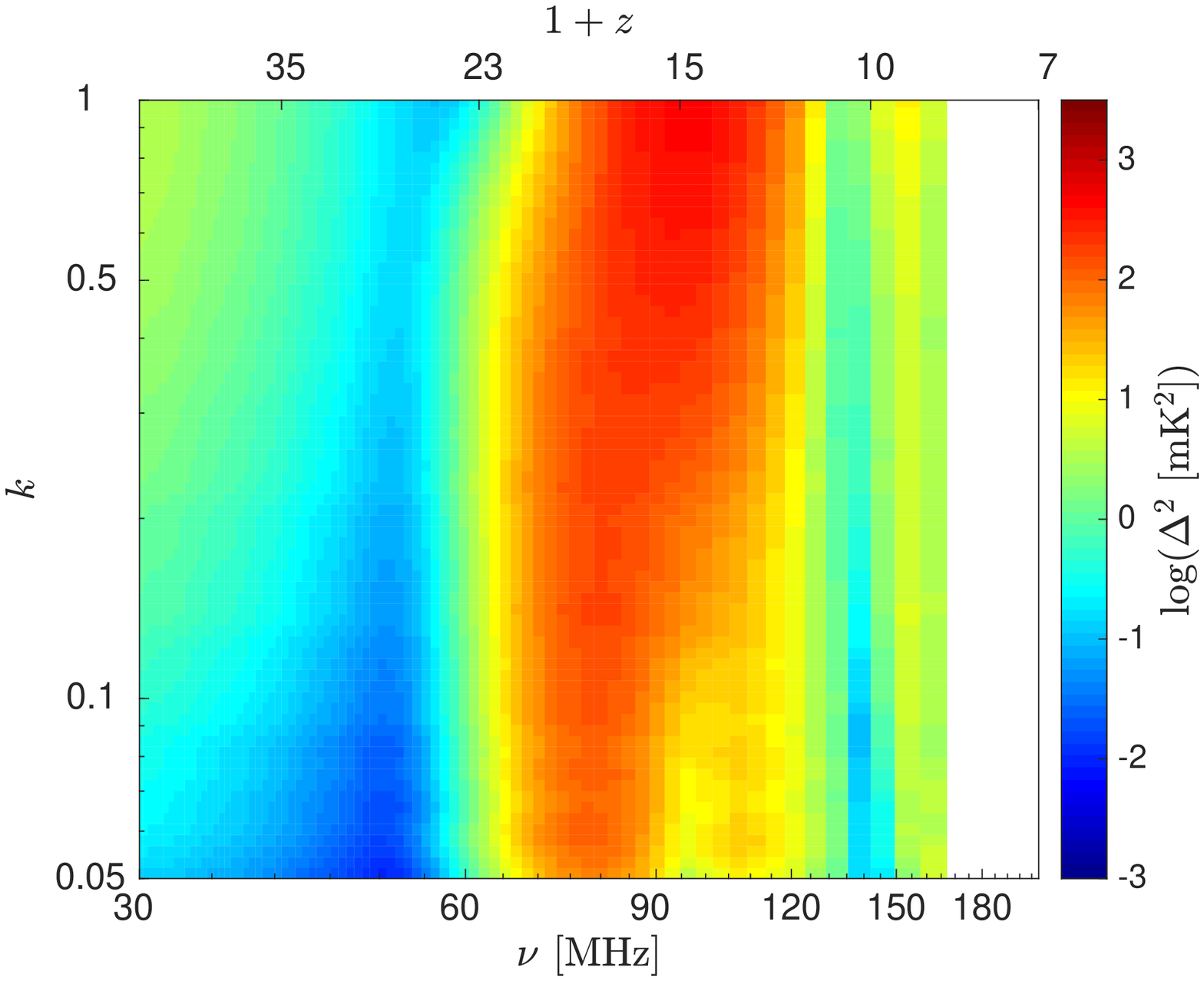}
		\vspace{0.1in}
	\end{subfigure}
	\begin{subfigure}[b]{0.4\textwidth}
		\includegraphics[width=3.1in]{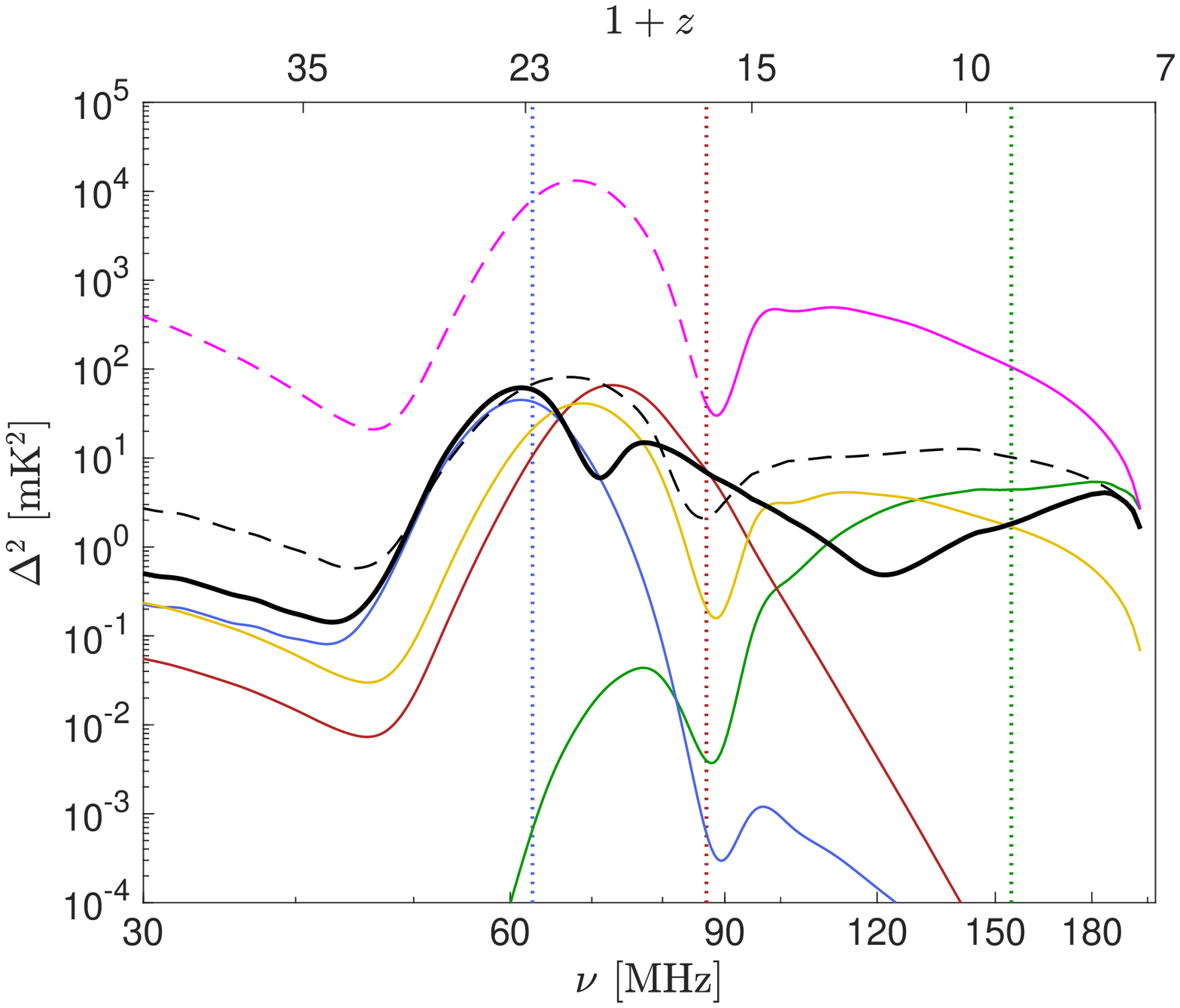}
		\vspace{0.1in}
	\end{subfigure}
	\hspace{0.5in}
	\begin{subfigure}[b]{0.4\textwidth}
		\includegraphics[width=3.1in]{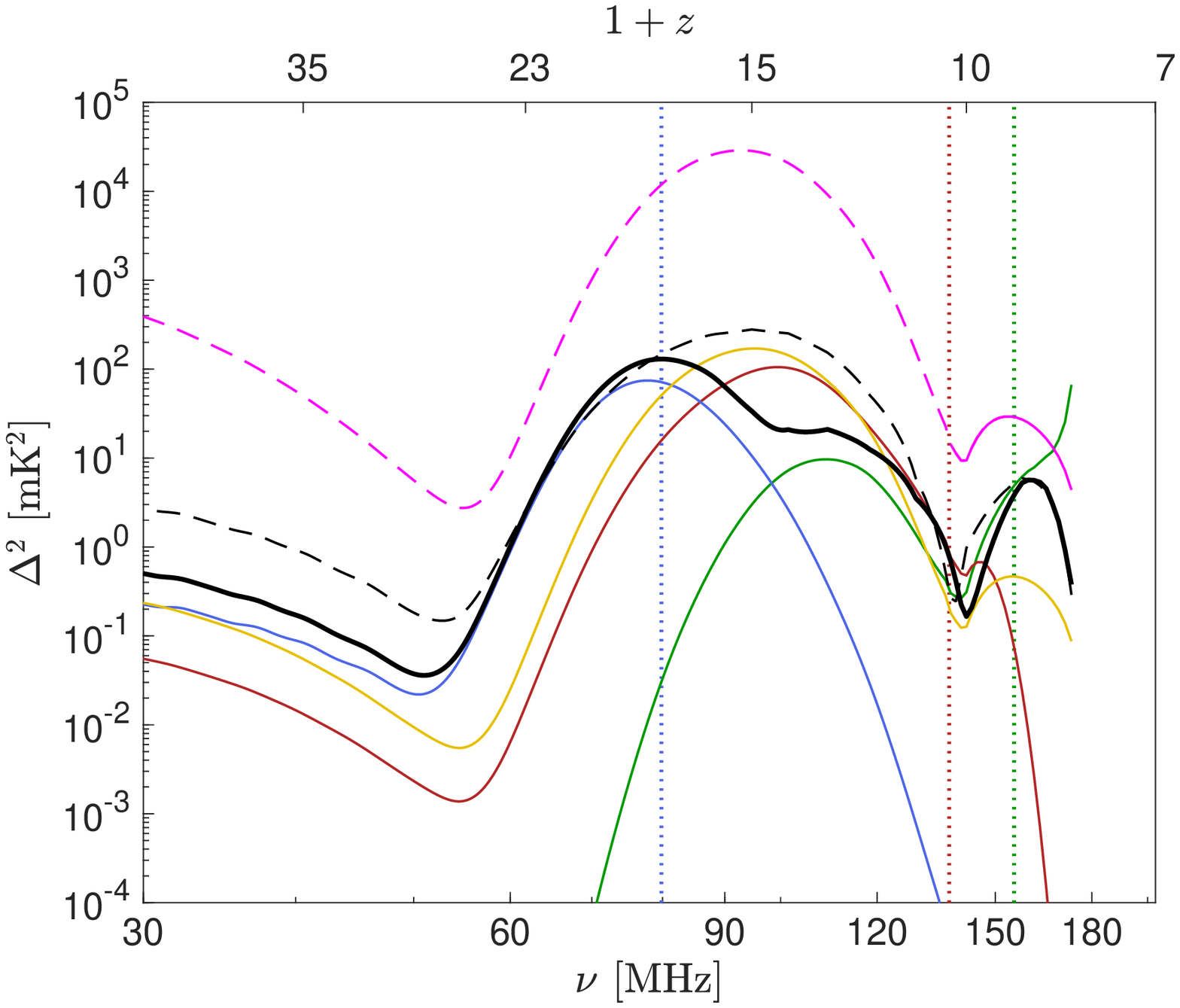}
		\vspace{0.1in}
	\end{subfigure}
	\begin{subfigure}[b]{0.4\textwidth}
		\includegraphics[width=3.1in]{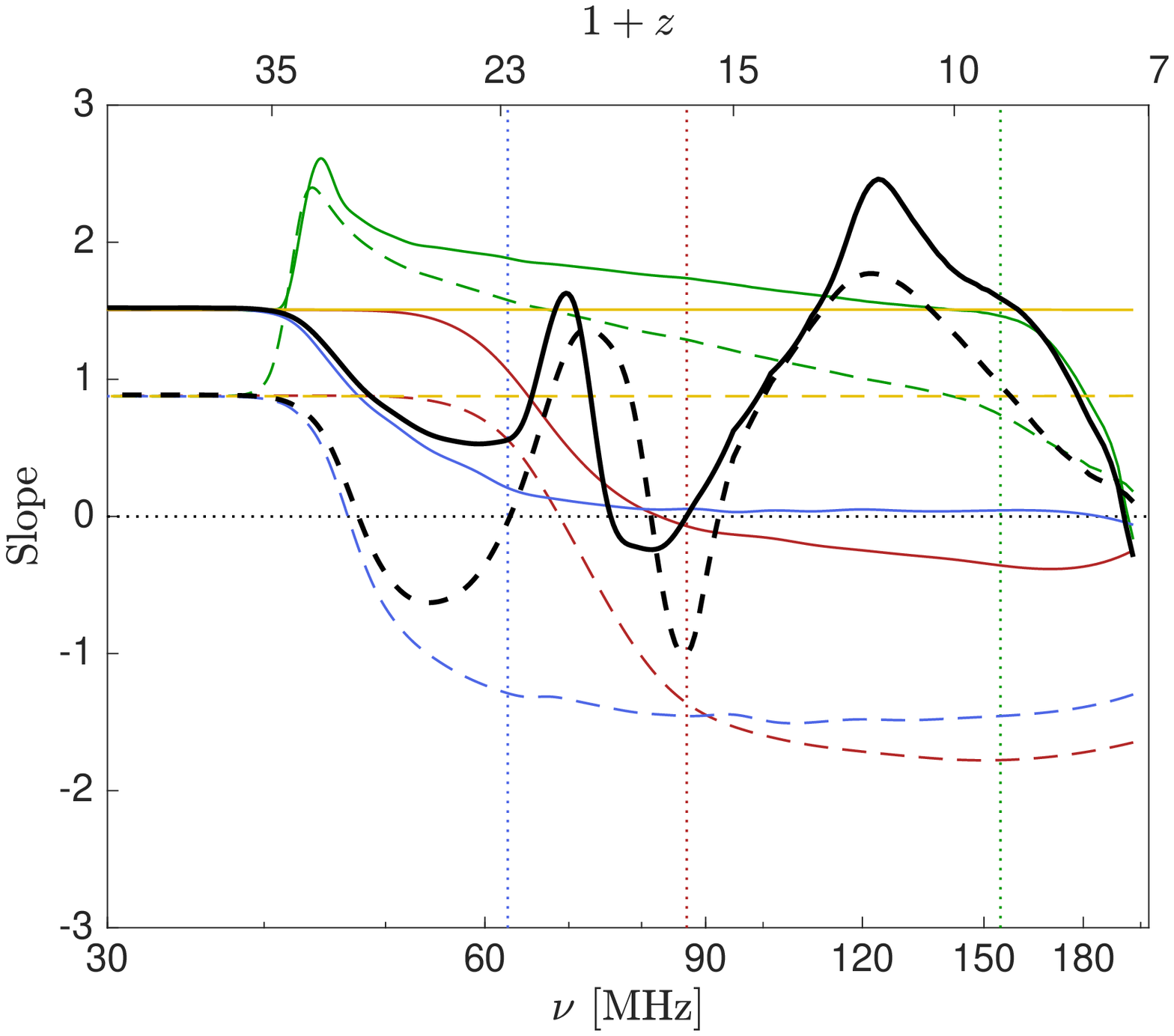}
	\end{subfigure}
	\hspace{0.5in}
	\begin{subfigure}[b]{0.4\textwidth}
		\includegraphics[width=3.1in]{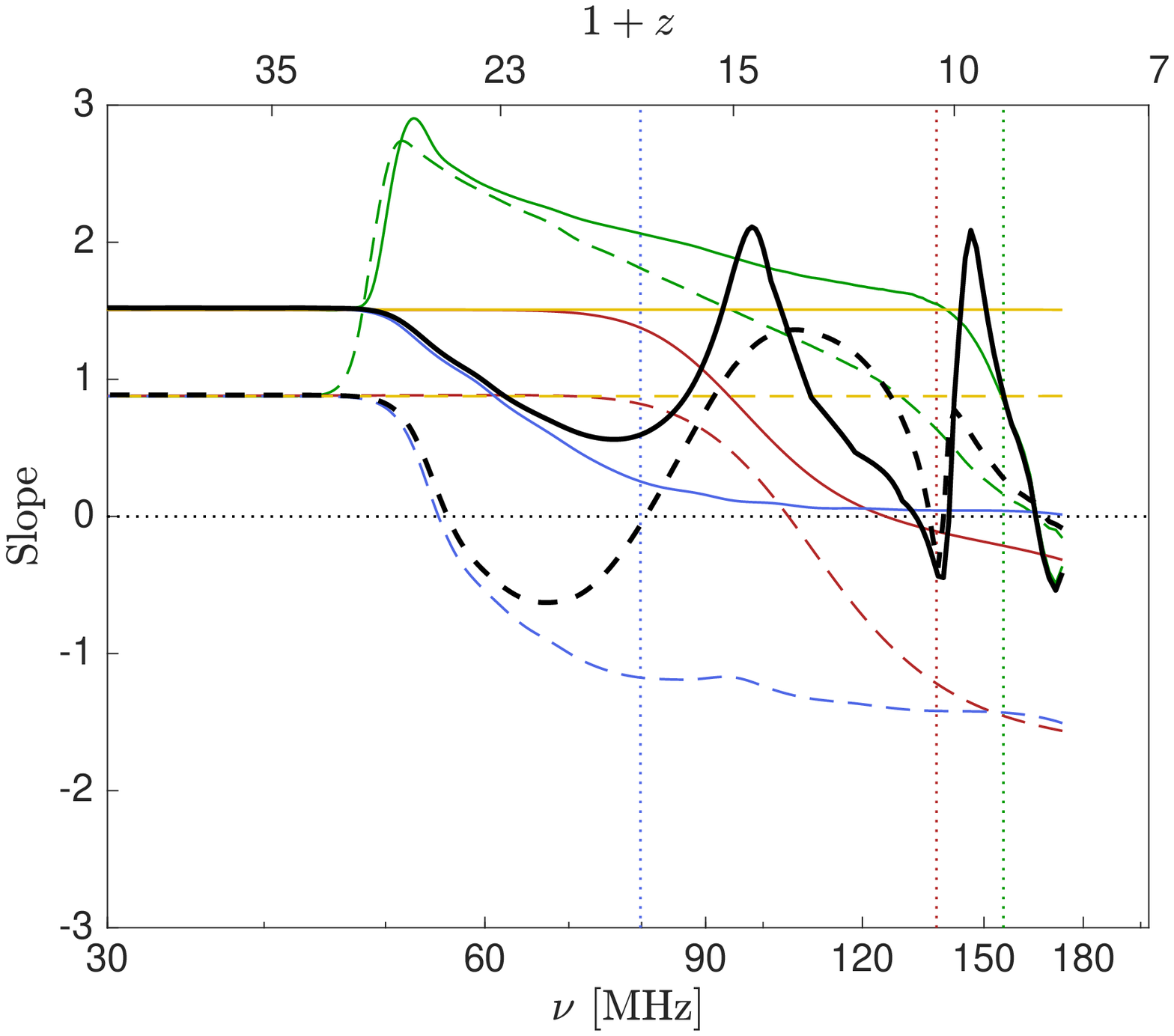}
	\end{subfigure}
	\caption{\label{secondA} Variation of Figures~\ref{fig:Cases1}
          and \ref{fig:Slopes1}, same line colors and styles. {\bf
            Left:} The standard case except that $f_X=8$, model
          \#57. {\bf Right:} The standard case except that $V_c=35.5$
          km s$^{-1}$, model \#88.}
	\label{fig:Cases3}
\end{figure*}

First we show variations of our standard case, altering one
astrophysical parameter at a time. The left column of
Fig.~\ref{fig:Cases3} presents the case of strong X-rays with $f_X=8$
instead of $f_X=1$ (\#57 in Table~\ref{Table:List}). Because of the
stronger heating, the Ly$\alpha$ peak becomes narrower compared to the
standard case. This is because in model \# 57 heating starts earlier
and X-rays (which anti-correlate with Ly$\alpha$) cut the Ly$\alpha$
peak off. For the same reason, the features of the slope (high-$z$ max
and low-$z$ min) shift to higher redshifts. Compared to the standard
case, the heating peak in this case is more prominent and occurs
earlier, but is still relatively small as expected with hard
X-rays. Next, we show a case with more massive halos (\#88, right
column of Fig.~\ref{fig:Cases3}), setting the minimum $V_c=35.5$ km
s$^{-1}$ instead of $V_c=16.5$ km s$^{-1}$ in the standard case. In
this case all the cosmological milestones are shifted to lower
redshifts, as heavier halos can only form stars somewhat later (though
the timing of reionization is fixed since we have not changed
$\tau$). The fluctuations are higher in this case because of the
higher bias of the more massive galactic halos, and reionization
occurs over a smaller redshift interval.

\begin{figure*}
	\centering
	\begin{subfigure}[b]{0.4\textwidth}
		\begin{center} 	\hspace{0.25in} \textbf{Low SFR}\par\medskip \end{center}
		\includegraphics[width=3.1in]{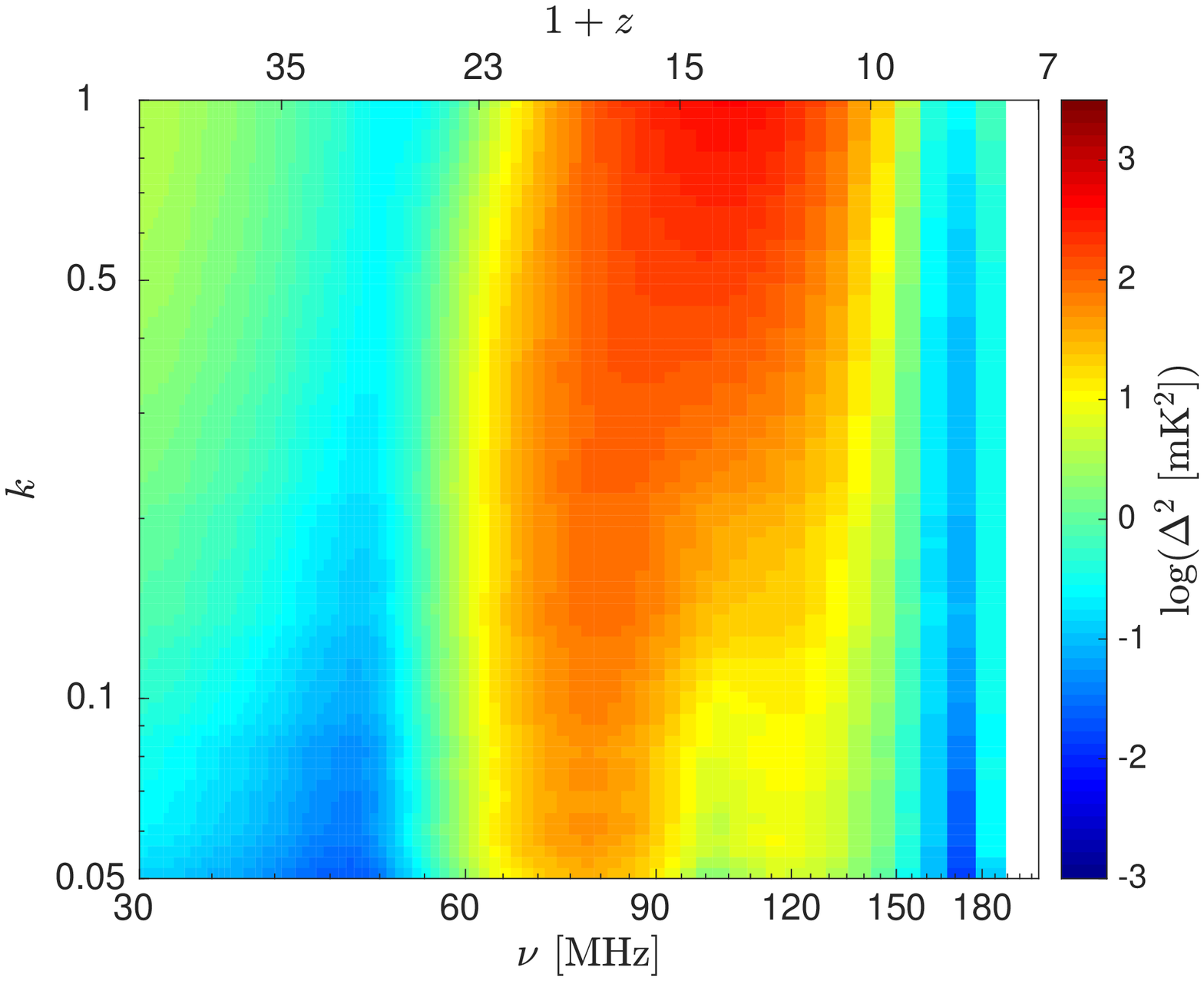}
		\vspace{0.1in}
	\end{subfigure}
	\hspace{0.5in}
	\begin{subfigure}[b]{0.4\textwidth}
		\begin{center}  \hspace{0.25in} \textbf{Very low mass halos}\par\medskip \end{center}
		\includegraphics[width=3.1in]{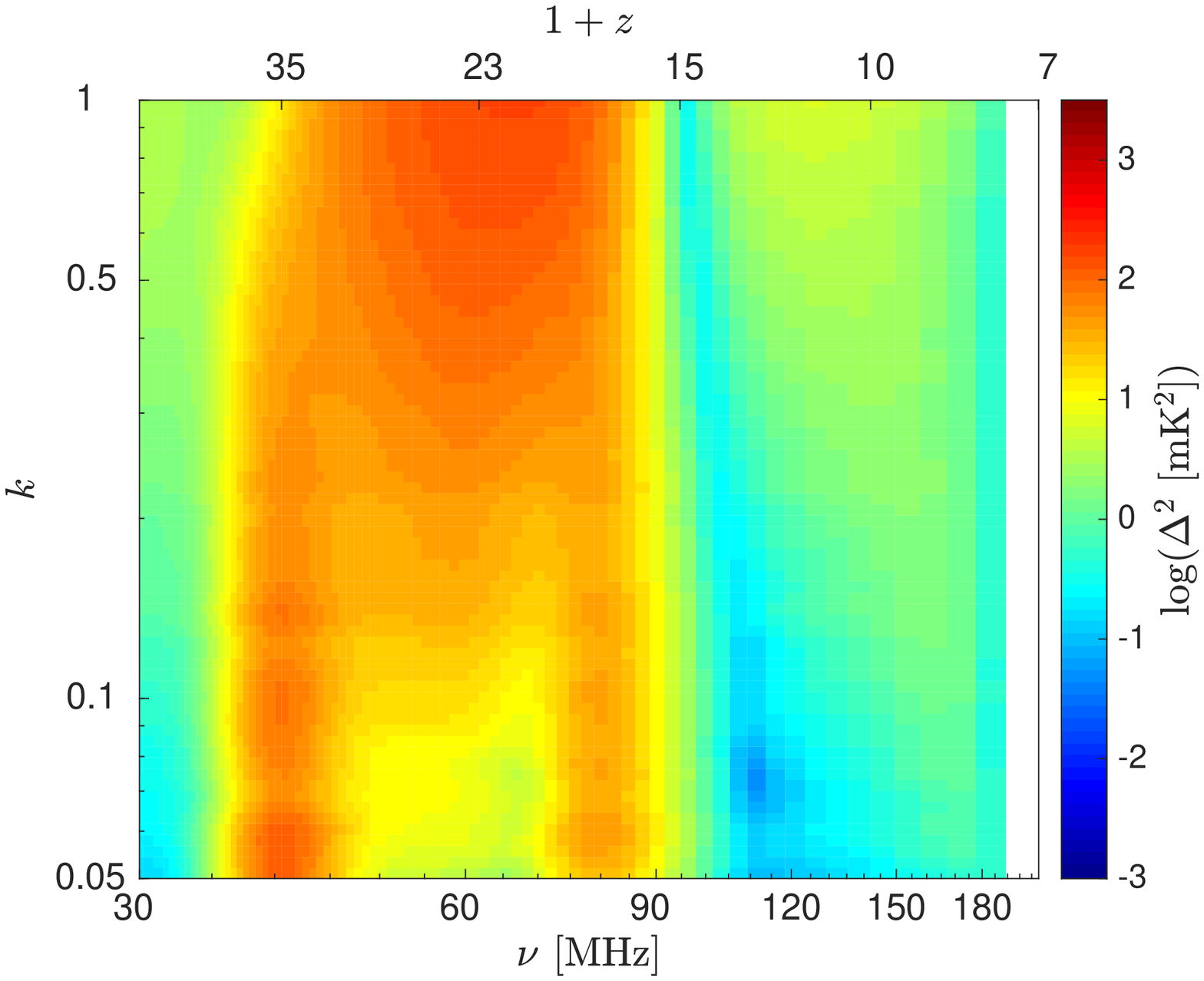}
		\vspace{0.1in}
	\end{subfigure}
	\begin{subfigure}[b]{0.4\textwidth}
		\includegraphics[width=3.1in]{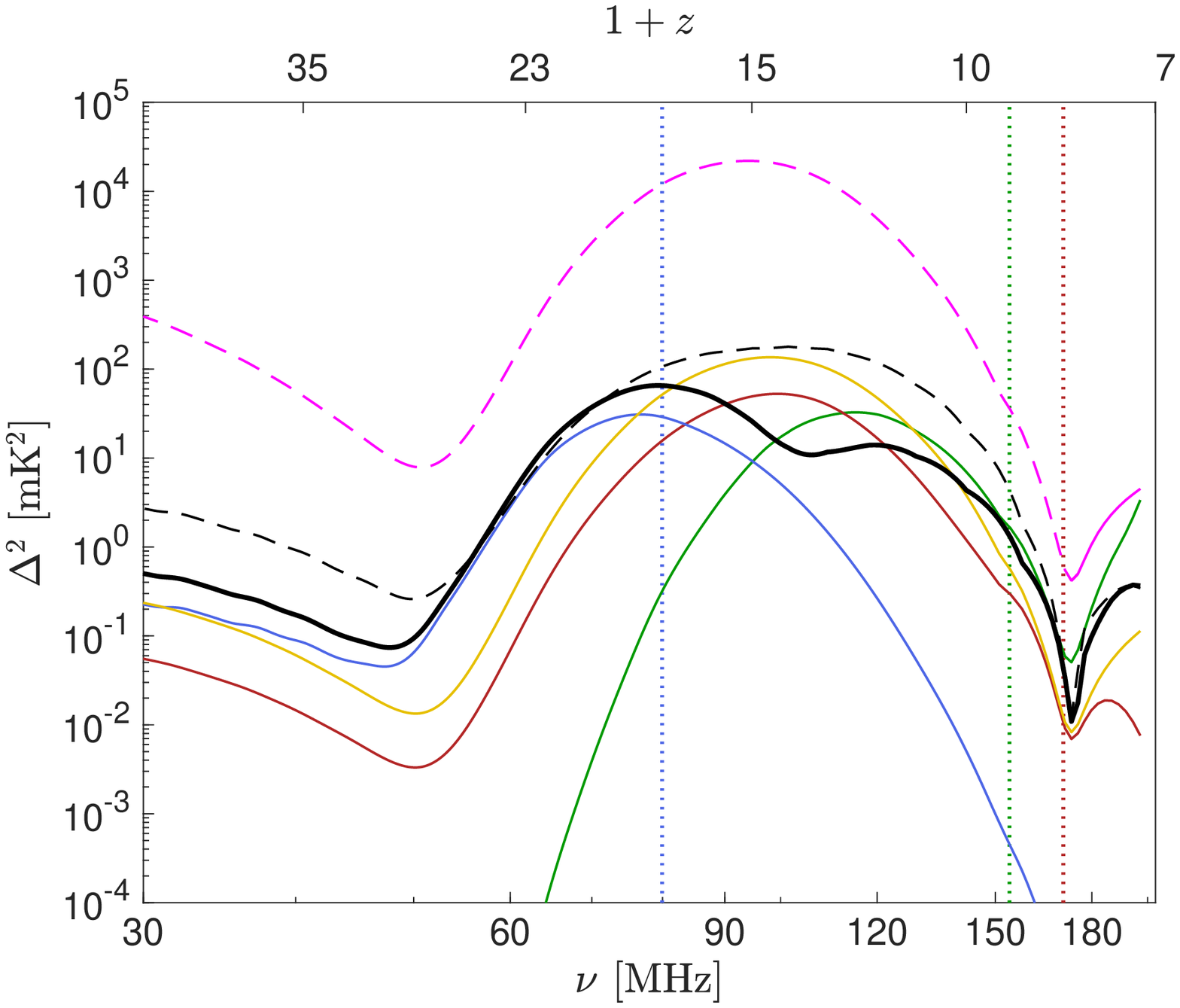}
		\vspace{0.1in}
	\end{subfigure}
	\hspace{0.5in}
	\begin{subfigure}[b]{0.4\textwidth}
		\includegraphics[width=3.1in]{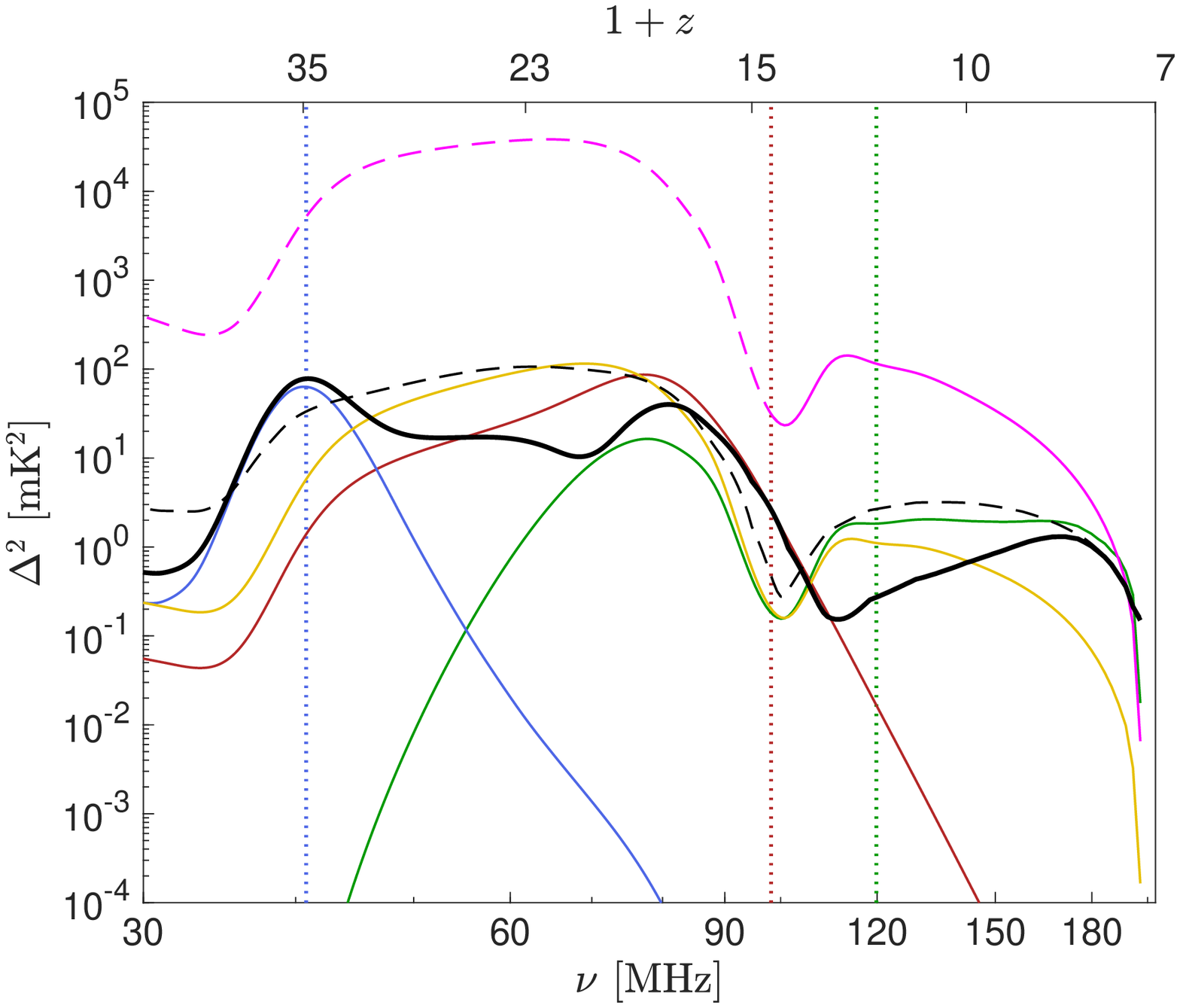}
		\vspace{0.1in}
	\end{subfigure}
	\begin{subfigure}[b]{0.4\textwidth}
		\includegraphics[width=3.1in]{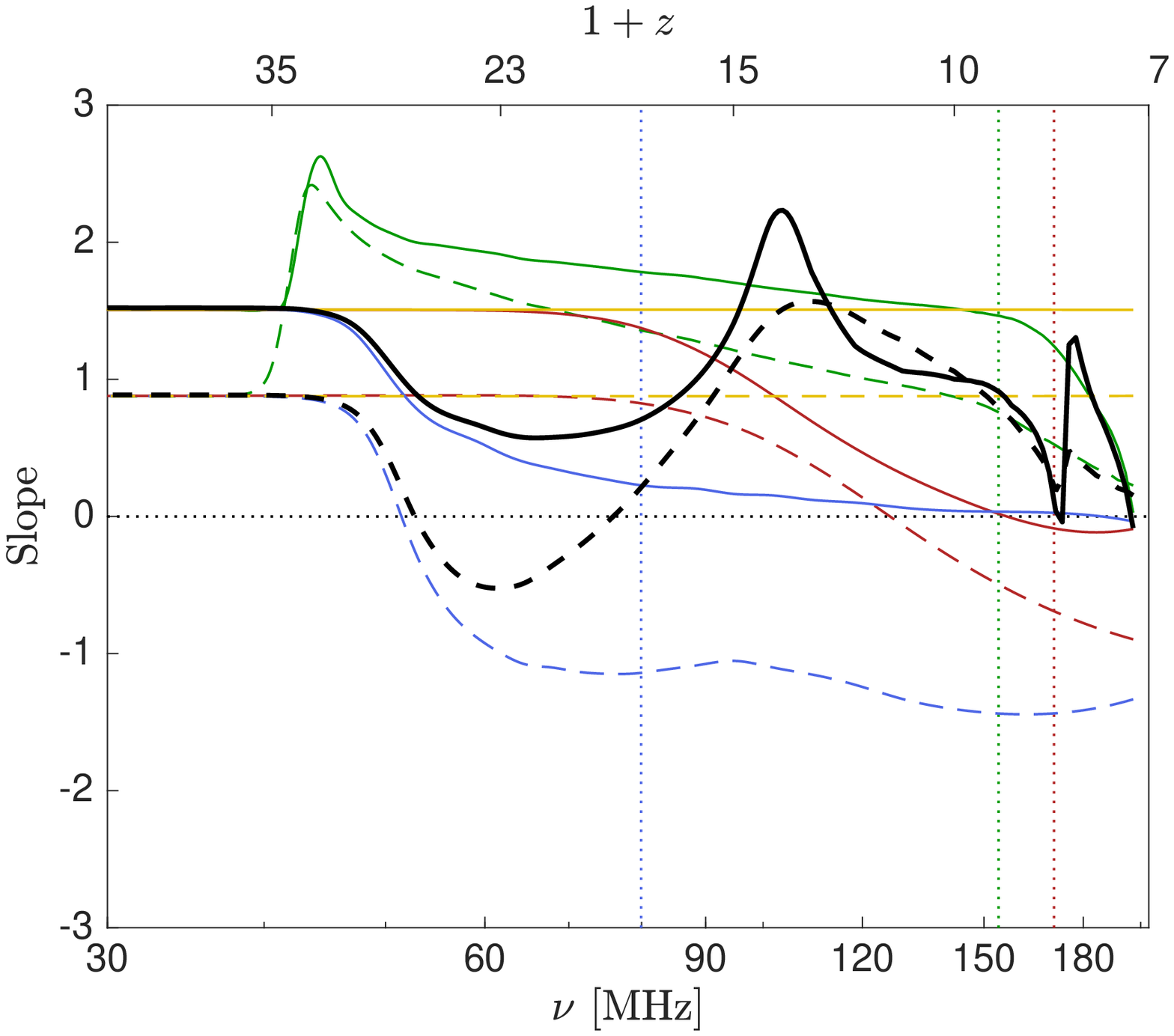}
	\end{subfigure}
	\hspace{0.5in}
	\begin{subfigure}[b]{0.4\textwidth}
		\includegraphics[width=3.1in]{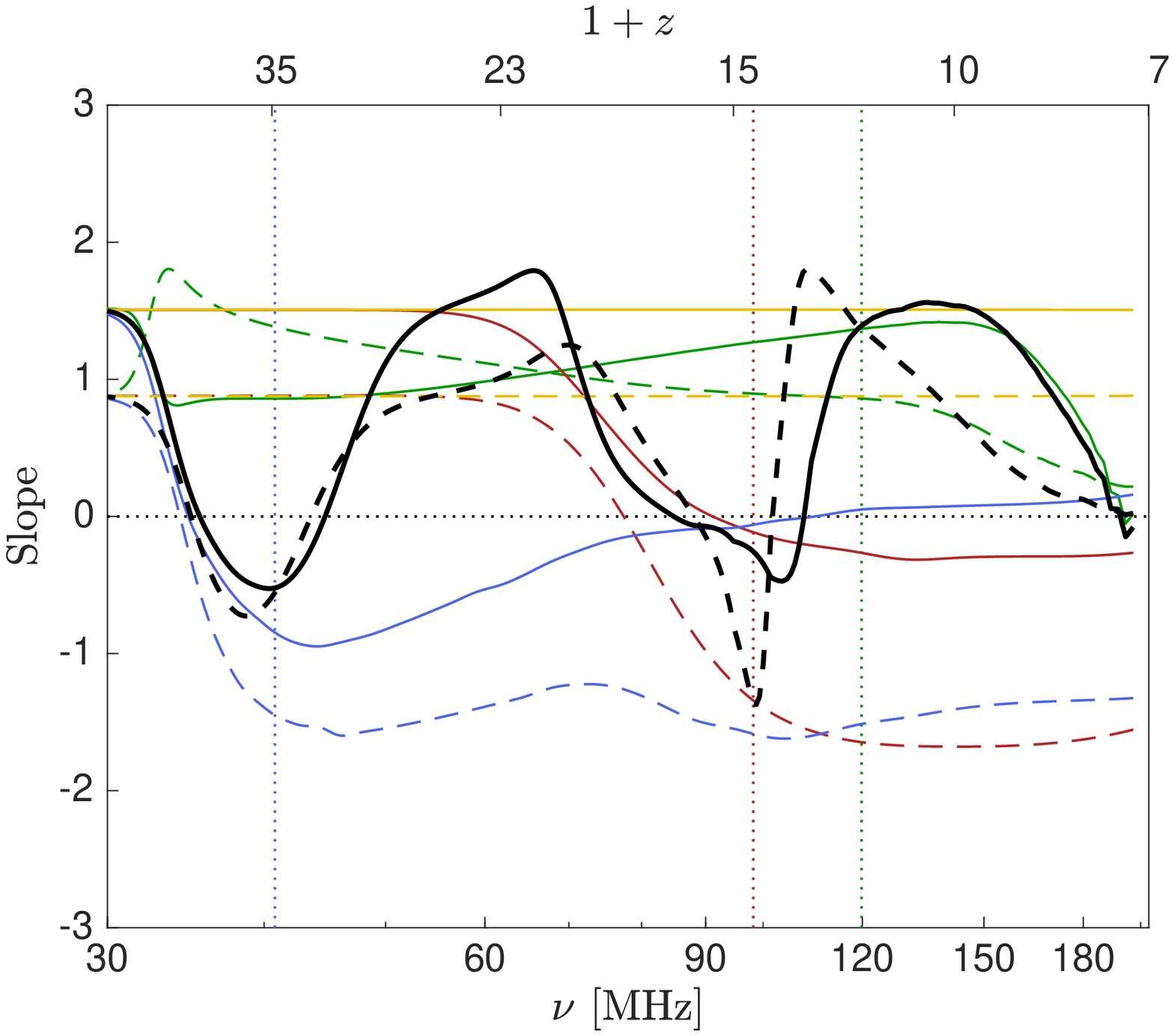}
	\end{subfigure}
	\caption{\label{secondB} Variation of Figures \ref{fig:Cases1}
          and \ref{fig:Slopes1}, same line colors and styles. {\bf
            Left:} The standard case except that $f_*=0.005$, model
          \#41. {\bf Right:} High-redshift star formation dominated by
          very low mass halos, and X-rays produced only by
          miniquasars, model \#108.}
	\label{fig:Cases2}
\end{figure*}

The left column of Fig.~\ref{fig:Cases2} shows a case with a low SFR
($f_*=0.005$, \#41, compared to $f_*=0.05$ in the standard case). In
this case, because of the inefficient star formation, it takes longer
to build up the radiative backgrounds. Halo formation advances more
rapidly at higher redshifts, so while Ly$\alpha$ coupling is delayed,
cosmic heating is delayed far more. As a result, the power spectrum
peaks are wider. Moreover, in this case density fluctuations dominate
the high-redshift peak on both large and small scales. The right
column of Fig.~\ref{fig:Cases2} shows an unusual case (\#108) in which
the fluctuation level remains consistenly high over a wide range of
high redshifts. This case has the earliest possible high star
formation, in that the star formation efficiency is very high ($f_* =
50\%$) and stars are assumed to form in the lowest possible halo
masses (corresponding to the molecular cooling threshold). This leads
to an early saturation of Ly$\alpha$ coupling. However, this model
assumes mini-quasars as the only X-ray heating source, so heating
occurs relatively late. As a result, the Ly$\alpha$ and heating era
are well separated in time allowing density fluctuations to dominate
the signal at intermediate redshifts (where the global signal shows
strong absorption and thus the fluctuations are strong as well). Note,
though, that this case has an optical depth of 0.098 to the CMB, which
is ruled out at about the $3-\sigma$ level \citep{Planck:2016a}. 

\begin{figure*}
	
	\centering
	
	\begin{subfigure}[b]{0.4\textwidth}
		\begin{center} \hspace{0.25in} 	\textbf{One-peak slope}\par\medskip \end{center}
		\includegraphics[width=3.1in]{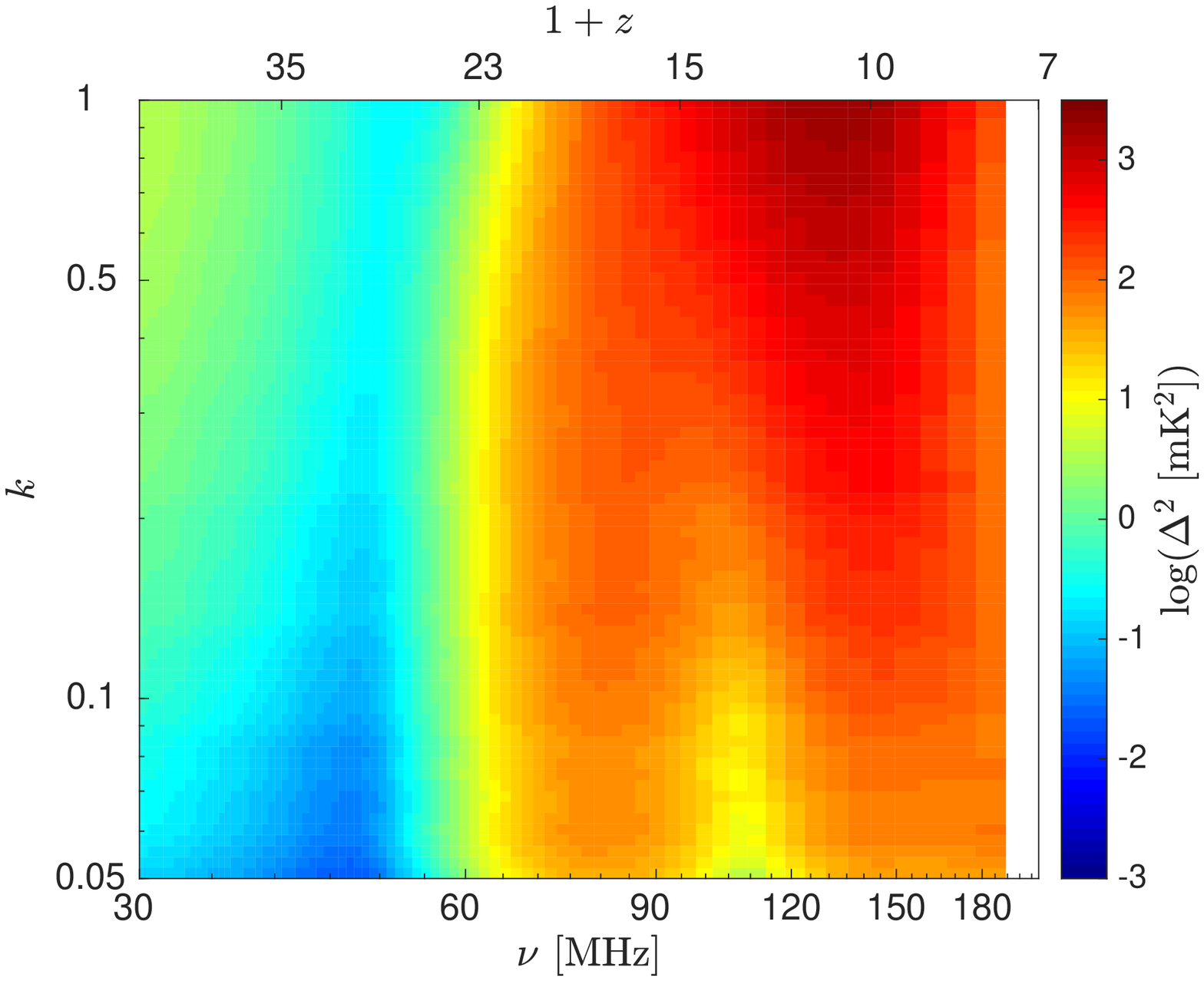}
		\vspace{0.1in}
	\end{subfigure}
	\hspace{0.5in}
	\begin{subfigure}[b]{0.4\textwidth}
		\begin{center} \hspace{0.25in} \textbf{Slope with a bump}\par\medskip \end{center}
		\includegraphics[width=3.1in]{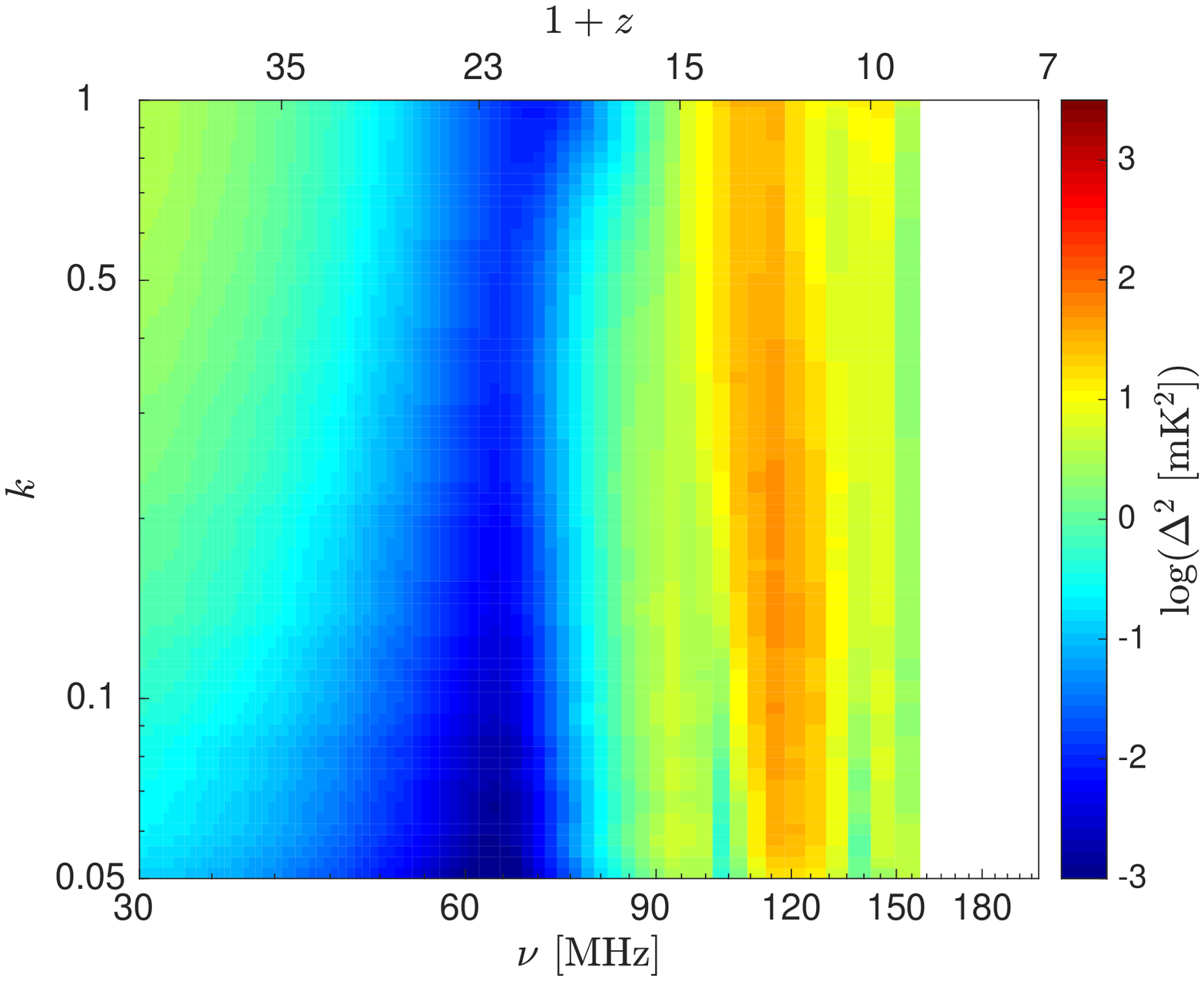}
		\vspace{0.1in}
	\end{subfigure}
	\begin{subfigure}[b]{0.4\textwidth}
		\includegraphics[width=3.1in]{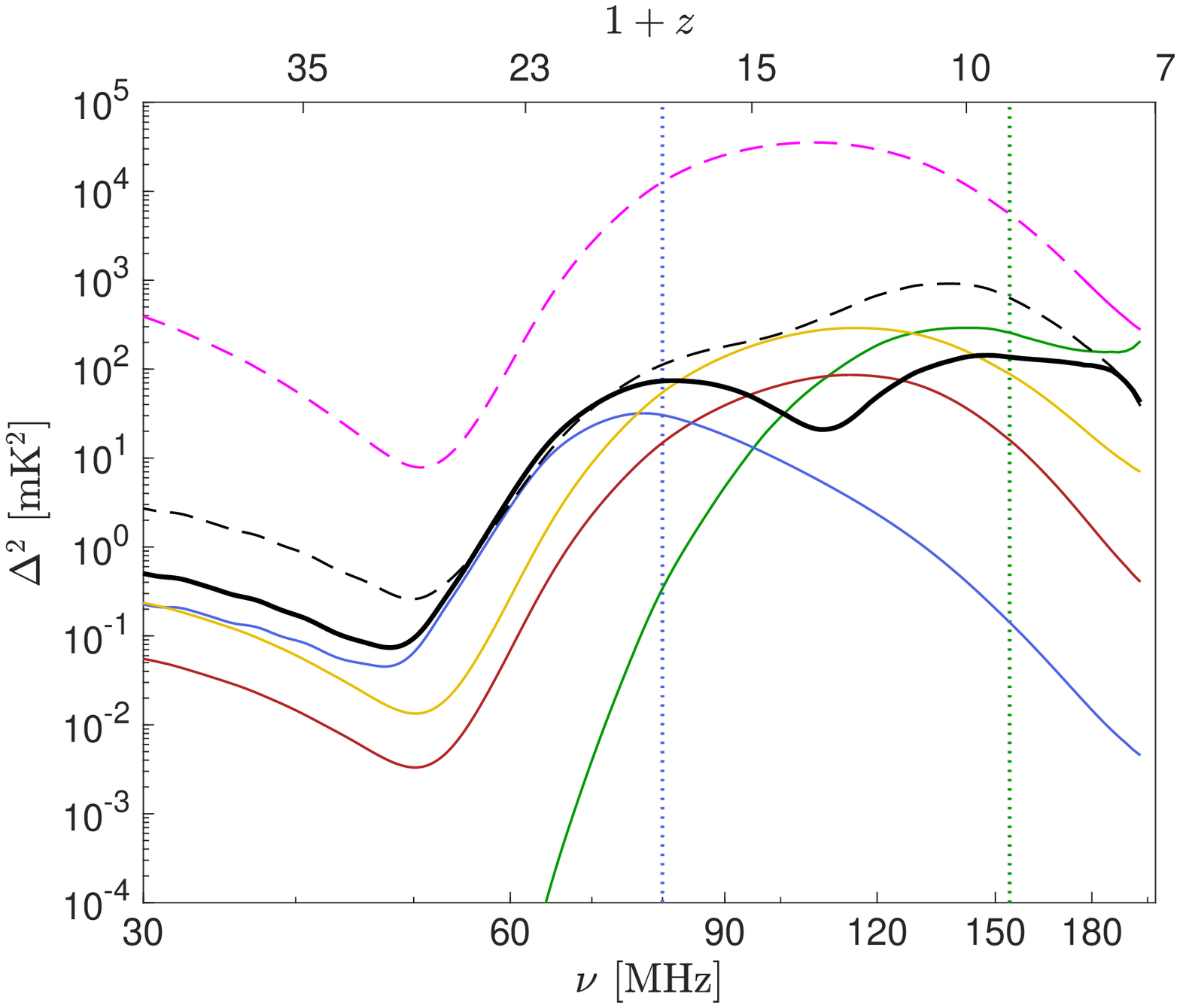}
		\vspace{0.1in}
	\end{subfigure}
	\hspace{0.5in}
	\begin{subfigure}[b]{0.4\textwidth}
		\includegraphics[width=3.1in]{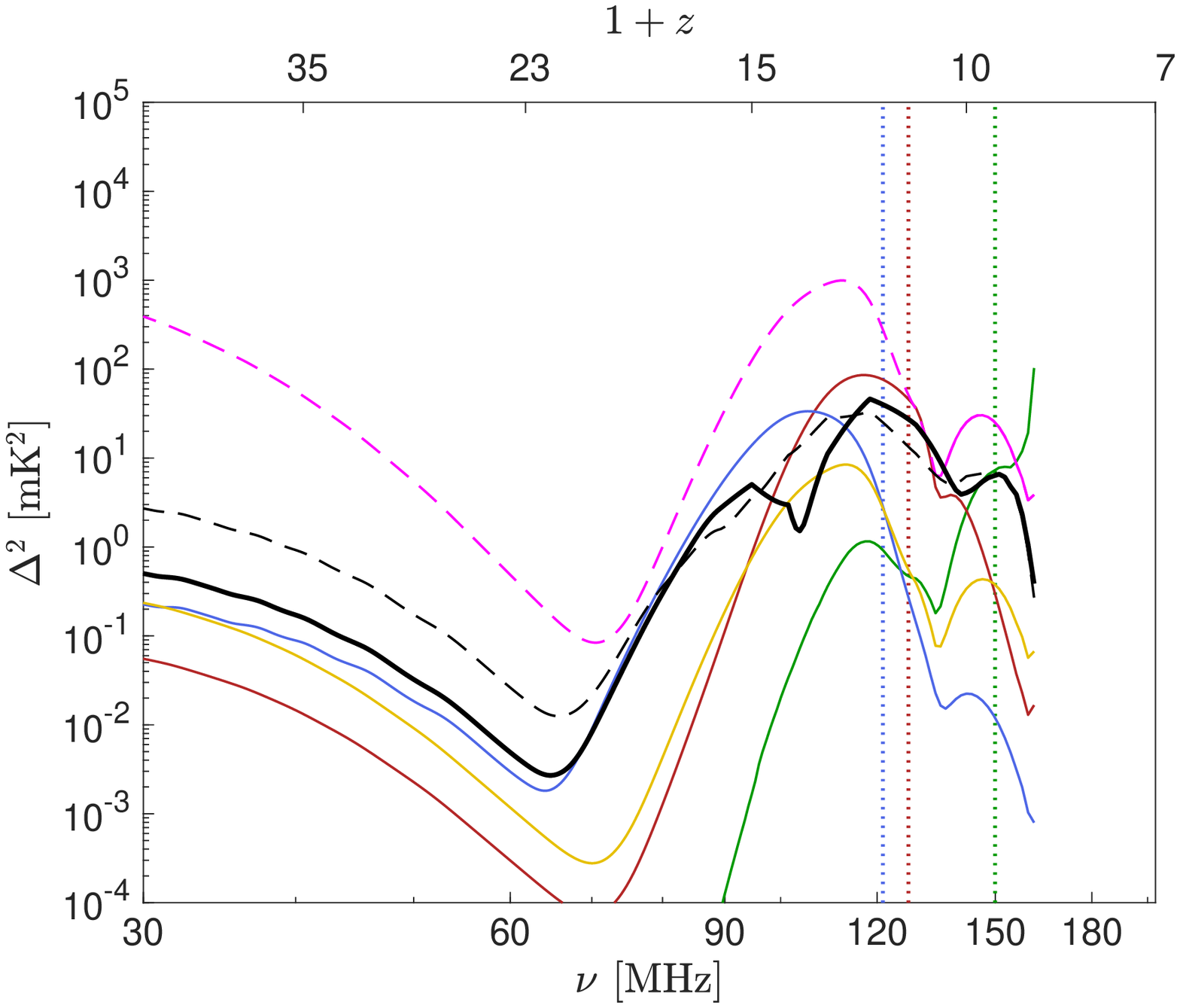}
		\vspace{0.1in}
	\end{subfigure}
	\begin{subfigure}[b]{0.4\textwidth}
		\includegraphics[width=3.1in]{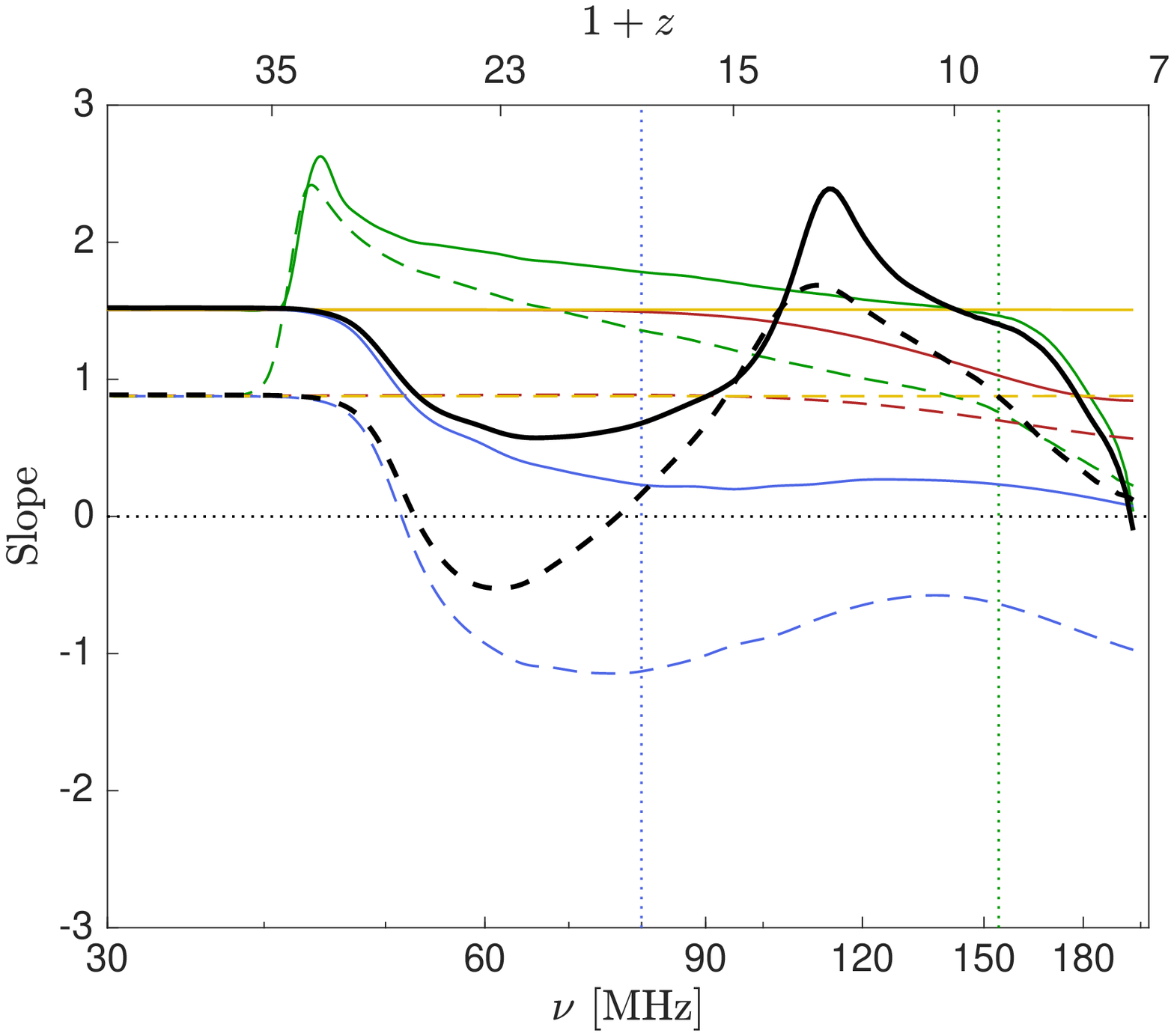}
	\end{subfigure}
	\hspace{0.5in}
	\begin{subfigure}[b]{0.4\textwidth}
		\includegraphics[width=3.1in]{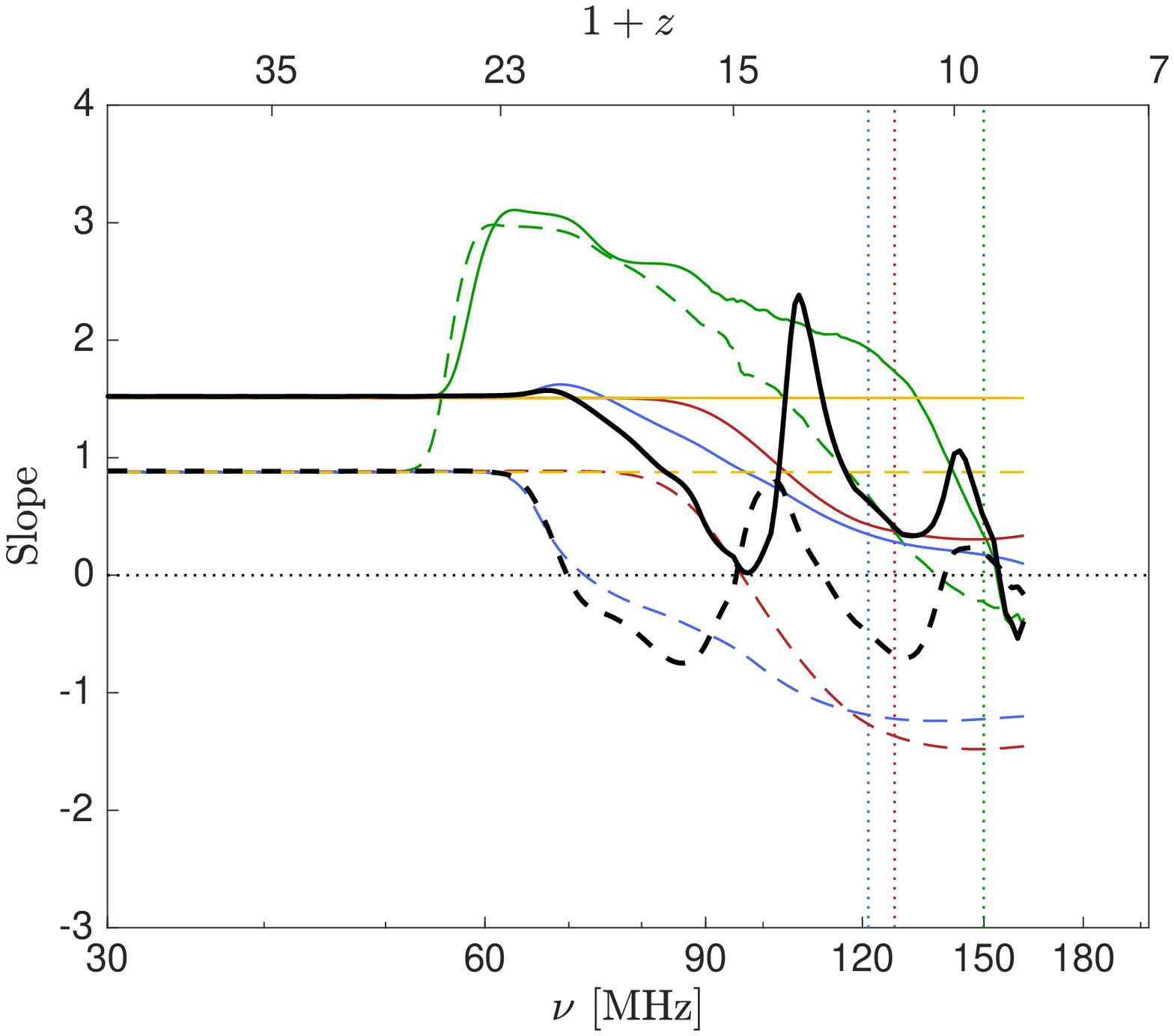}
	\end{subfigure}
	\caption{\label{secondC} Variation of Figures \ref{fig:Cases1}
          and \ref{fig:Slopes1}, same line colors and styles. {\bf
            Left:} A case with one minimum and one maximum in the
          slope, model \#37. {\bf Right:} A case with an extra bump at
          very high redshifts due to late Ly$\alpha$ coupling, model
          \#122.}
	\label{fig:Cases4}
\end{figure*}

Fig.~\ref{fig:Cases4} shows two examples of cases with slopes that
have unusual structure (i.e., different from the typical shape of two
minima and two maxima). The left panel presents a case (\#37) with
very low efficiencies of SFR and heating ($f_*=0.005$ and $f_X=0.1$),
so that there is no heating transition before the end of
reionization. As a result, the high-redshift maximum and low-redshift
minimum of the slope disappear. Note that in this type of case, the
low-redshift maximum occurs when ionization and heating fluctuations
together cancel out density+velocity on large scales. Finally, we show
a case (right panel of Fig.~\ref{fig:Cases4}, \#122) with very weak
star formation (as it is limited to very massive halos, with $V_c =
76.5$ km/s, and even there it has a low efficiency). This case has an
extra bump (i.e., small maximum) at very high redshifts due to late
Ly$\alpha$ coupling.

\end{document}